\documentclass[useAMS,usenatbib,usegraphicx]{mn2e}
\usepackage{psfig}   
\usepackage{graphicx}
\usepackage{subfigure}

\title[Kinematic evolution of star-forming regions]{Dynamical evolution of star forming regions - II. Basic kinematics}

\author[R.~J.~Parker \& N.~J.~Wright]{
  Richard J.~Parker$^{1}$\thanks{E-mail: R.J.Parker@ljmu.ac.uk} and Nicholas J.~Wright$^{2,3}$
  \vspace*{0.1cm}\\
   $^1$ Astrophysics Research Institute, Liverpool John Moores University, 146 Brownlow Hill, Liverpool, L3 5RF, UK \\
   $^2$ Centre for Astrophysics Research, Science and Technology Research Institute, University of Hertfordshire, Hatfield, AL10 9AB, UK\\
   $^3$ Astrophysics Group, Keele University, Keele, Staffordshire, ST5 5BG, UK
}

\begin{document}

\date{}
                             
\pagerange{\pageref{firstpage}--\pageref{lastpage}} \pubyear{2016}

\maketitle

\label{firstpage}

\def\mnras{MNRAS}
\def\apj{ApJ}
\def\aj{AJ}
\def\aap{A\&A}
\def\apjl{ApJL}
\def\apjs{ApJS}
\def\araa{ARA\&A}
\def\pasj{PASJ}

\begin{abstract}
We follow the dynamical evolution of young star-forming regions with a wide range of initial conditions and examine how the radial velocity dispersion, $\sigma$, evolves over time. We compare this velocity dispersion to the theoretically expected value for the velocity dispersion if a region were in virial equilibrium, $\sigma_{\rm vir}$ and thus assess the virial state ($\sigma / \sigma_{\rm vir}$) of these systems. We find that in regions that are initially subvirial, or in global virial equilibrium but subvirial on local scales, the system relaxes to virial equilibrium within several million years, or roughly  25 -- 50 crossing times, according to the measured virial ratio. However, the measured velocity dispersion, $\sigma$, appears to be a bad diagnostic of the current virial state of these systems as it suggests that they become supervirial when compared to the velocity dispersion estimated from the virial mass, $\sigma_{\rm vir}$. We suggest that this discrepancy is caused by the fact that the regions are never fully relaxed, and that the early non-equilibrium evolution is imprinted in the one-dimensional velocity dispersion at these early epochs. If measured early enough ($<$2\,Myr in our simulations, or ~$\sim$20 crossing times), the velocity dispersion can be used to determine whether a region was highly supervirial at birth without the risk of degeneracy. We show that combining $\sigma$, or the ratio of $\sigma$ to the interquartile range (IQR) dispersion, with measures of spatial structure, places stronger constraints on the dynamical history of a region than using the velocity dispersion in isolation. 
\end{abstract}


\section{Introduction}

The physics of the star formation process results in stars grouped together in regions whose density exceeds the mean density of the Galactic disc by several orders of magnitude \citep{Blaauw64,Lada03,Porras03,Bressert10}. Depending on the initial density of the system, and whether or not it is gravitationally bound (subvirial or virialised), a star-forming region will evolve into either a dense star cluster, or a low-density association \citep{Gieles11,Kruijssen12b,Parker14b}.

Determining whether a star-forming region is gravitationally bound or unbound is not just important for understanding whether it will form a long-lived cluster, but also has implications for many other areas of astrophysics. If stars spend a significant amount of time in dense systems then dynamical interactions can play an important role in the disruption of protoplanetary disks and fledgling planetary systems \citep[e.g.][]{Bonnell01b,Hurley02,Adams06,Spurzem09,deJuan12,Olczak12,Parker12a,Craig13,Hao13,Rosotti14,Vincke15} and the processing of primordial binary systems \citep[e.g.][]{Kroupa95a,Kroupa99,Kaczmarek11,Marks11,Parker12b,Geller13a,Leigh13}. The densely clustered environment can also affect the evolution of protoplanetary disks and planetary systems due to the extreme radiation fields from massive stars in such environments \citep{Armitage00,Scally01,Adams04,Adams06,Thompson13}.

However, at early stages ($<$10\,Myr) it is often difficult to determine the past structural and dynamical conditions in a region, and whether it will go on to form a long-lived (open) cluster or an unbound association--like complex. In earlier work \citep{Parker12d,Parker14b,Parker14e,Wright14} we showed that it is possible to distinguish between different initial virial ratios (i.e.\,\,whether a region was bound or unbound to begin with) using spatial information for the entire region, including the relative spatial distribution of the most massive stars compared to lower mass stars. However, certain initial conditions produce degeneracies in the spatial distributions at later times, and so extra diagnostic information is required.

The most obvious next step in this analysis is to include information from the velocity distribution of the stars. This is especially pertinent in the era of the {\it Gaia} astrometric satellite, which will provide proper motion velocities for stars in and around unobscured clusters and associations by the completion of the mission. However, regions that suffer high and/or variable extinction will be missed, and as a result considerable effort is being invested in obtaining radial velocities using complementary ground-based programmes, such as the Gaia--European Southern Observatory Survey \citep[GES,][]{Gilmore12}, the Apache Point Observatory Galactic Evolution Experiment \citep[APOGEE,][]{Zasowski13} and its associated INfrared Spectra of Young Nebulous Clusters (IN-SYNC) survey \citep{Cottaar14b,Foster15}.

These surveys are obtaining radial velocities for hundreds of stars in young star-forming regions with accuracies down to $\sim$0.25\,km\,s$^{-1}$. By collecting large enough samples of radial velocities the velocity dispersion, $\sigma$, can be measured, which can then be compared to the expected velocity dispersion if the region were in virial equilibrium. This allows the virial state of these regions to be assessed and hence whether they will remain as (or form) bound clusters, or disperse as unbound associations that will dissolve into the Galactic field within a few Myr. Early results from these surveys suggest that several regions are consistent with being bound, including the nearby dense region NCG\,1333 \citep{Foster15} and the more diffuse $\rho$~Oph (Rigliaco, in prep.). Furthermore, several regions (Vela OB2) appear to have multiple star formation events with strongly offset radial velocities, and different velocity dispersions between events \citep{Jeffries14,Sacco15}.

Given the wealth of observational data that will soon exist, a parameter-space study of the formation and evolution of star-forming regions investigating how the radial velocity dispersion evolves as a function of initial conditions (density, virial ratio, spatial structure)  and how such measurements can be effectively used seems very timely. In this paper, we use the pure $N$-body simulations from \citet{Parker14b} to determine the evolution of the radial velocity dispersion with time. The paper is organised as follows. In Section~\ref{method} we discuss the initial conditions of the $N$-body simulations, and in Section~\ref{results} we present our results. We provide a discussion in Section~\ref{discuss} and we conclude in Section~\ref{conclude}.

\section{Method}
\label{method}

The star-forming regions we simulate have either 1500 members, which corresponds to a total mass of $\sim 500$ M$_\odot$, or 150 members, 
corresponding to a mass of $\sim 50$ M$_\odot$. All the simulations have a radius of 1\,pc and therefore the latter set of simulations start with a lower spatial density than the simulations with 1500 members. For each set of initial conditions we run an ensemble of 20 simulations, identical apart from the random number seed used to initialise the positions, masses and velocities of the stars. 

Our star-forming regions are set up as fractals; observations of young unevolved star-forming regions indicate a high level of substructure is present \citep*[i.e. they do not have a radially smooth profile, e.g.][and references therein]{Cartwright04,Schmeja08,Sanchez09,Gouliermis14}. The fractal distribution provides a way of creating substructure on all scales. Note that we are not claiming that young star clusters are fractal \citep[although they may be, e.g.][]{Elmegreen01}, but the fractal distribution is a relatively simple method of setting up substructured regions, as the level of substructure is described by just one parameter, the fractal dimension, $D$. In three dimensions, $D = 1.6$ indicates a highly substructured distribution, and $D = 3.0$ is a roughly uniform sphere.

We set up the fractals according to the method in \citet{Goodwin04a}. This begins by defining a cube of side $N_{\rm div}$ (we adopt $N_{\rm div} = 2.0$ 
throughout), inside of which the fractal is built. A first-generation parent is placed at the centre of the cube, which then spawns $N_{\rm div}^3$ subcubes, each containing a first generation child at its centre. The fractal is then built by determining which of the children themselves become parents, and spawn their own offspring. This is determined by the fractal dimension, $D$, where the probability that the child becomes a parent is given by $N_{\rm div}^{(D - 3)}$. For a lower fractal dimension fewer children mature and the final distribution contains more substructure. Any children that do not become parents in a given step are removed, along with all of their parents. A small amount of noise is then added to the positions of the remaining children, preventing the cluster from having a gridded appearance and the children become parents of the next generation. Each new parent then spawns $N_{\rm div}^3$ second-generation children in $N_{\rm div}^3$ sub-subcubes, with each second-generation child having a $N_{\rm div}^{(D - 3)}$ probability of becoming a second generation parent. This process is repeated until there are substantially more children than required. The children are pruned to produce a sphere from the cube and are then randomly removed (so maintaining the fractal dimension) until the required number of children is left. These children then become stars in the cluster. 

To determine the velocity structure of the cloud, children inherit their parent's velocity plus a random component that decreases with each generation of the fractal where the random component scales with $N_{\rm div}^{(D - 3)}$, as for the spatial distribution.  The children of the first generation are given random velocities from a Gaussian of mean zero. Each new generation inherits their parent's velocity plus an extra random component that becomes smaller with each generation. This results in a velocity structure in which nearby stars have similar velocities, but distant stars can have very different velocities. The velocity of every star is scaled to obtain the desired virial ratio of the star-forming region. 

We vary the initial virial ratio, $\alpha_{\rm vir} = T/|\Omega|$, where $T$ and $|\Omega|$ are the total kinetic energy and total potential energy of the stars, respectively. A star-forming region is in virial equilibrium if $\alpha_{\rm vir} = 0.5$. We adopt three different virial ratios for our regions; $\alpha_{\rm vir} = 0.3$ (subvirial, or `cool'), $\alpha_{\rm vir} = 0.5$ (virialised, or `tepid') and $\alpha_{\rm vir} = 1.5$ (supervirial, or `hot').

By construction, the local virial ratio can often be very different to the global virial ratio. Because velocities are correlated on local scales, the local virial ratio is usually subvirial, whereas the global ratio can be supervirial (if so defined in the initial conditions). The set-up facilitates violent relaxation and the rapid dynamical evolution of the substructure \citep[see][]{Allison10}. The correlation of velocities on local scales is partly informed by the \citet{Larson81} relation for the dependence of velocity dispersion $\sigma$ (in km\,s$^{-1}$) on the size $L$ (in pc) of molecular clouds
\begin{equation}
\sigma = 1.1L^{0.38}, 
\end{equation}
although the fractals do not follow the exact relation. Removing this velocity correlation on local scales generally results in the faster removal of substructure \citep[see fig.~9 in][]{Parker14b} and likely inhibits dynamical mass segregation \citep{Allison10}. In future papers, we will explore different initial velocity distributions and their influence on the dynamical evolution of star-forming regions in more detail. 

The regions are set up with fractal dimensions of $D = 1.6$ (very clumpy), $D = 2.0$ and $D = 3.0$ (a roughly uniform sphere), in order to investigate 
the full parameter space. 

The regions contain 1500 or 150 stars each and have initial radii of 1\,pc with no primordial binaries or gas potential. We draw stellar masses from the recent probability density function for the Initial Mass Function (IMF) by \citet{Maschberger13} which has the form:
\begin{equation}
p(m) \propto \left(\frac{m}{\mu}\right)^{-\alpha}\left(1 + \left(\frac{m}{\mu}\right)^{1 - \alpha}\right)^{-\beta}
\label{imf}.
\end{equation}
Here, $\mu = 0.2$\,M$_\odot$ is the average stellar mass, $\alpha = 2.3$ is the \citet{Salpeter55} power-law exponent for higher mass stars, and $\beta = 1.4$ is used to describe the slope of the 
IMF for low-mass objects \citep*[which also deviates from the log-normal form;][]{Bastian10}. Finally, we sample from this IMF within the mass range $m_{\rm low} = 0.01$\,M$_\odot$ to $m_{\rm up} = 50$\,M$_\odot$.

Each version of the initial conditions is run 20 times, with the random number seed changed each time to gauge the effects of stochasticity in the evolution of the star-forming regions. We run the simulations for 10\,Myr using the \texttt{kira} integrator in the Starlab package \citep{Zwart99,Zwart01}. We do not include stellar evolution in the simulations. A summary of the simulation parameter space is given in Table~\ref{cluster_setup}. Examples of the initial spatial distributions adopted for these simulations, and their subsequent appearance following dynamical evolution are presented in \citet{Parker11c,Parker12d,Parker14a,Parker14b} and we refer the interested reader to those papers for further information.

\begin{table}
\caption[bf]{A summary of the different star-forming region properties adopted for the simulations.
The values in the columns are: the number of stars in each region ($N_{\rm stars}$), 
the typical mass of this region ($M_{\rm region}$),  the initial virial ratio of the region ($\alpha_{\rm vir}$), and the initial fractal dimension ($D$).}
\begin{center}
\begin{tabular}{|c|c|c|c|}
\hline 
$N_{\rm stars}$ & $M_{\rm region}$  &  $\alpha_{\rm vir}$ & $D$ \\
\hline
1500 & $\sim 500$\,M$_\odot$ & 0.3  & 1.6 \\
1500 & $\sim 500$\,M$_\odot$ & 0.3  & 2.0 \\
1500 & $\sim 500$\,M$_\odot$ & 0.3  & 3.0 \\
\hline 
1500 & $\sim 500$\,M$_\odot$ & 0.5  & 1.6 \\
1500 & $\sim 500$\,M$_\odot$ & 0.5  & 2.0 \\
1500 & $\sim 500$\,M$_\odot$ & 0.5  & 3.0 \\
\hline
1500 & $\sim 500$\,M$_\odot$ & 1.5  & 1.6 \\
1500 & $\sim 500$\,M$_\odot$ & 1.5  & 2.0  \\
1500 & $\sim 500$\,M$_\odot$ & 1.5  & 3.0 \\
\hline
150 & $\sim 50$\,M$_\odot$ & 0.3  & 1.6 \\
150 & $\sim 50$\,M$_\odot$ & 0.3  & 2.0 \\
150 & $\sim 50$\,M$_\odot$ & 0.3  & 3.0 \\
\hline 
150 & $\sim 50$\,M$_\odot$ & 0.5  & 1.6 \\
150 & $\sim 50$\,M$_\odot$ & 0.5  & 2.0 \\
150 & $\sim 50$\,M$_\odot$ & 0.5  & 3.0 \\
\hline
150 & $\sim 50$\,M$_\odot$ & 1.5  & 1.6 \\
150 & $\sim 50$\,M$_\odot$ & 1.5  & 2.0  \\
150 & $\sim 50$\,M$_\odot$ & 1.5  & 3.0 \\
\hline
\end{tabular}
\end{center}
\label{cluster_setup}
\end{table}
 
\subsection{Comparison of velocity dispersions}
\label{comparison}

\begin{figure*}
  \begin{center}
\setlength{\subfigcapskip}{10pt}
\hspace*{-1.cm}\subfigure[]{\label{cool-a}\rotatebox{270}{\includegraphics[scale=0.28]{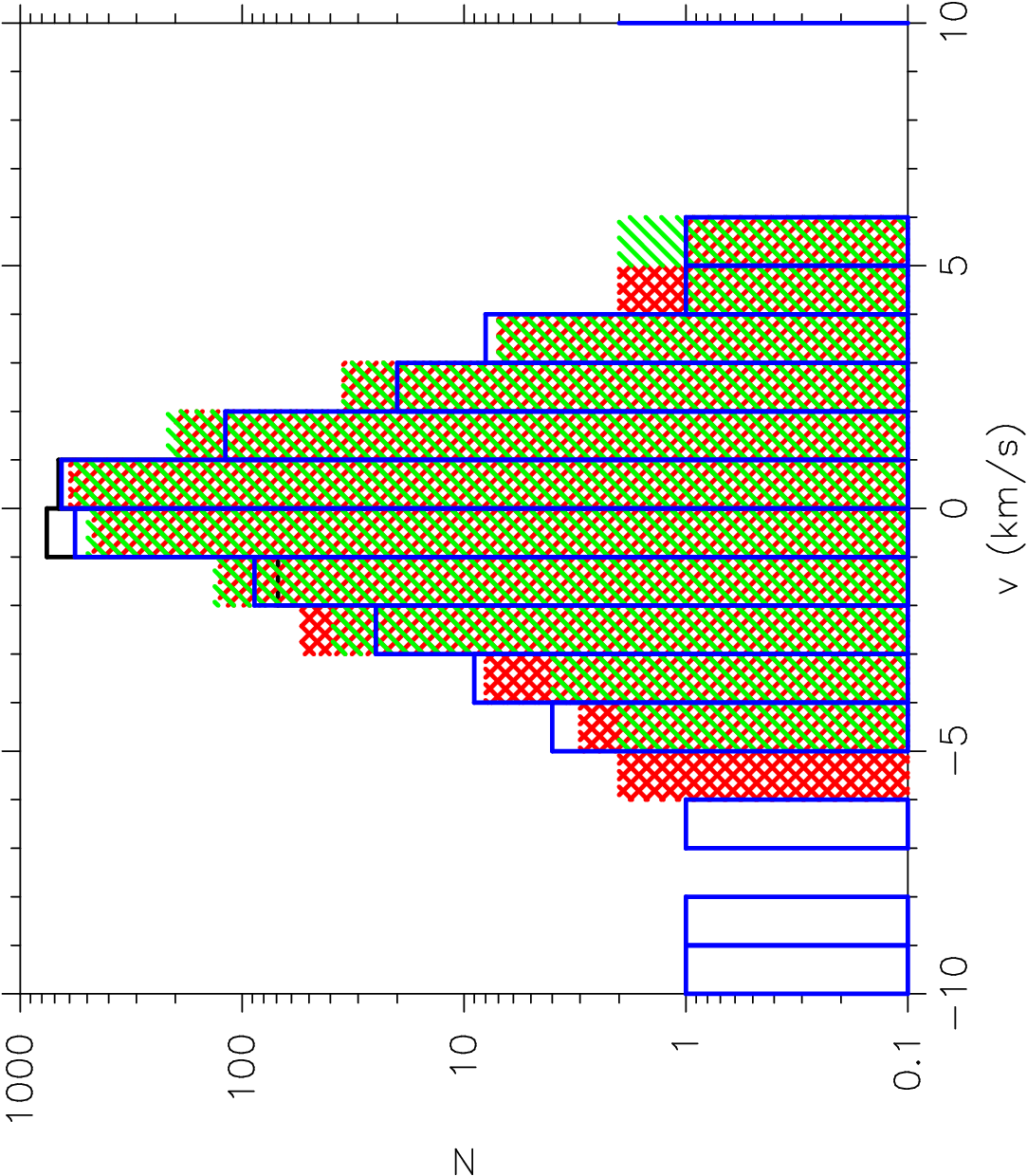}}} 
\hspace*{0.3cm} 
\subfigure[]{\label{cool-b}\rotatebox{270}{\includegraphics[scale=0.28]{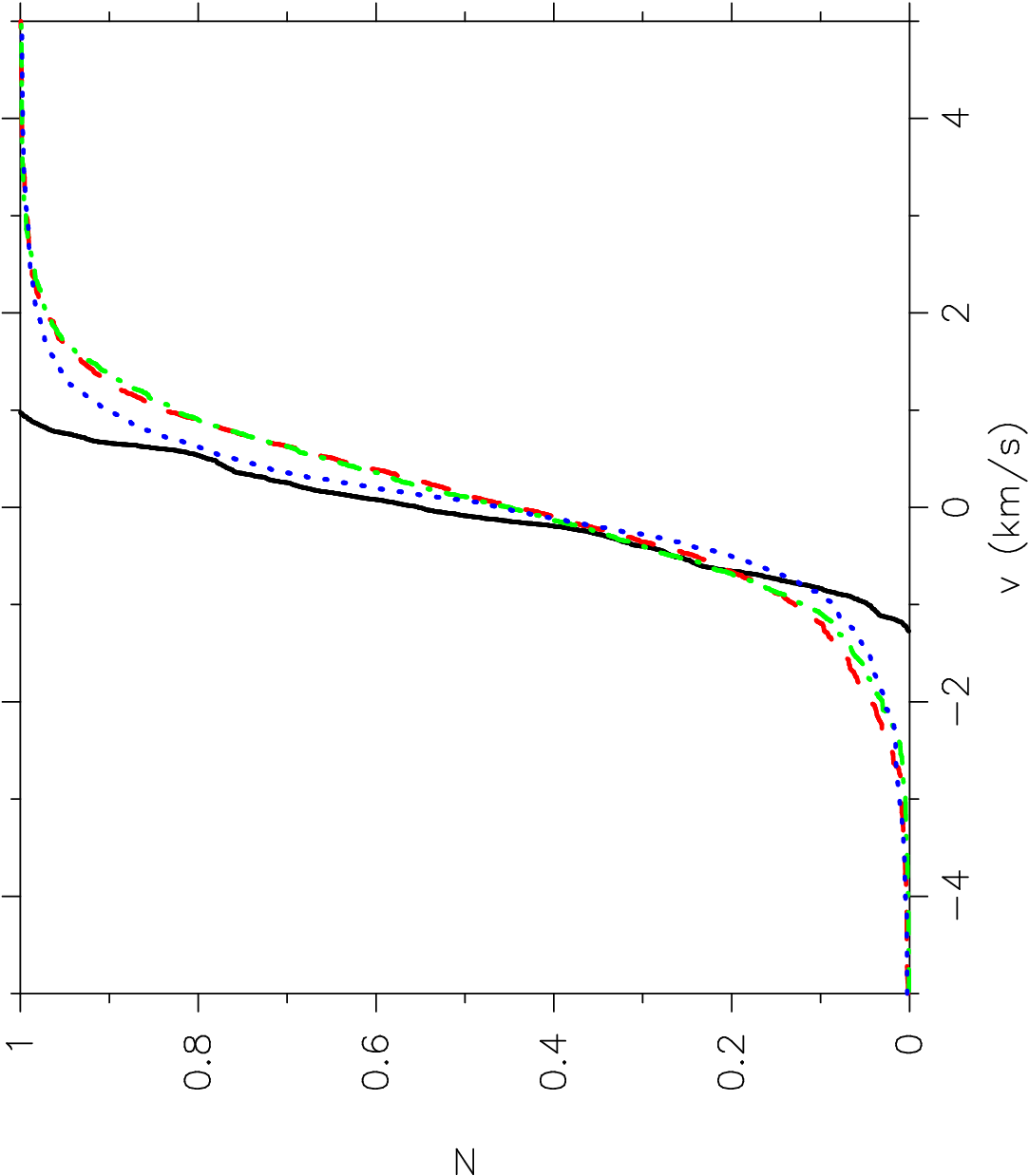}}} 
\hspace*{0.3cm} 
\subfigure[]{\label{cool-c}\rotatebox{270}{\includegraphics[scale=0.28]{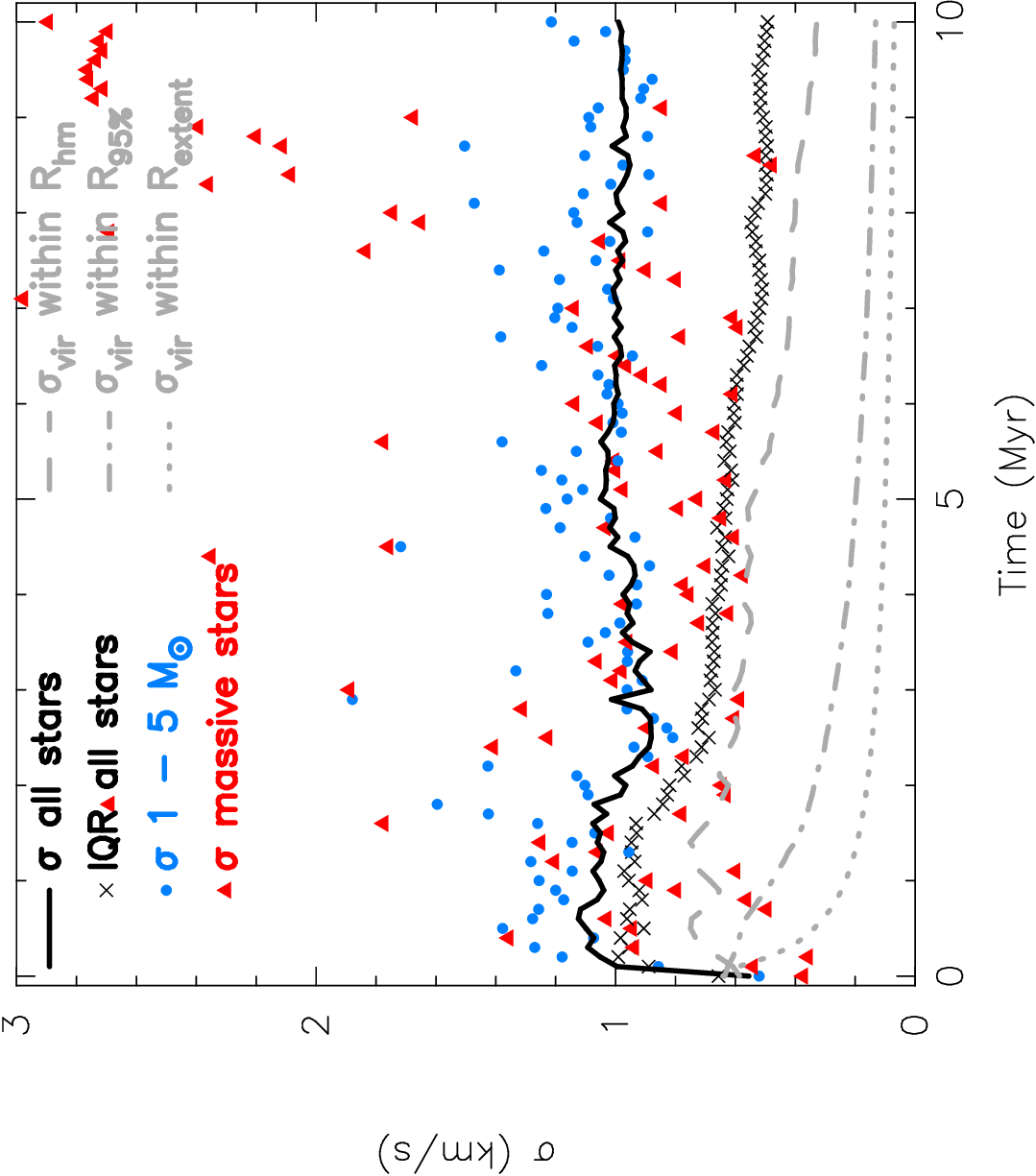}}}
\caption[bf]{Evolution of the velocity dispersion in a subvirial ($\alpha_{\rm vir} = 0.3$), substructured ($D = 1.6$) star-forming region that collapses to form a cluster. In panel (a) we show a raw histogram of radial velocities for all stars at 0\,Myr (the black open histogram), 0.5\,Myr (red cross-hatched histogram), 1\,Myr (green hatched histogram) and at 5\,Myr (blue open histogram). In panel (b) we show the corresponding cumulative distributions; the initial (0\,Myr) distribution is shown by the solid (black) line, the distribution at 0.5\,Myr is shown by the red dashed line, the distribution at 1\,Myr is shown by the dot-dashed green line and the distribution at 5\,Myr is shown by the dotted blue line. Note the different x-axis scale between panels (a) and (b). In panel (c) we show the evolution of this velocity dispersion for all stars by the solid black line, the velocity dispersion for stars with masses 1 -- 5\,M$_\odot$ by the blue circles, and the velocity dispersion for the ten most massive stars (usually $m > 5$\,M$_\odot$) by the red triangles. The black crosses show the velocity dispersion calculated from the IQR for all stars. Finally, we show the maximum velocity dispersion for the region to still be in virial equilibrium by the broken grey lines; the dashed line is for objects within the half-mass radius, the dot-dashed line is for all stars within 95\,per cent of the extent of the region, and the dotted line is for the full region in the simulation (i.e.\,\,all stars in the simulations, including those that have been dynamically ejected).}
\label{cool_indiv}
  \end{center}
\end{figure*}

At each snapshot in the simulation, we compare the dispersion of the z-component of the velocity vector (i.e.\,\,the radial velocity) to the virial mass velocity dispersion, i.e.\,\,the velocity dispersion for the region to be in virial equilibrium. The total mass, $M$ is related to the mean three-dimensional velocity dispersion $\sigma$ thus \citep[e.g.][and references therein]{Fleck06}: 
\begin{equation}
M = \frac{r_g\sigma^2}{G},
\label{fleck}
\end{equation}
where $G$ is the gravitational constant and $r_g$ is the radius within which the mass in enclosed. The three dimensional velocity dispersion is related to the line-of-sight radial velocity dispersion, $\sigma_{\rm los}$ by
\begin{equation}
\sigma^2 \simeq 3\sigma_{\rm los}^2.
\label{3ds}
\end{equation}
The adopted radius, $r_g$, depends on the density distribution of the region. For a smooth, centrally concentrated cluster various authors \citep{Spitzer87,McCrady03} find
\begin{equation}
r_g \sim \frac{5}{2} \times \frac{4}{3}R_{\rm hl},
\label{rg}
\end{equation}
where $R_{\rm hl}$ is the half-light radius. Substituting Eqns.~\ref{rg}~and~\ref{3ds} into Eqn.~\ref{fleck} gives:
\begin{equation}
M = \frac{10\sigma_{\rm los}^2R_{\rm hl}}{G},
\end{equation}
where `10' is the structure parameter, $\eta$. Depending on the assumed geometry, $\eta$ can be anywhere between 1 and 12 \citep*{Elson87,Zwart10}. For example, a smooth, centrally concentrated Plummer sphere \citep{Plummer11} has $\eta = 10$. If we assume the `cluster radius', $R$, is $2 \times R_{\rm hl}$, then the virial radial velocity dispersion within $R$ for a mass $M$ is 
\begin{equation}
\sigma_{\rm vir} = \sqrt{\frac{2GM}{\eta R}}.
\label{virial_mass}
\end{equation}
In order to compare the velocity dispersion in our simulations to that from the virial mass estimate, we must make choices for $M$, $R$ and $\eta$. In the following section we will present different estimates of $\sigma_{\rm vir}$ for a variety of enclosed masses (and hence radii). For the remainder of the paper, we assume that $\eta = 10$ because the initially (sub)virial simulations rapidly lose substructure and evolve to a smooth, centrally concentrated profile \citep{Parker14b}. However, it is worth bearing in mind that -- especially at early stages in the simulations -- $\eta$ may be as low as unity. For regions that retain spatial (and hence kinematic) substructure such as those that are initially supervirial, the determination of $\eta$ is more problematic. However, as we will see, the initially supervirial simulations always remain supervirial, and so $\sigma_{\rm vir}$ does not need to be as stringently defined.

\section{Results}
\label{results}

In this section we first examine the evolution of the velocity dispersion in typical examples of a substructured, subvirial (cool) star-forming region and a substructured, supervirial (warm) star-forming region, before comparing the average evolution of all of the regions in our chosen parameter space. 

\subsection{Evolution of a substructured, subvirial star-forming region}

In Fig.~\ref{cool_indiv} we show the evolution of a `typical' subvirial ($\alpha_{\rm vir} = 0.3$), substructured ($D = 1.6$) star-forming region, which undergoes cool collapse to form a cluster \citep{Parker14b}. In panel (a) we show the histogram of radial velocities (i.e.\,\,the $z$-component of the velocity vector for each star in the simulation) at various times. %The black open histogram is the distribution at 0\,Myr, the red cross-hatched histogram is the distribution at 0.5\,Myr, the green hatched histogram is at 1\,Myr and the blue open histogram is at 5\,Myr. In panel (b) we show the cumulative distributions of these data; the solid black line is the initial (0\,Myr distribution), the  dashed red line is the distribution at 0.5\,Myr, the dot-dashed green line is the distribution at 1\,Myr and the dotted blue line is the distribution at 5\,Myr. 
The general trend is for the distribution to widen within the first 0.5\,Myr as dynamical interactions inflate the distribution, but then the first ejections of stars leads to several outliers with high velocities and the distribution deflates slightly, and continues to do so past 1\,Myr (the green dot-dashed and blue dotted lines).

In panel (c) we show the evolution of the radial velocity dispersion (the statistical dispersion about the mean) for all stars in this typical simulation by the solid black line. The initial dispersion is $\sigma = 0.55$\,km\,s$^{-1}$, but as the region evolves and collapses to form a cluster, the velocity distribution inflates and so the dispersion increases to $\sigma = 1.1$\,km\,s$^{-1}$. In order to determine whether the cluster is bound at a given epoch, we compare this velocity dispersion to the virial mass velocity dispersion (Eq.~\ref{virial_mass} in Section~\ref{comparison}). In Fig.~\ref{cool-c} we show this for several different definitions of the region mass/radius. %The dotted grey line is for the entire extent of the region (i.e.\,\,all stars in the simulations, including those that have been dynamically ejected). The dot-dashed line is the virial velocity dispersion estimate for the mass of the most central 95\,per cent of stars in the region, and the dashed line is the virial velocity dispersion estimate within the half-mass radius.

\begin{figure}
\begin{center}
\rotatebox{270}{\includegraphics[scale=0.35]{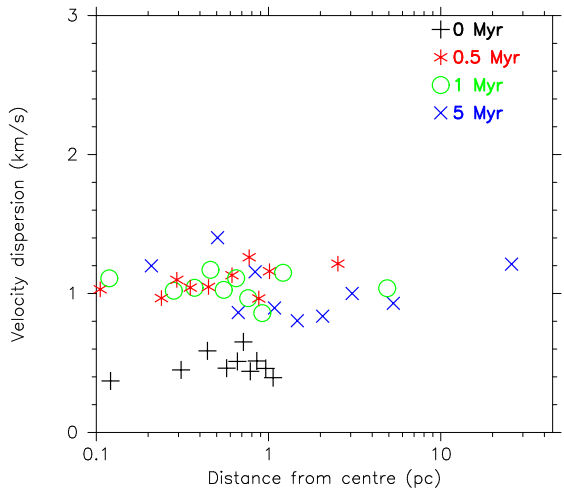}}
\end{center}
\caption[bf]{Velocity dispersion, $\sigma$, as a function of radius for bins containing a fixed number of stars (150 in each bin), for the subvirial region shown in Fig.~\ref{cool_indiv}. The black plus signs show the initial distribution (0\,Myr), the red asterisks are for 0.5\,Myr, the green circles are for 1\,Myr and the blue crosses are for 5\,Myr.}
\label{radial_sigma}
\end{figure}

In theory, the velocity dispersion may change as a function of distance from the cluster centre. However, in all of our simulations we searched for radial dependence and found no significant trends. Fig.~\ref{radial_sigma} shows the velocity dispersion as a function of distance from the centre-of-mass of the simulation at 0\,Myr (black plus signs), 0.5\,Myr (red asterisks), 1\,Myr (green circles) and 5\,Myr (blue crosses) for the simulation shown in Fig.~\ref{cool_indiv}. The data are binned so that 150 stars are contained in each bin. Therefore, comparing the velocity dispersion of all stars to estimates of the virial mass velocity dispersion within various radii is valid in our simulations. We note however the recent result by \citet{Rathborne15}, who find an increase in velocity dispersion (and therefore an increase in virial ratio) as a function of radius in the G0.253+0.016 molecular clump. If this is a general result, and the velocity dispersion of molecular gas directly maps to the velocity dispersion of stars that subsequently form, this is an important constraint on the initial conditions of future $N$-body simulations.

If we compare the dispersion for all stars in Fig.~\ref{cool-c} (the solid black line) to the virial mass estimates, we note that the region starts with a slightly lower velocity dispersion than any of the three virial estimates ($\sigma = 0.55$\,km\,s$^{-1}$ versus $\sigma_{\rm vir} = 0.6$\,km\,s$^{-1}$). We note here that $\sigma_{\rm vir}$ has been calculated assuming $\eta = 10$, whereas $\eta$ is unlikely to be this high for a substructured distribution, meaning that the virial velocity dispersion will actually be slightly higher at this point. During the collapse  of the region to form a star cluster, and its subsequent expansion the velocity dispersion increases and the estimate of $\sigma_{\rm vir}$ implies that the cluster is supervirial after 1\,Myr ($\sigma > \sigma_{\rm vir}$). At this stage the region has attained a centrally concentrated and clustered distribution \citep[see fig.~1 of][for the spatial distribution of stars in this simulation]{Parker14b} and therefore $\eta = 10$ is a reasonable approximation to the true value. The implications of this discrepancy between the observed virial state and the actual virial state are important for the interpretation of observations, and are discussed further in Section~4.

We also examine the velocity dispersion for the 10 most massive stars, shown by the red triangle symbols in Fig.~\ref{cool-c}, and the stars with masses in the range 1 -- 5\,M$_\odot$ by the blue circles. In this simulation, there appears to be no statistically significant difference in the velocities of the most massive stars, and the intermediate mass stars, as a function of these subvirial, substructured initial conditions. 

Finally, instead of using the statistical dispersion to estimate the velocity dispersion, we show the velocity dispersion derived from the  (outlier-resistant) interquartile range (IQR), for all stars by the black crosses. The IQR velocity dispersion is given by 
\begin{equation}
IQR = 0.741 \, (q_{75} - q_{25}),
\end{equation}
 where $q_{25}$ and $q_{75}$ are the 25th and 75th percentiles of the velocity distributions, and 0.741 normalises the IQR to the same scale as the velocity dispersion given by $\sigma$. Interestingly, the IQR velocity dispersion is similar to the statistical dispersion for the first 3 -- 4 \,Myr, before decreasing below the statistical dispersion. This is due to the ejection of stars at high velocity from the centre of the cluster as the cluster evolves. These ejected stars lead to high velocity tails in the velocity distribution that causes the statistical dispersion to be larger than the IQR dispersion.

\subsection{Evolution of a substructured, supervirial star-forming region}

\begin{figure*}
  \begin{center}
\setlength{\subfigcapskip}{10pt}
\hspace*{-1.cm}\subfigure[]{\label{hot-a}\rotatebox{270}{\includegraphics[scale=0.28]{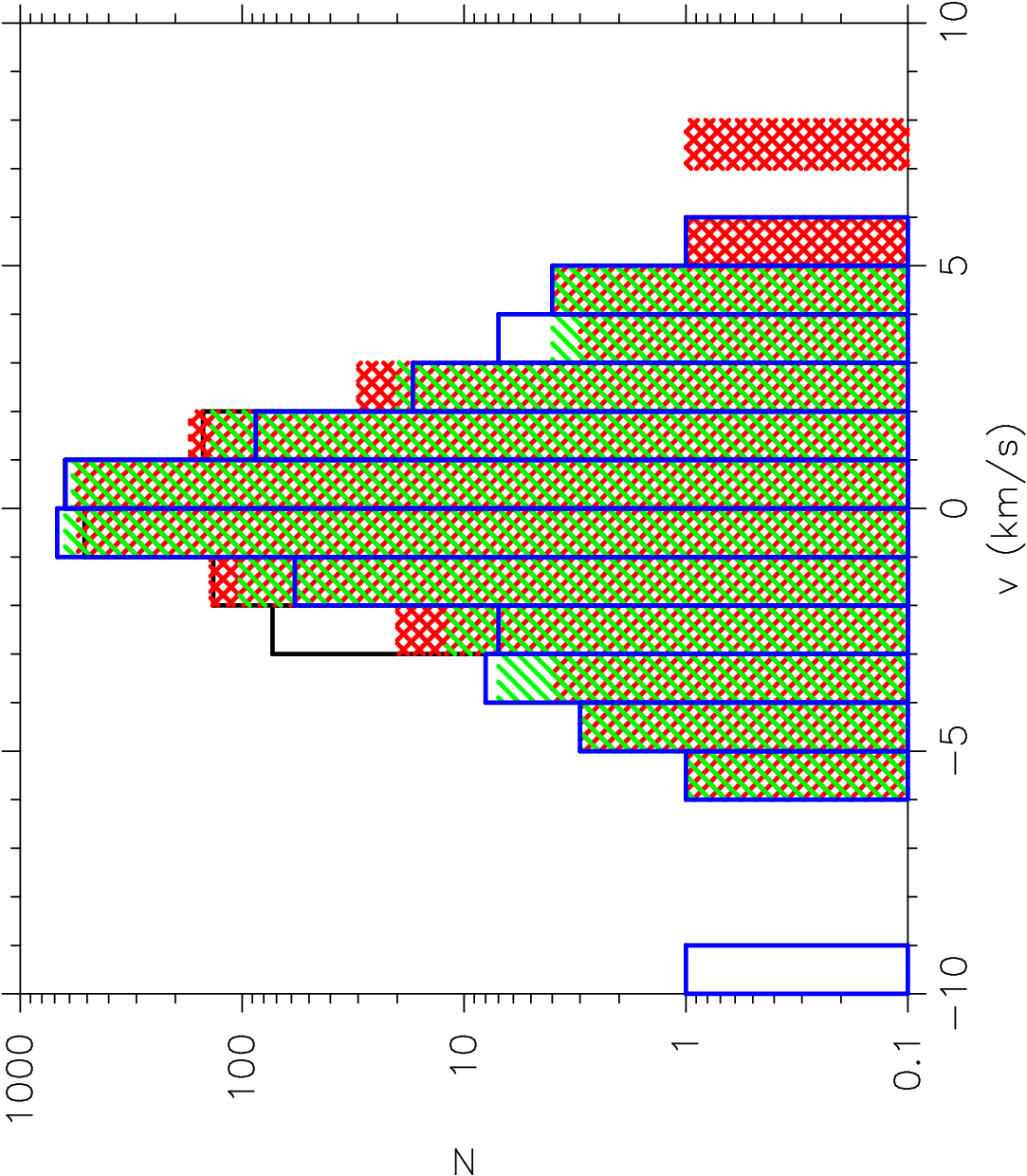}}} 
\hspace*{0.3cm} 
\subfigure[]{\label{hot-b}\rotatebox{270}{\includegraphics[scale=0.28]{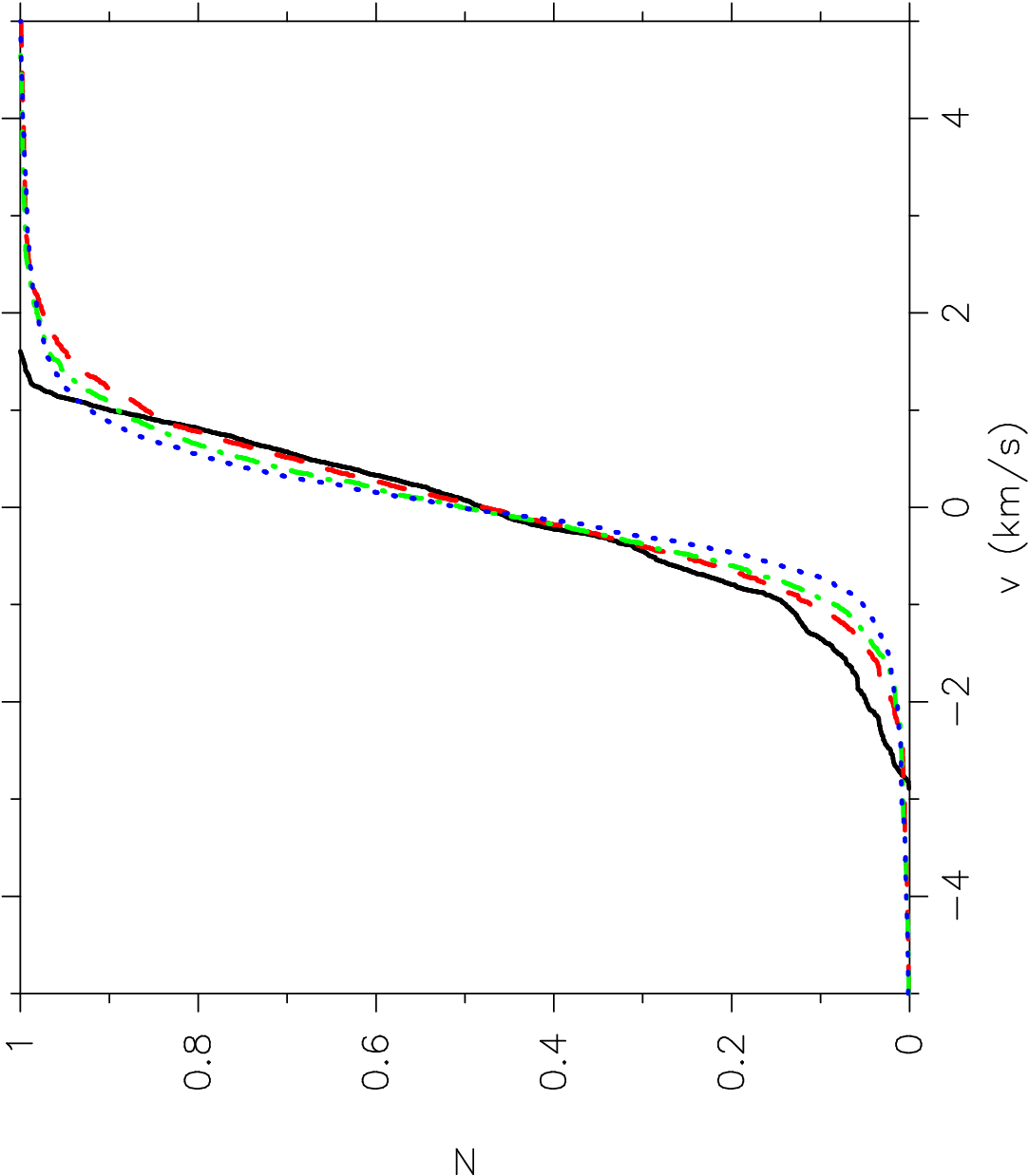}}} 
\hspace*{0.3cm} 
\subfigure[]{\label{hot-c}\rotatebox{270}{\includegraphics[scale=0.28]{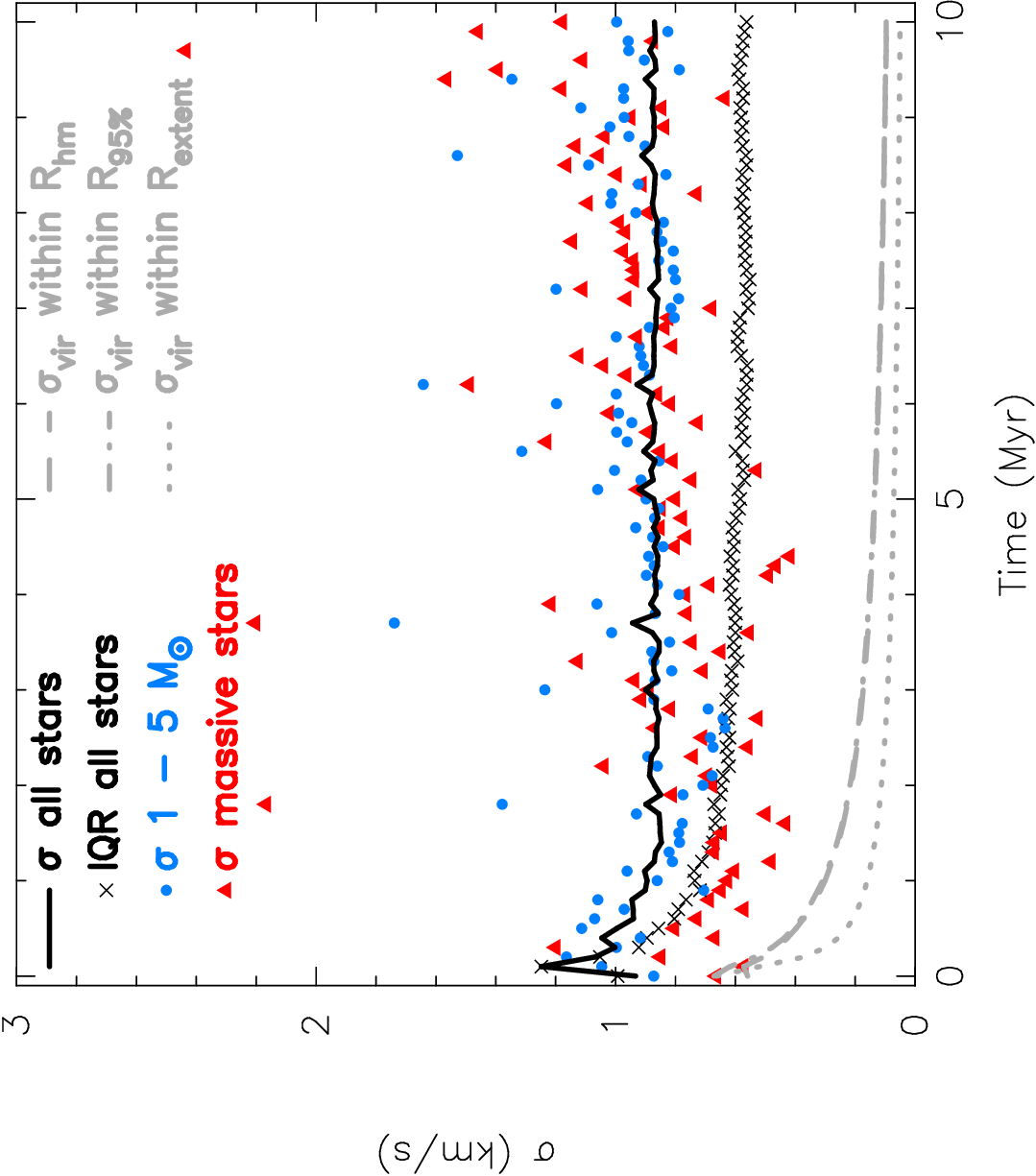}}}
\caption[bf]{Evolution of the velocity dispersion in a supervirial ($\alpha_{\rm vir} = 1.5$), substructured ($D = 1.6$) star-forming region that expands to form an association. In panel (a) we show the raw histogram for the radial velocities of all stars at 0\,Myr (black open histogram), 0.5\,Myr (red cross-hatched histogram), 1\,Myr (green hatched histogram) and at 5\,Myr (blue open histogram). In panel (b) we show the corresponding cumulative distributions; the initial (0\,Myr) distribution is shown by the solid (black) line, the distribution at 0.5\,Myr is shown by the red dashed line, the distribution at 1\,Myr is shown by the dot-dashed green line and the distribution at 5\,Myr is shown by the dotted blue line. Note the different x-axis scale between panels (a) and (b). In panel (c) we show the evolution of this velocity dispersion for all stars by the solid black line, the velocity dispersion for stars with masses 1 -- 5\,M$_\odot$ by the blue circles, and the velocity dispersion for the ten most massive stars (usually $m > 5$\,M$_\odot$) by the red triangles. The black crosses show the velocity dispersion calculated from the IQR for all stars. Finally, we show the maximum velocity dispersion for the region to still be in virial equilibrium by the broken grey lines; the dashed line is for objects within the half-mass radius, the dot-dashed line is for all stars within 95\,per cent of the extent of the region, and the dotted line is for the full region in the simulation (the half-mass and 95\,per cent lines are virtually overlaid).}
\label{hot_indiv}
  \end{center}
\end{figure*}

In Fig.~\ref{hot_indiv} we show the evolution of a `typical' supervirial ($\alpha_{\rm vir} = 1.5$), substructured ($D = 1.6$) region which undergoes warm expansion to form an association \citep{Parker14b}. In panel (a) we show a histogram of radial velocities (i.e.\,\, the $z$-component of the velocity vector for each star in the simulation) at various  times. %The black open histogram is the distribution at 0\,Myr, the red cross-hatched histogram is the distribution at 0.5\,Myr, the green hatched histogram is at 1\,Myr and the blue open histogram is at 5\,Myr. In panel (b) we show the cumulative distributions of these data; the solid black line is the initial (0\,Myr) distribution, the  dashed red line is the distribution at 0.5\,Myr, the dot-dashed green line is the distribution at 1\,Myr and the dotted blue line is the distribution at 5\,Myr.  
As the star-forming region expands the region relaxes slightly and the distribution after 5\,Myr (the blue dotted line) has narrowed considerably when compared with the initial distribution (the solid black line).

In panel (c) we show the evolution of the velocity dispersion for all stars in this typical simulation by the solid black line. The initial velocity dispersion rises slightly due to relaxation in the substructure before decreasing as the region expands and the stars start to slow down. After 0.5\,Myr the velocity dispersion of the stars is higher than any estimate of the virial mass velocity dispersion, $\sigma_{\rm vir}$ (the broken grey lines in panel (c)), showing that the region remains supervirial for its entire evolution. 

We show the evolution of the velocity dispersion for the 10 most massive stars by the red triangle symbols, and the dispersion for stars with masses 1 -- 5\,M$_\odot$ by the blue circles. Both subsets attain very similar (albeit more noisy) velocity dispersions to the velocity dispersion for all stars. The IQR velocity dispersion is lower than the statistical dispersion in these initial conditions, which we again attribute to the early ejection of stars in the dense substructure.

\subsection{Evolution of all regions}
\label{results:all}

The star-forming regions presented in Figs.~\ref{cool_indiv}~and~\ref{hot_indiv} were chosen as `typical' examples from each suite of 20 (initially) identical simulations. However, the dynamical evolution of regions can be highly stochastic \citet{Allison10,Parker12b,Parker14b}. In this section we will present the evolution of the velocity dispersions in either 10, or all 20 realisations of the same initial conditions. 

\subsubsection{Evolution of the absolute velocity dispersion}

\begin{figure*}
  \begin{center}
\setlength{\subfigcapskip}{10pt}
\vspace*{-0.3cm}
\hspace*{-1.cm}\subfigure[$\alpha_{\rm vir} = 0.3, D = 1.6$]{\label{vd_lines-a}\rotatebox{270}{\includegraphics[scale=0.27]{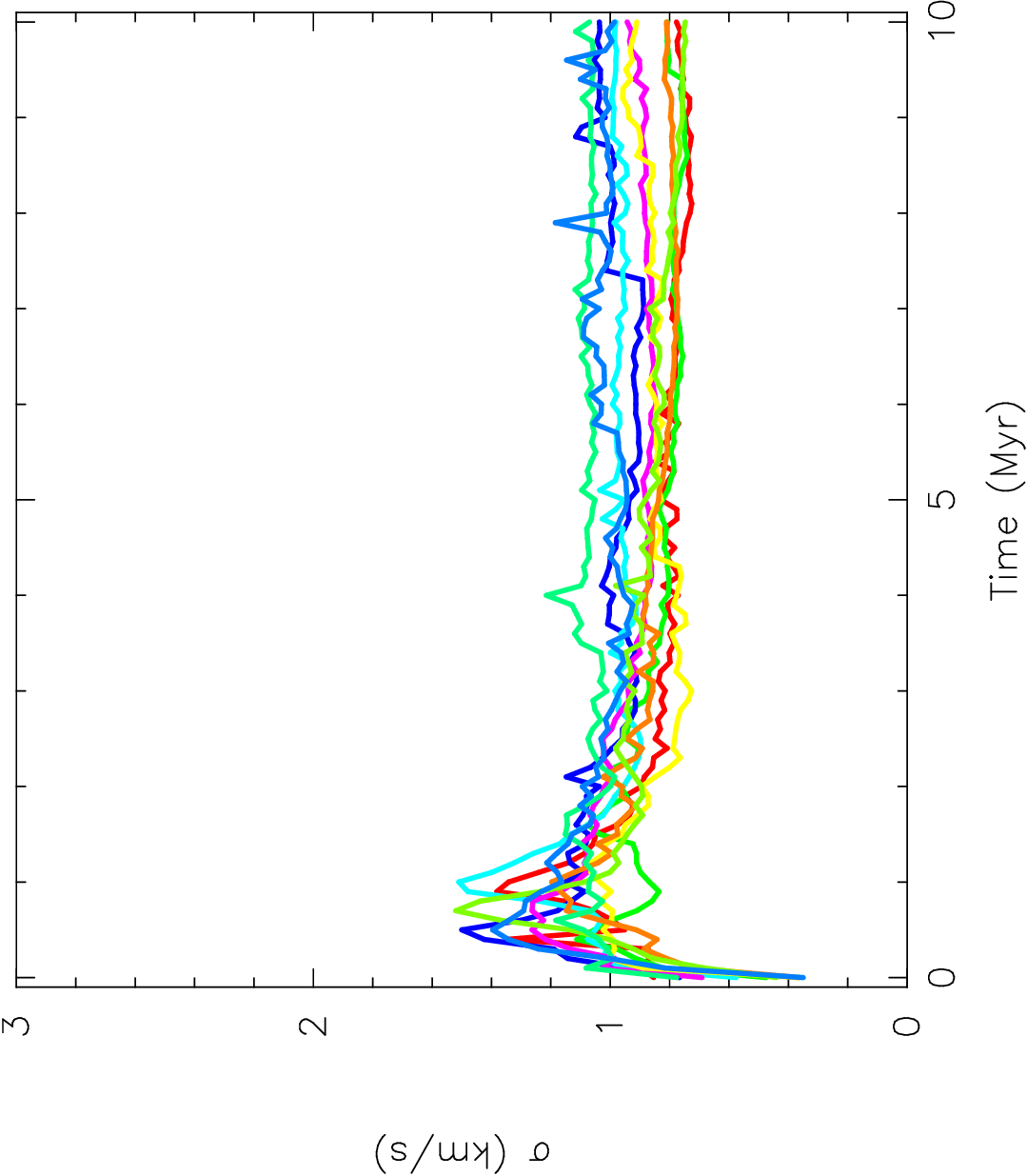}}}
\hspace*{0.3cm} 
\subfigure[$\alpha_{\rm vir} = 0.3, D = 2.0$]{\label{vd_lines-b}\rotatebox{270}{\includegraphics[scale=0.27]{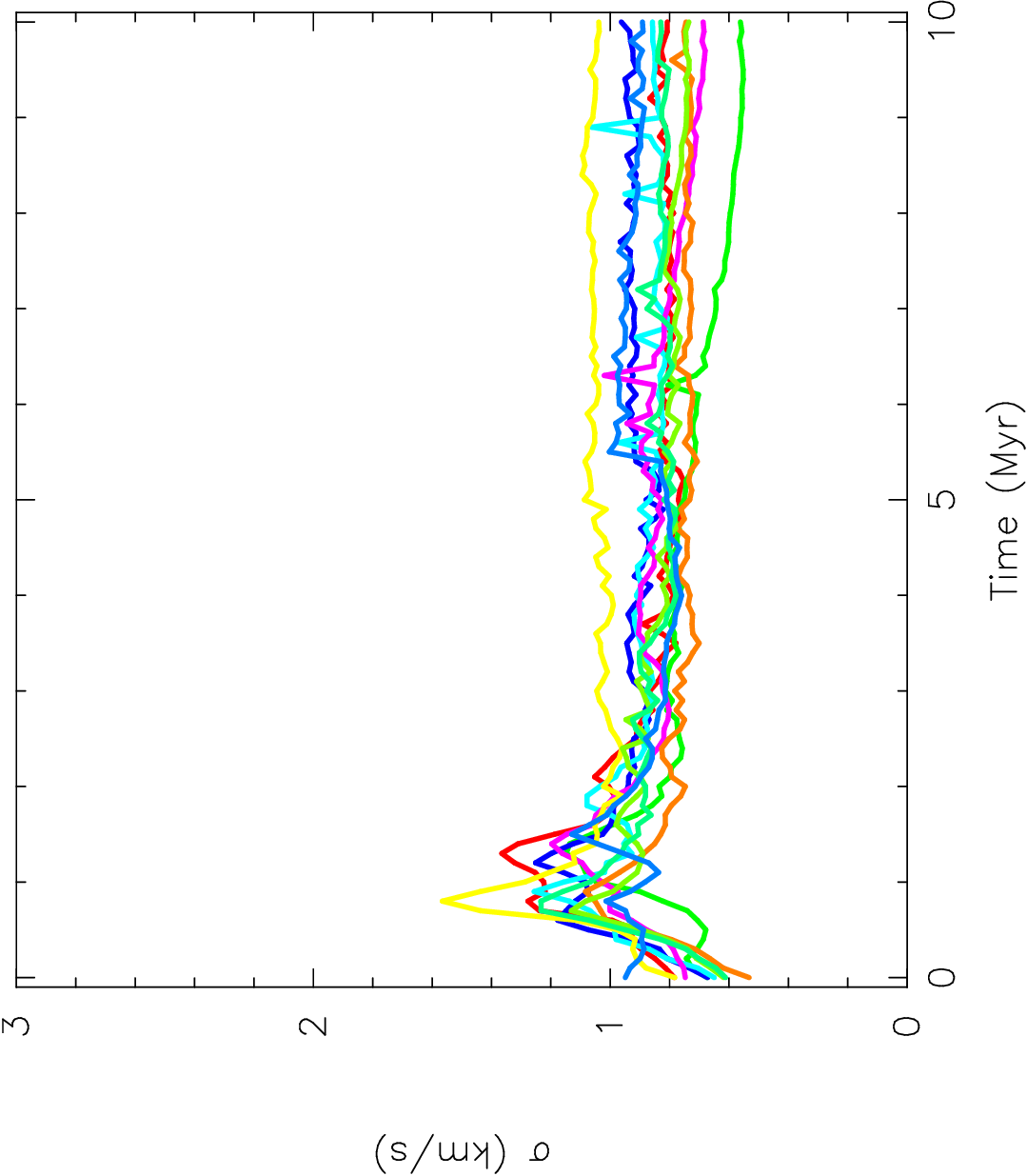}}}
\hspace*{0.3cm} 
\subfigure[$\alpha_{\rm vir} = 0.3, D = 3.0$]{\label{vd_lines-c}\rotatebox{270}{\includegraphics[scale=0.27]{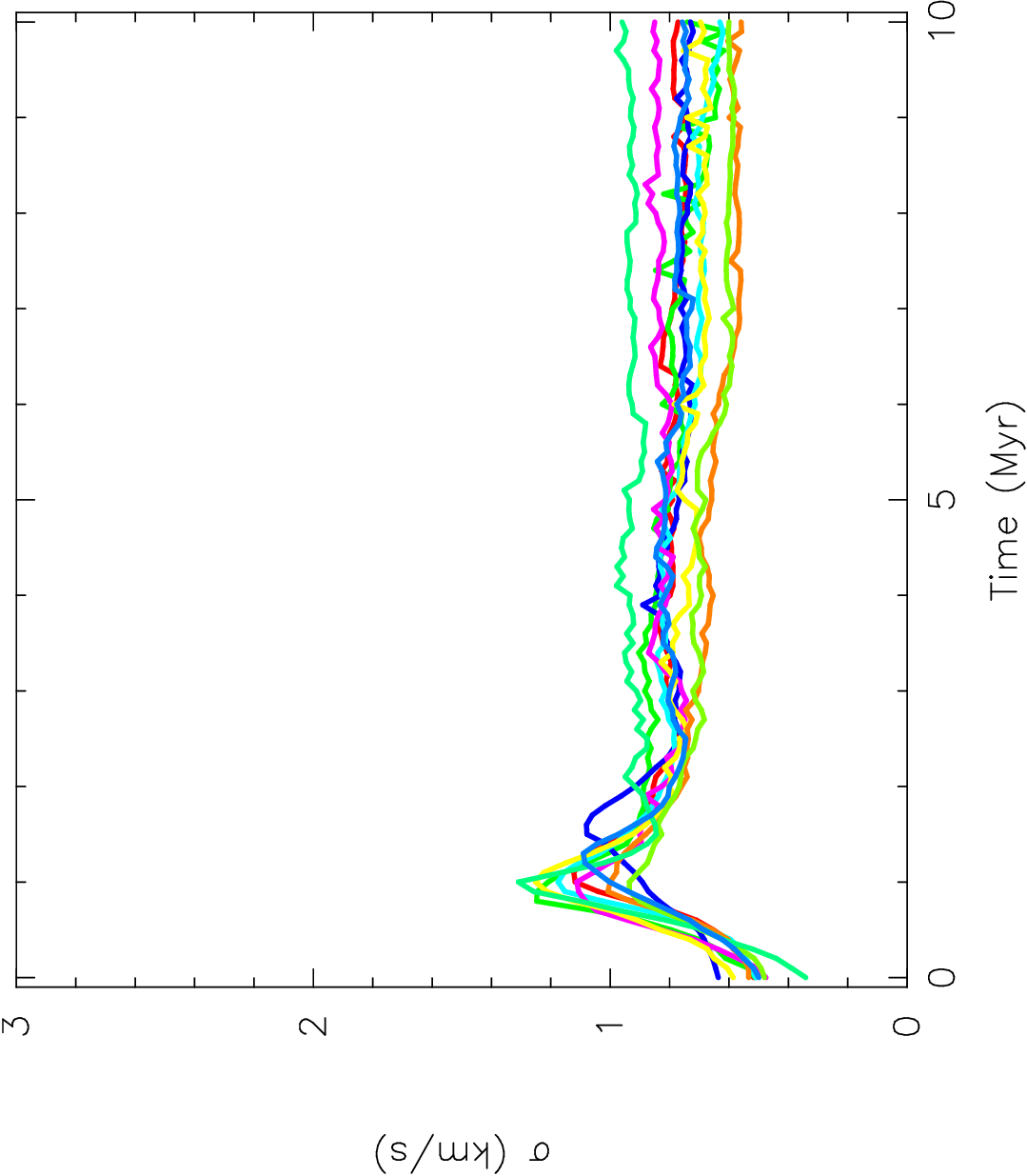}}}
%\vspace*{-0.3cm}
\hspace*{-1.cm}
\subfigure[$\alpha_{\rm vir} = 0.5, D = 1.6$]{\label{vd_lines-d}\rotatebox{270}{\includegraphics[scale=0.27]{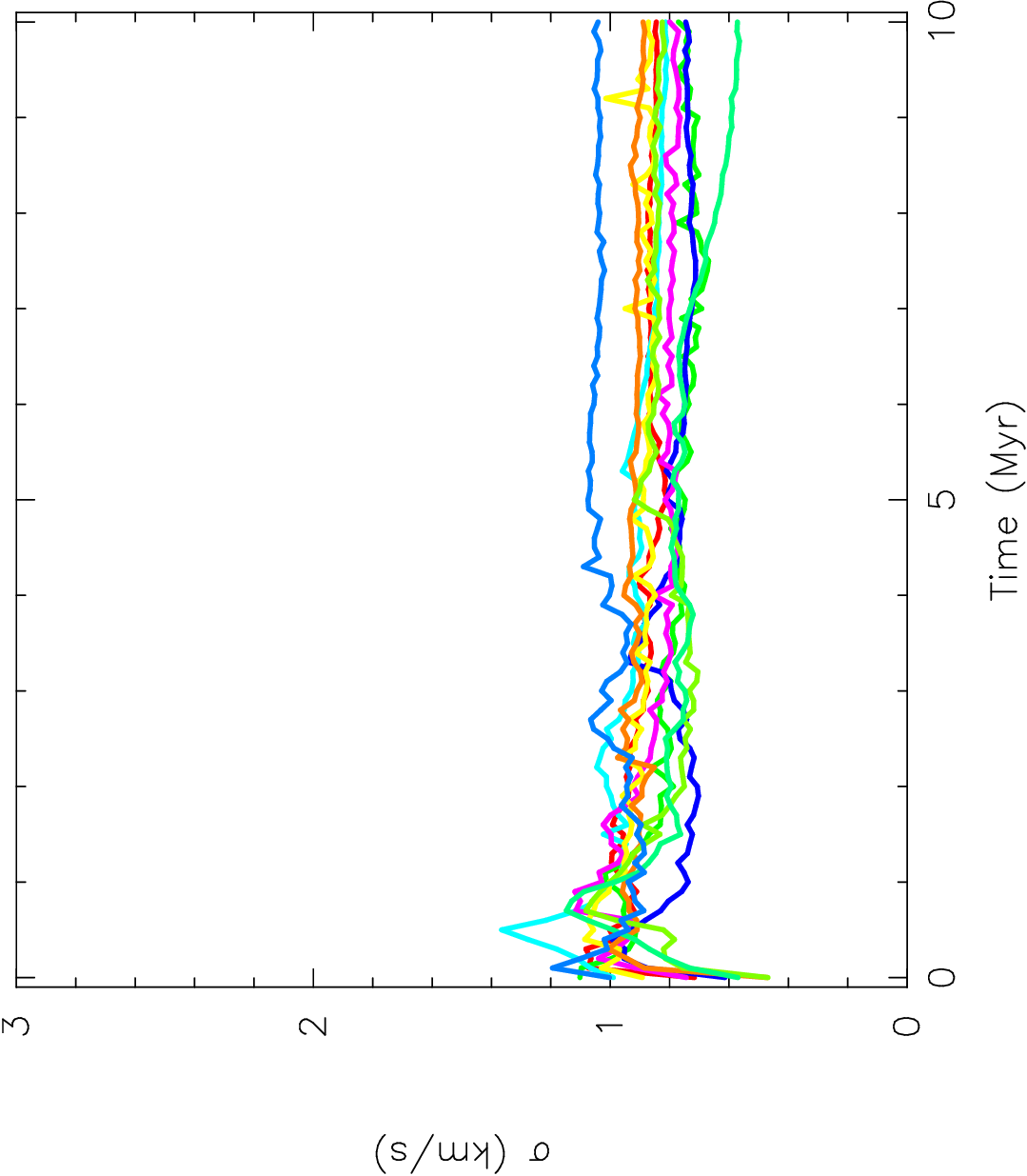}}}
\hspace*{0.3cm} 
\subfigure[$\alpha_{\rm vir} = 0.5, D = 2.0$]{\label{vd_lines-e}\rotatebox{270}{\includegraphics[scale=0.27]{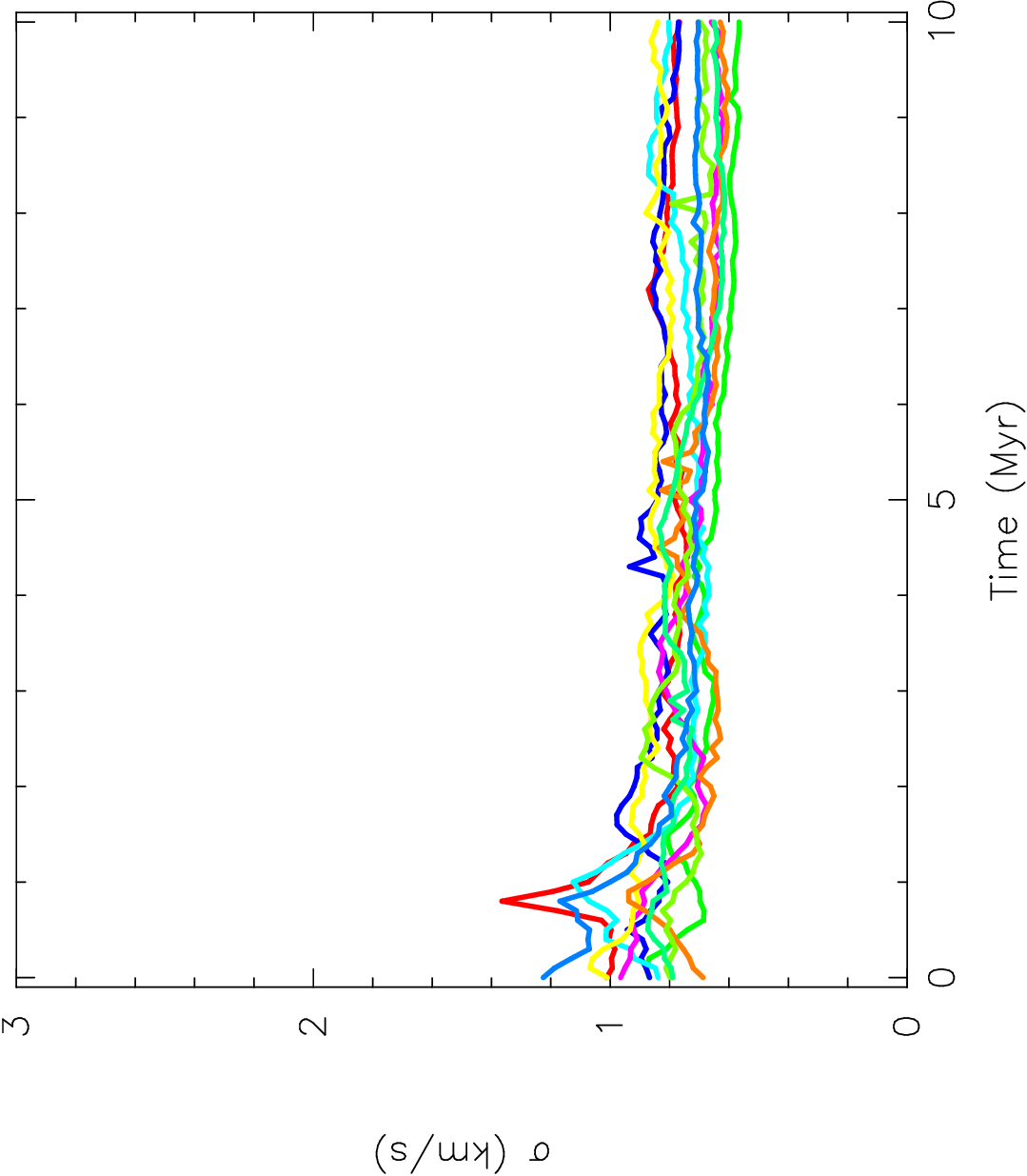}}}
\hspace*{0.3cm} 
\subfigure[$\alpha_{\rm vir} = 0.5, D = 3.0$]{\label{vd_lines-f}\rotatebox{270}{\includegraphics[scale=0.27]{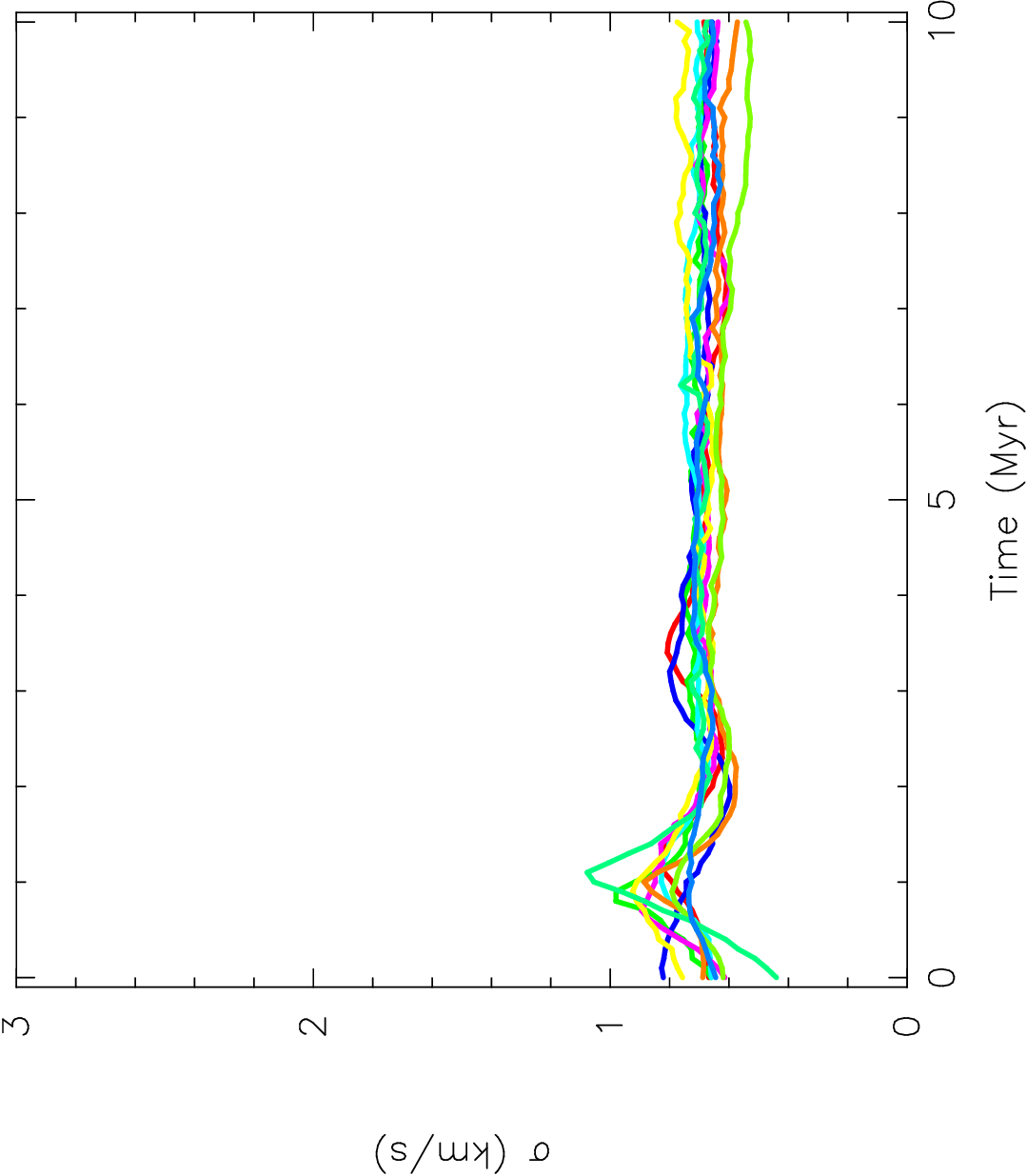}}}
\hspace*{-1.cm}
\subfigure[$\alpha_{\rm vir} = 1.5, D = 1.6$]{\label{vd_lines-g}\rotatebox{270}{\includegraphics[scale=0.27]{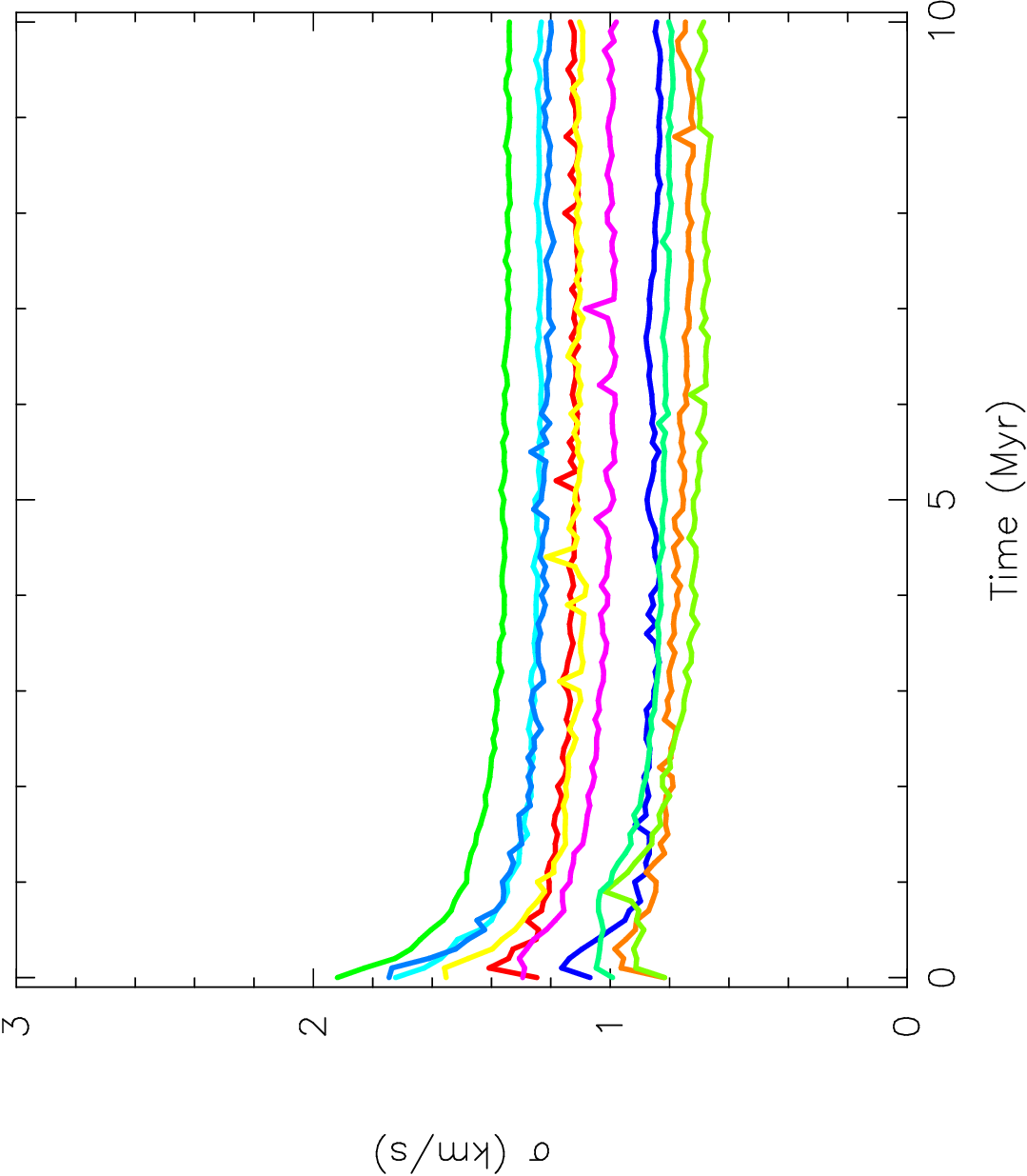}}}
\hspace*{0.3cm} 
\subfigure[$\alpha_{\rm vir} = 1.5, D = 2.0$]{\label{vd_lines-h}\rotatebox{270}{\includegraphics[scale=0.27]{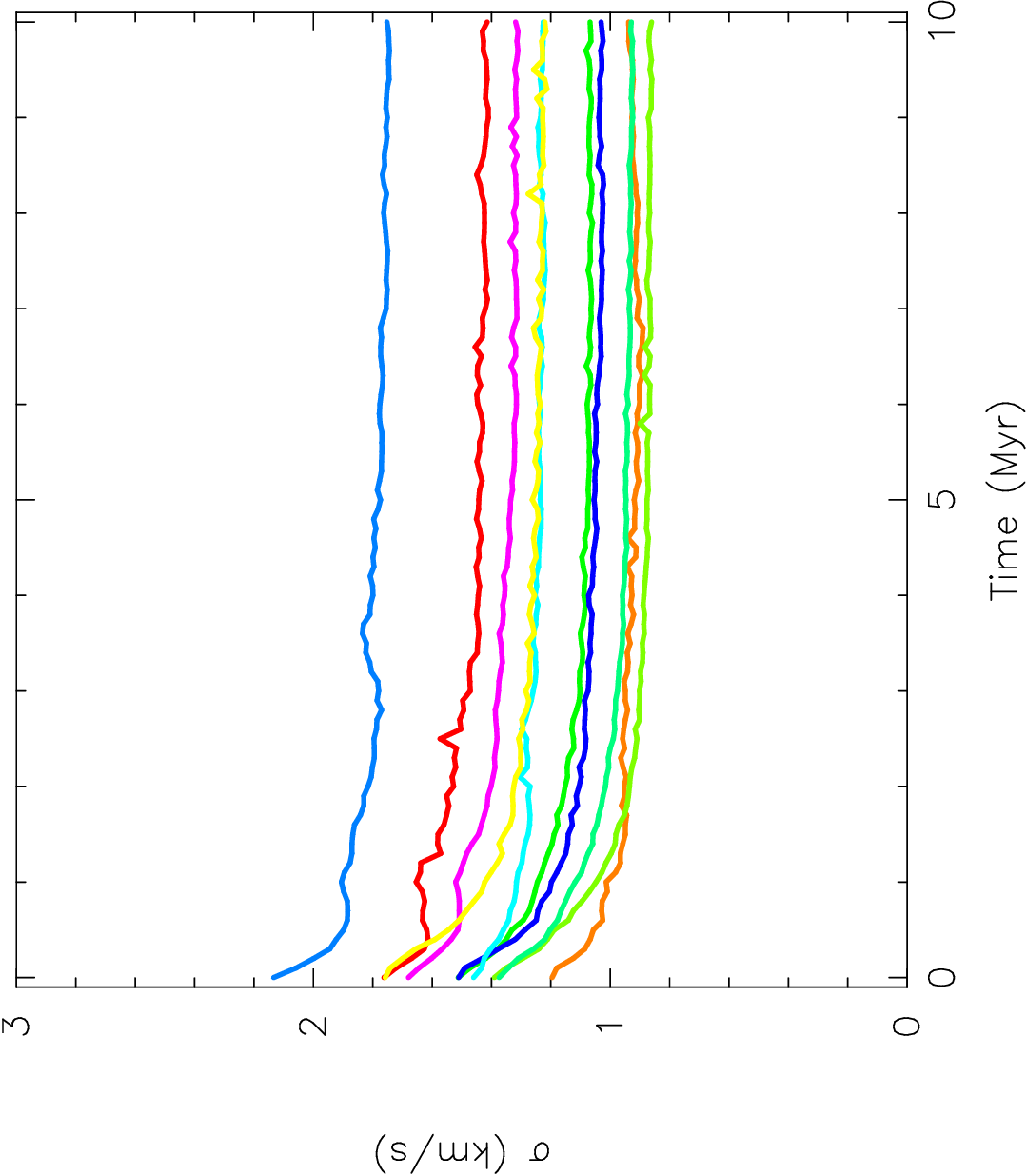}}}
\hspace*{0.3cm} 
\subfigure[$\alpha_{\rm vir} = 1.5, D = 3.0$]{\label{vd_lines-i}\rotatebox{270}{\includegraphics[scale=0.27]{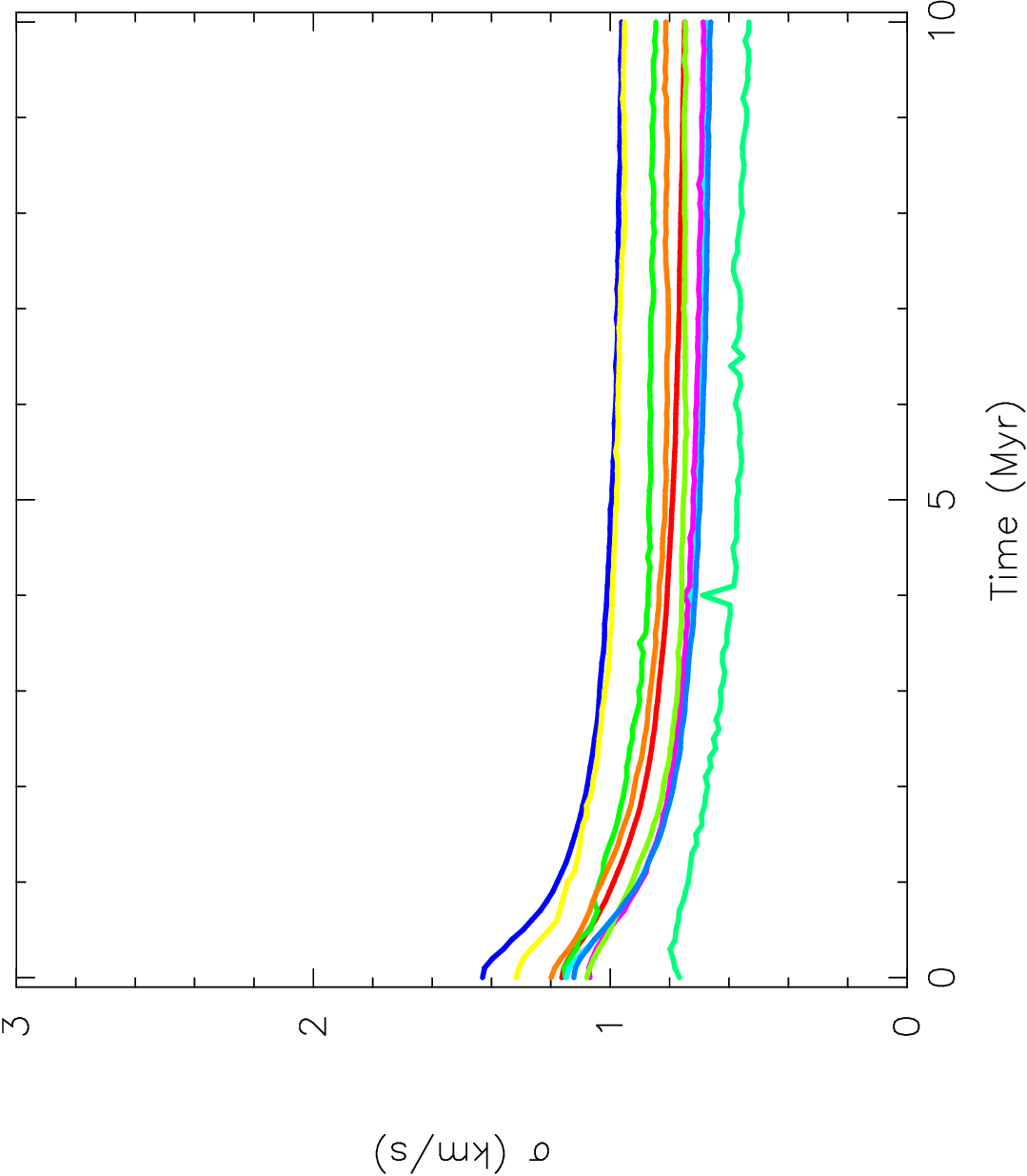}}}
\caption[bf]{Evolution of the velocity dispersion of all stars with time for all initial conditions. Each panel shows 10 individual simulations with identical initial conditions (we randomly omit the 10 remaining realisations of each simulation for clarity).}
\label{vd_lines}
  \end{center}
\end{figure*}

In Fig.~\ref{vd_lines} we show the evolution of the statistical velocity dispersion about the mean radial velocity for 10 randomly chosen realisations of our initial conditions. The top panels (a -- c) show simulations undergoing cool-collapse ($\alpha_{\rm vir} = 0.3$). For these initial conditions the velocity dispersion increases as the regions collapse, before the subsequent expansion causes the dispersion to decrease. The middle panels (d -- f) are simulations initially in virial equilibrium  ($\alpha_{\rm vir} = 0.5$); here the velocity dispersions remain roughly constant at around or just under $\sigma = 1$\,km\,s$^{-1}$. The bottom panels (g -- i) are simulations that are initially supervirial (expanding) ($\alpha_{\rm vir} = 1.5$) and display a larger spread than the virial or sub-virial simulations. 

\begin{figure*}
  \begin{center}
\setlength{\subfigcapskip}{10pt}
\vspace*{-0.3cm}
\hspace*{-1.cm}\subfigure[$\alpha_{\rm vir} = 0.3, D = 1.6$]{\label{vdA_med-a}\rotatebox{270}{\includegraphics[scale=0.27]{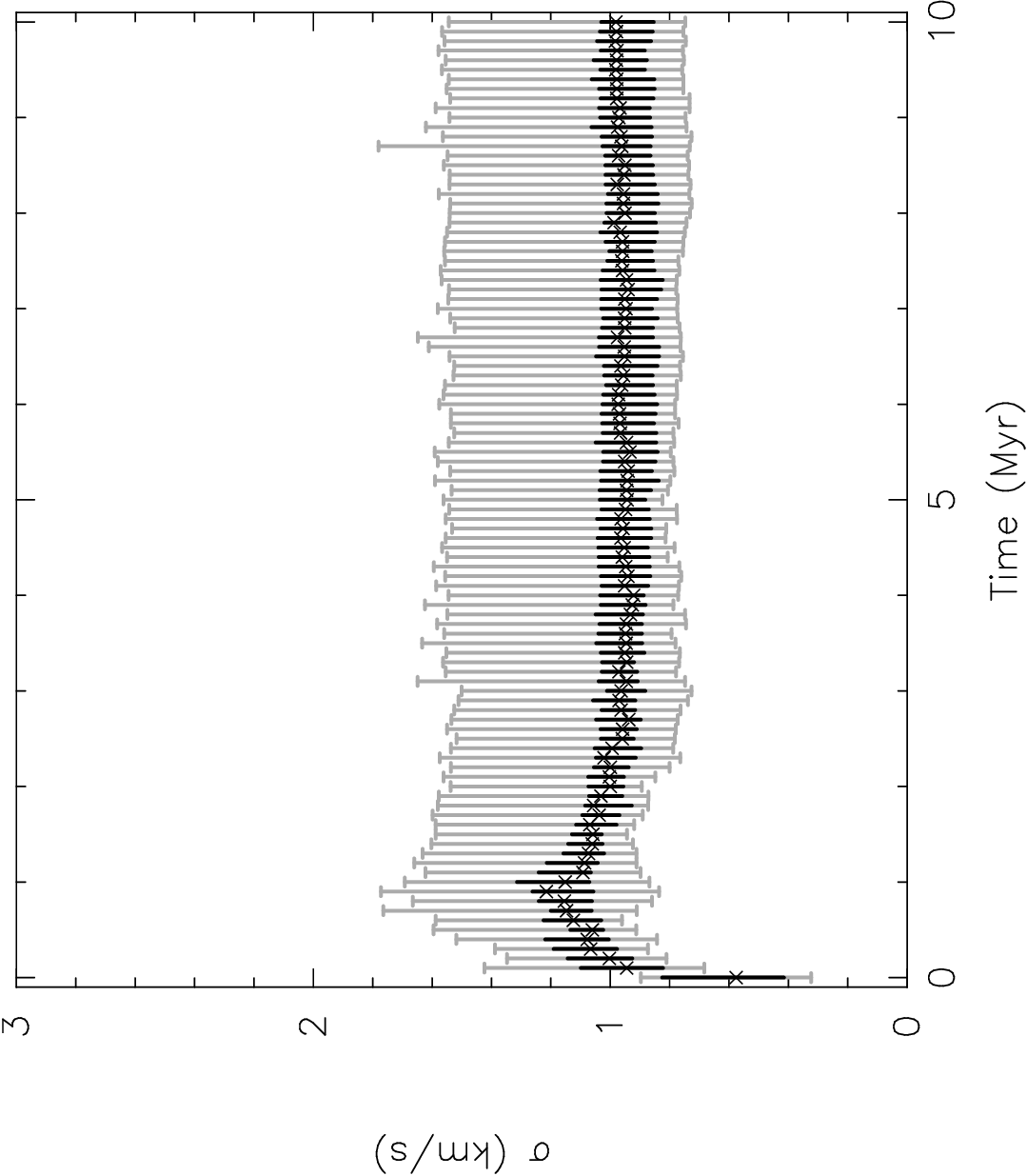}}}
\hspace*{0.3cm} 
\subfigure[$\alpha_{\rm vir} = 0.3, D = 2.0$]{\label{vdA_med-b}\rotatebox{270}{\includegraphics[scale=0.27]{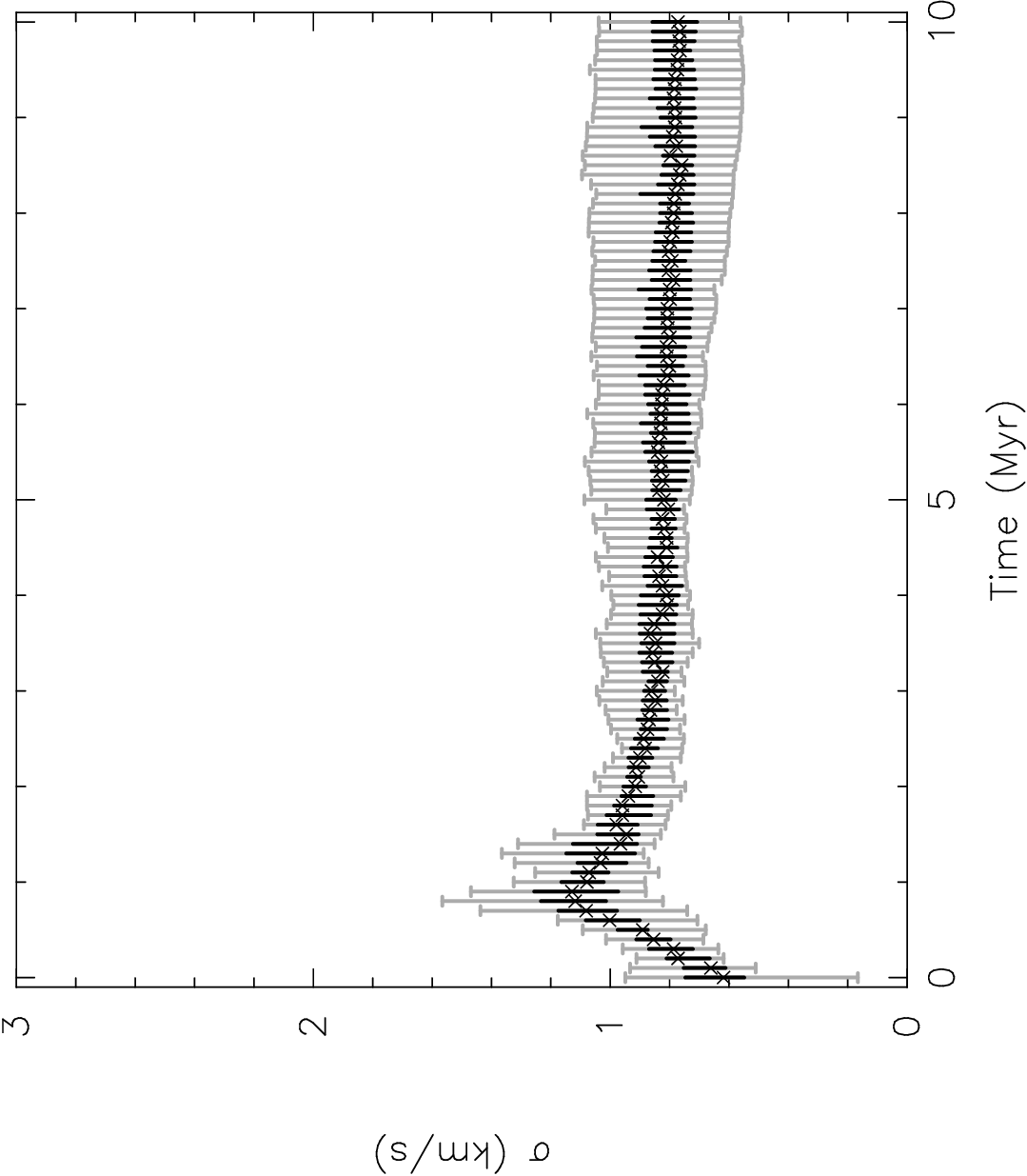}}}
\hspace*{0.3cm} 
\subfigure[$\alpha_{\rm vir} = 0.3, D = 3.0$]{\label{vdA_med-c}\rotatebox{270}{\includegraphics[scale=0.27]{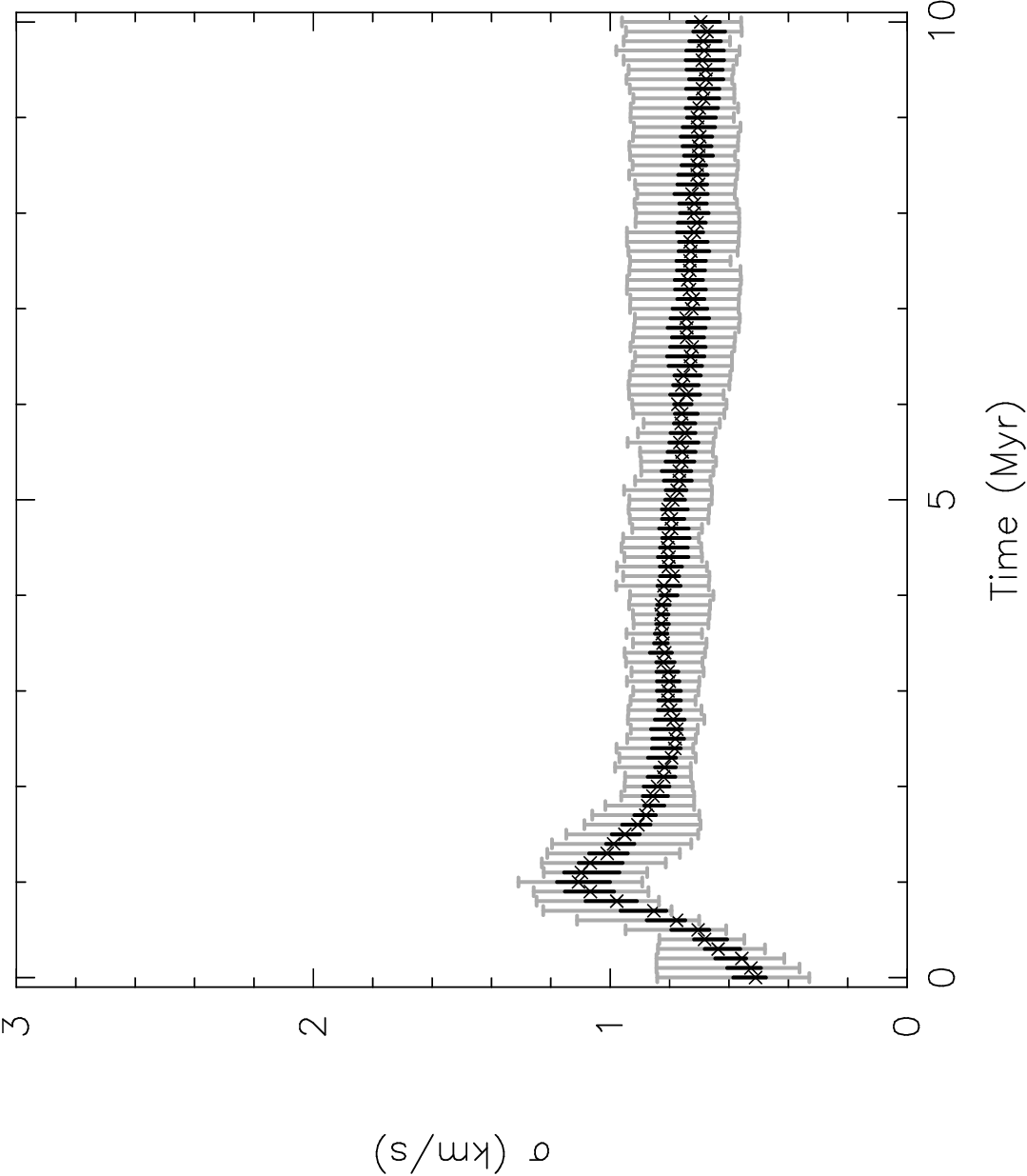}}}
%\vspace*{-0.3cm}
\hspace*{-1.cm}
\subfigure[$\alpha_{\rm vir} = 0.5, D = 1.6$]{\label{vdA_med-d}\rotatebox{270}{\includegraphics[scale=0.27]{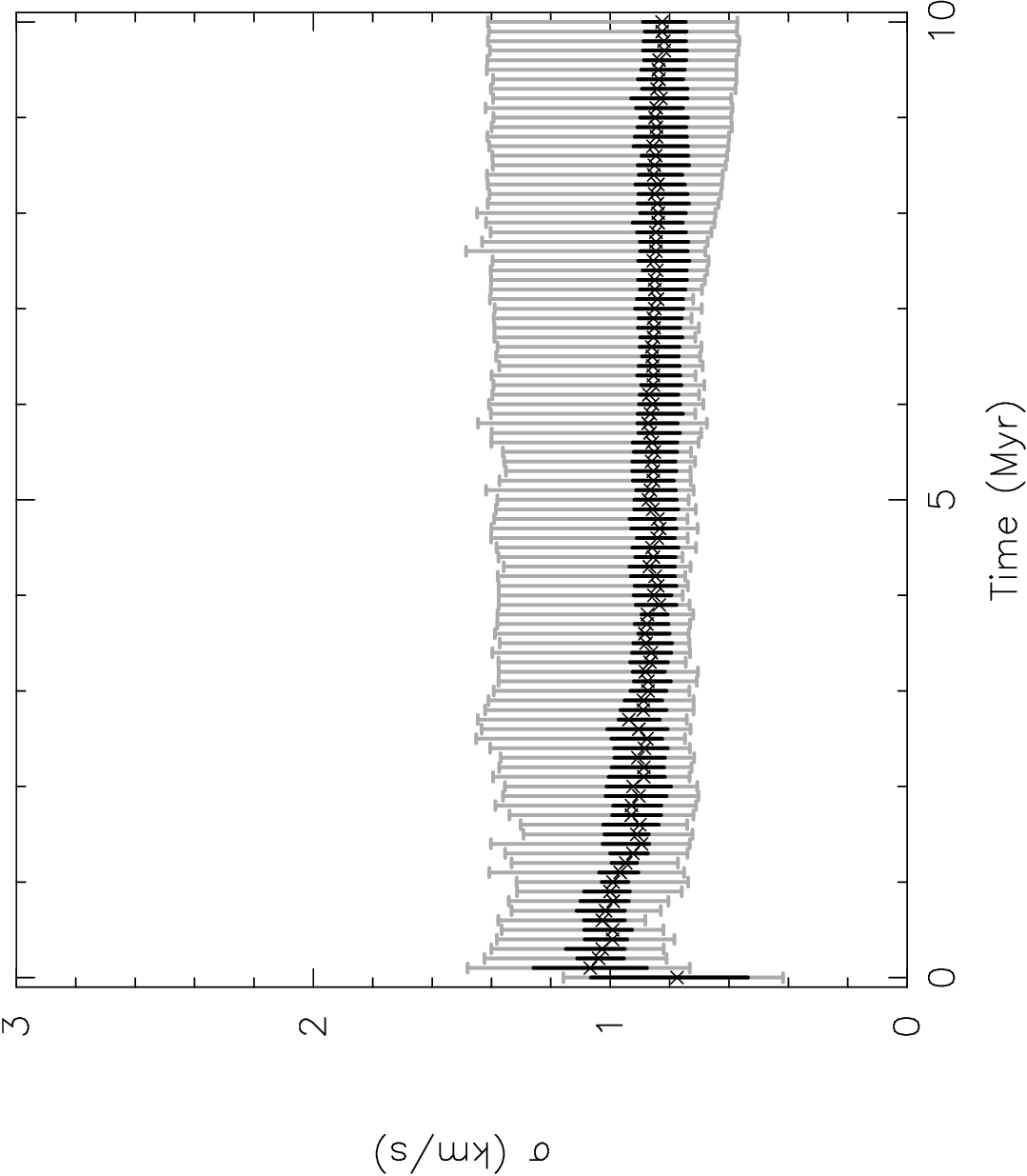}}}
\hspace*{0.3cm} 
\subfigure[$\alpha_{\rm vir} = 0.5, D = 2.0$]{\label{vdA_med-e}\rotatebox{270}{\includegraphics[scale=0.27]{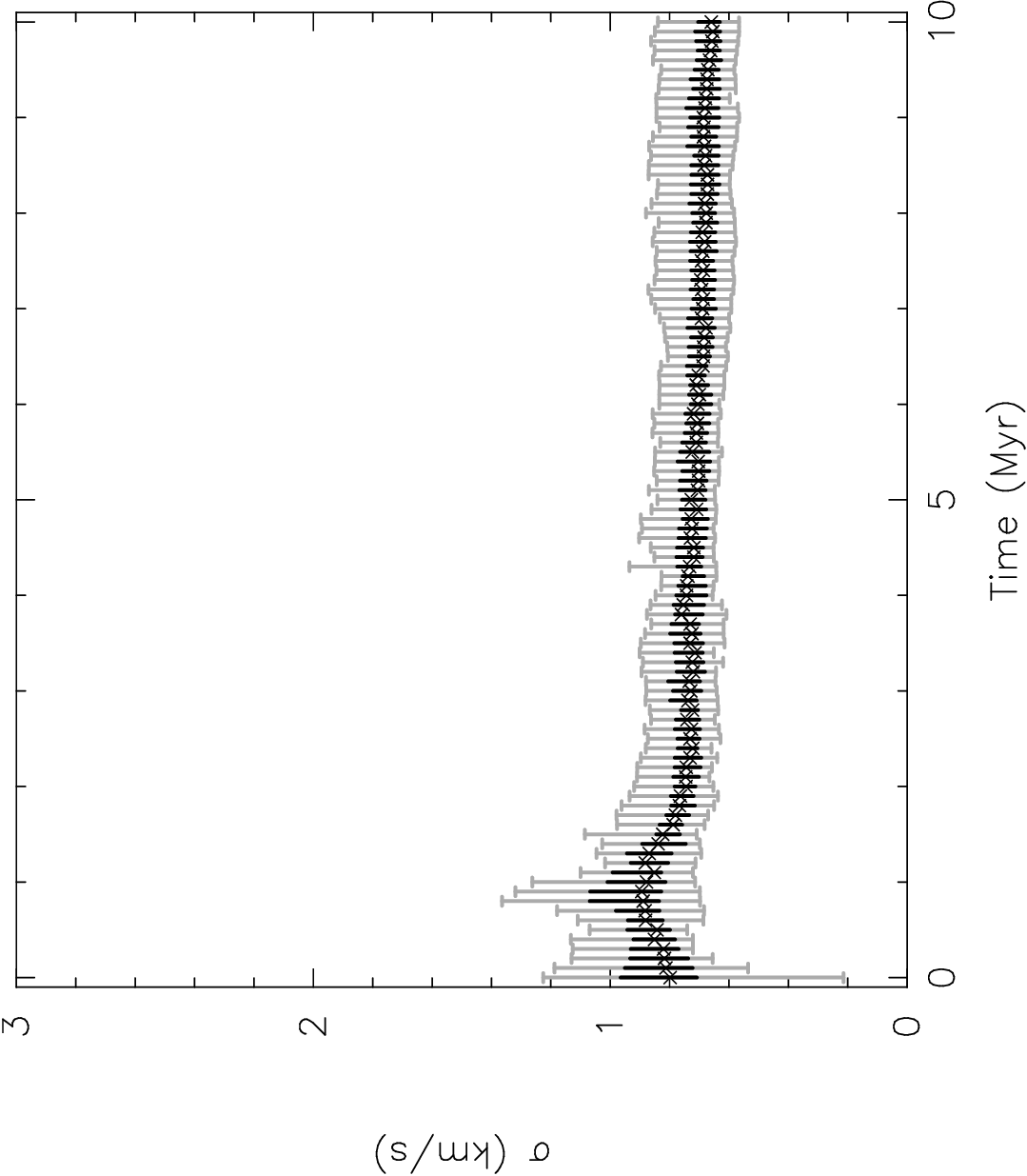}}}
\hspace*{0.3cm} 
\subfigure[$\alpha_{\rm vir} = 0.5, D = 3.0$]{\label{vdA_med-f}\rotatebox{270}{\includegraphics[scale=0.27]{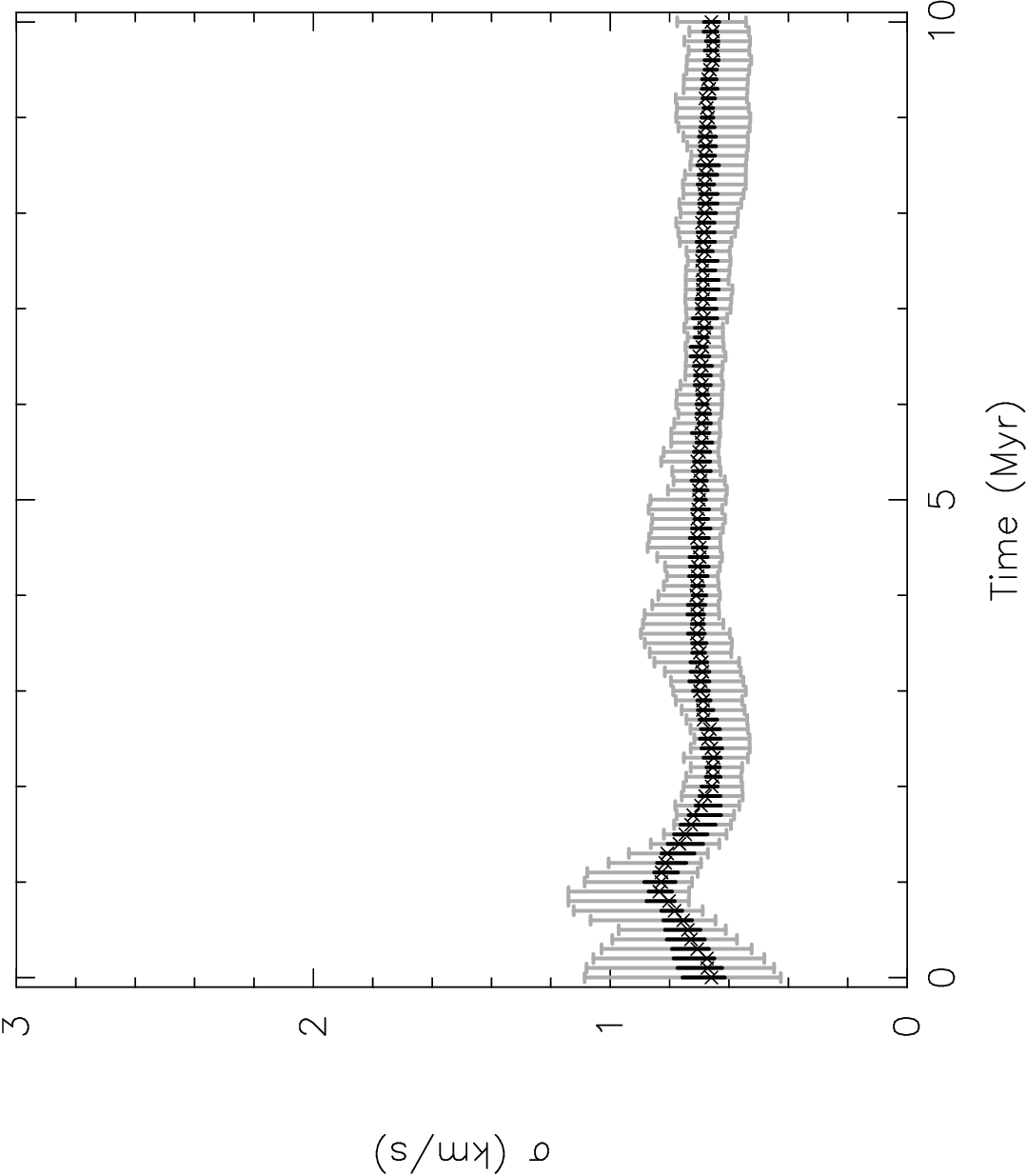}}}
\hspace*{-1.cm}
\subfigure[$\alpha_{\rm vir} = 1.5, D = 1.6$]{\label{vdA_med-g}\rotatebox{270}{\includegraphics[scale=0.27]{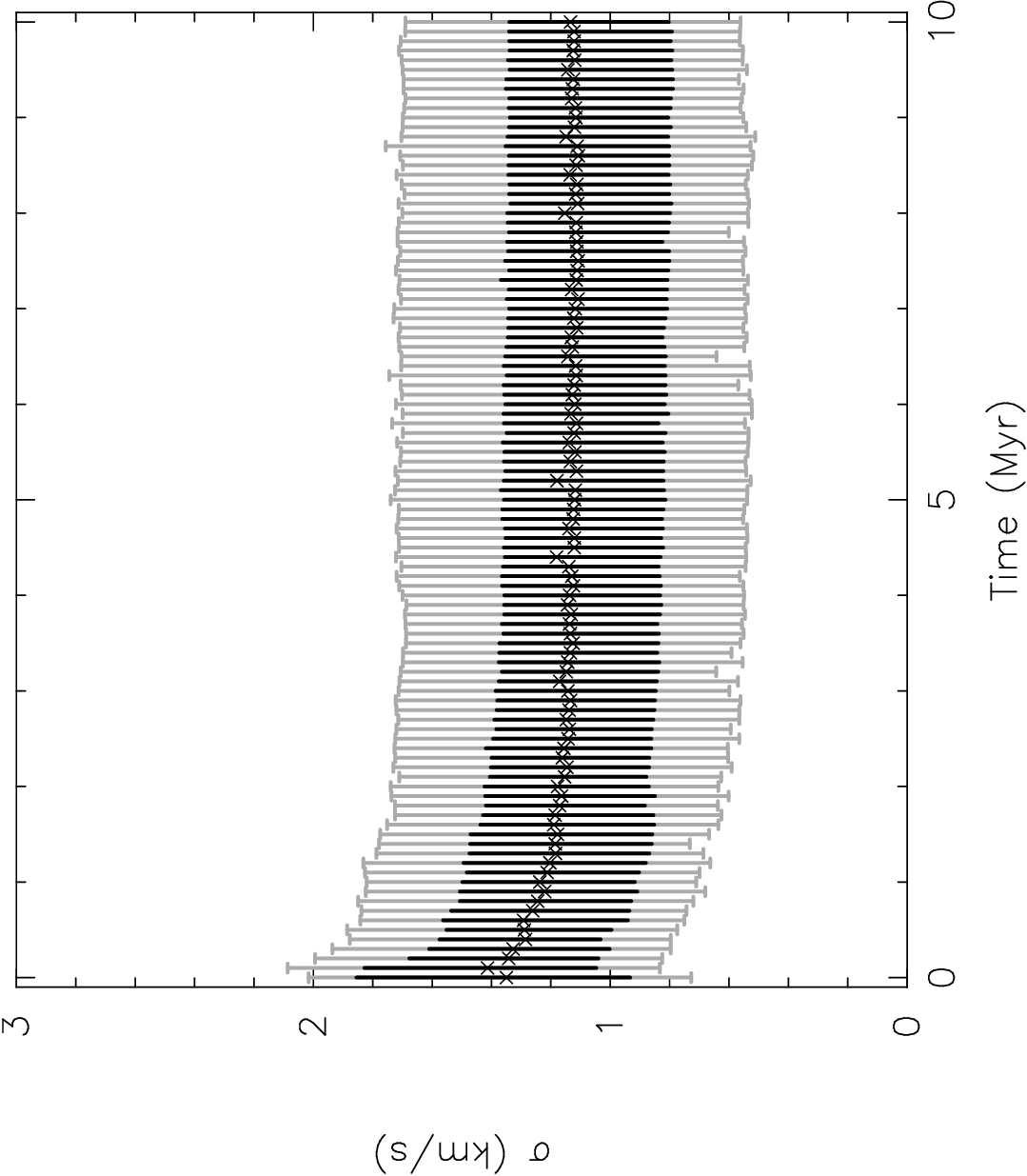}}}
\hspace*{0.3cm} 
\subfigure[$\alpha_{\rm vir} = 1.5, D = 2.0$]{\label{vdA_med-h}\rotatebox{270}{\includegraphics[scale=0.27]{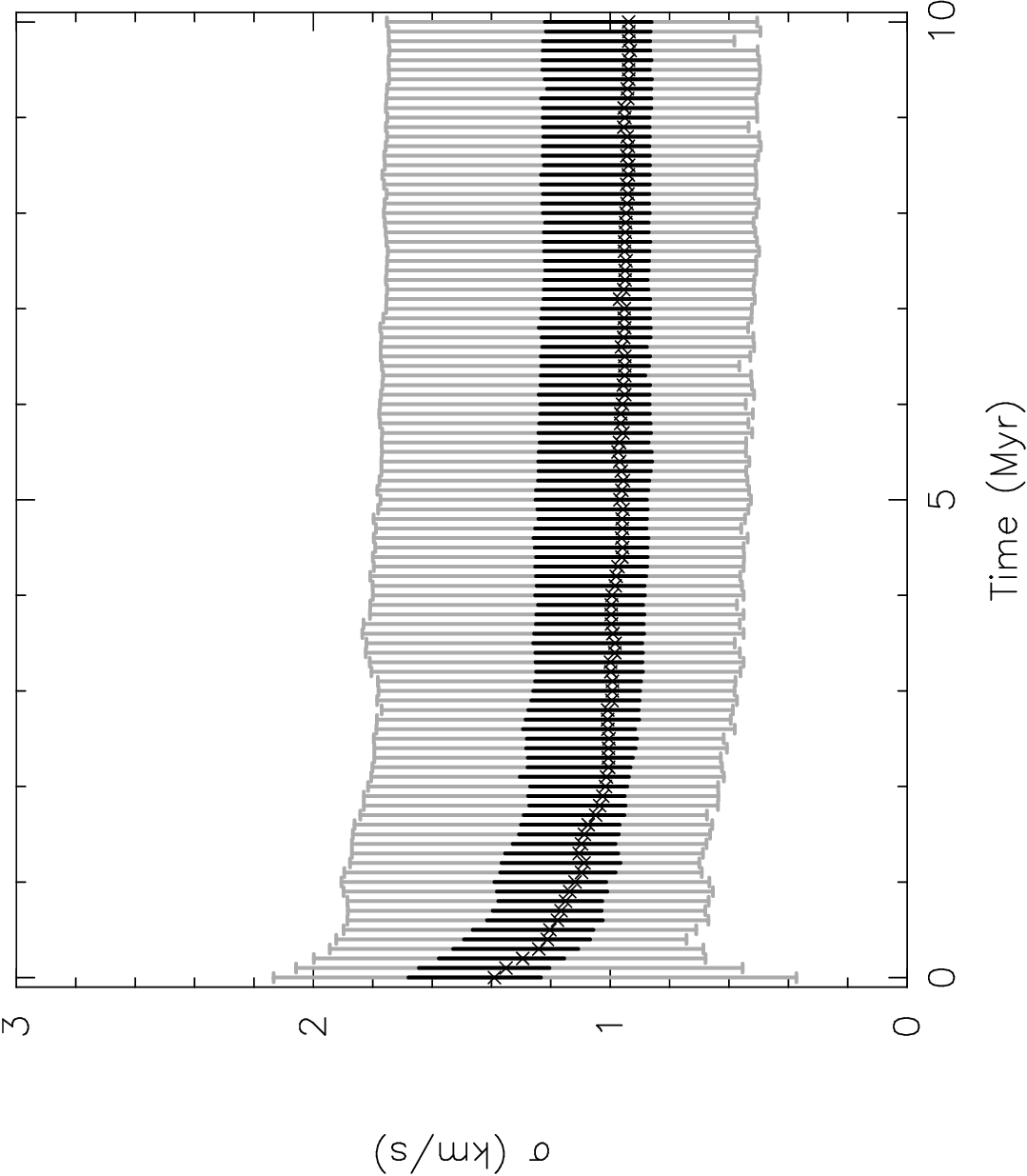}}}
\hspace*{0.3cm} 
\subfigure[$\alpha_{\rm vir} = 1.5, D = 3.0$]{\label{vdA_med-i}\rotatebox{270}{\includegraphics[scale=0.27]{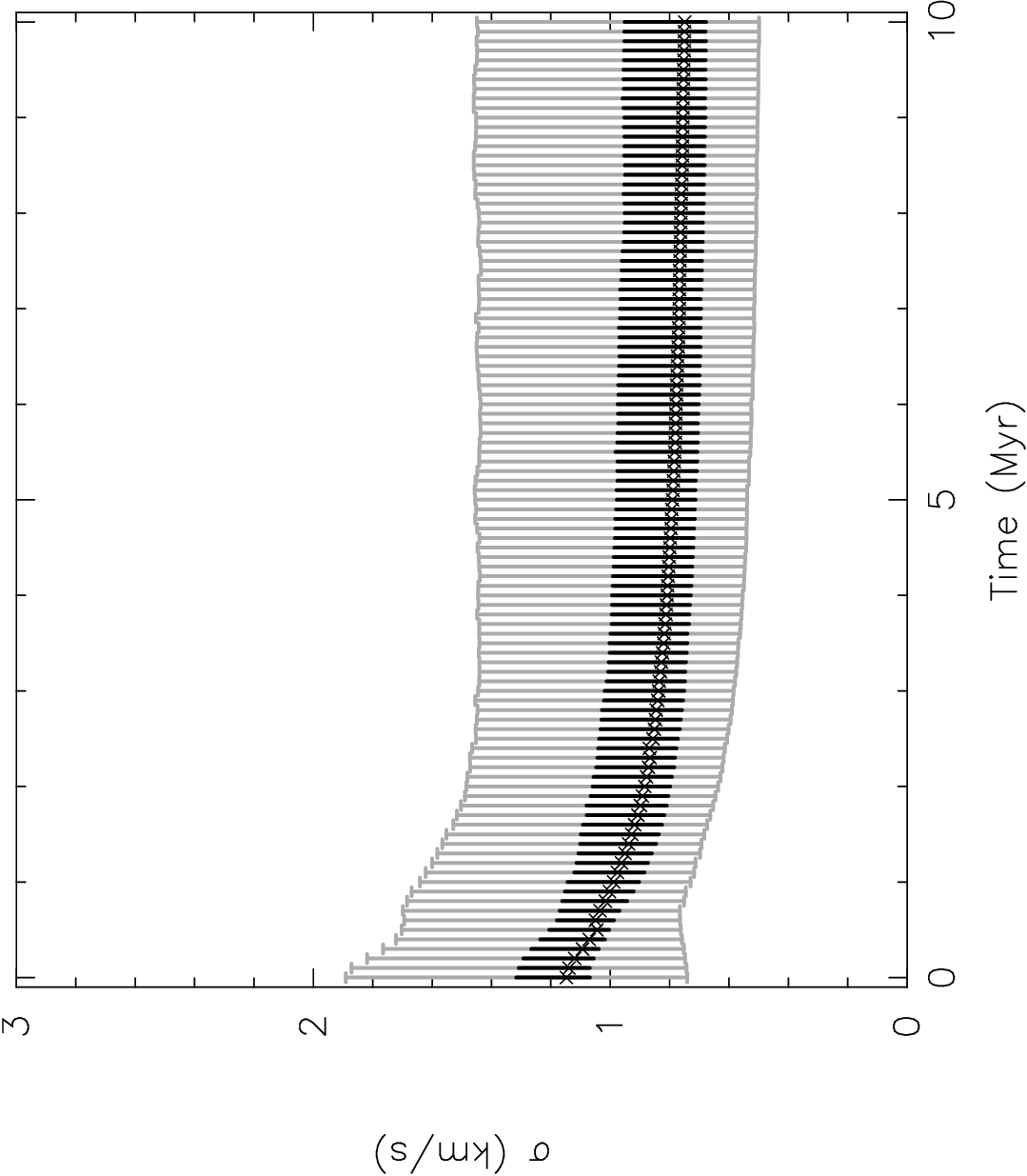}}}
\caption[bf]{Evolution of the velocity dispersion of all stars with time for all simulations. Each panel shows the median value of all 20 simulations with identical initial conditions (the crosses) and the darker error bars indicate 
25 and 75 percentile values. The entire range of possible values from the 20 sets of initial conditions is shown by the lighter error bars. }
\label{vdA_med}
  \end{center}
\end{figure*}

In Fig.~\ref{vdA_med} we show the range of values at any point in time for each set of initial conditions. At each time, the cross symbol indicates the median velocity dispersion from 20 simulations with identical initial conditions. The black `error bars' indicate the 25 -- 75\,percentile range in the simulations, and the grey `error bars' indicate the full range from 20 simulations. These are not error bars in the conventional sense, but serve to illustrate the large range possible due to stochasticity in the dynamical evolution. The simulations with the widest range of velocity dispersions are those with the most substructure ($D = 1.6$; panels a, d, and g), and with the highest virial ratios ($\alpha_{\rm vir} = 1.5$; panels g, h, and i). In both cases, this is a relic of our initial conditions set-up; in the case of the $D = 1.6$ models the velocity substructure correlates with the physical substructure \citep[c.f. Larson's relations,][]{Larson81}, and in the $\alpha_{\rm vir} = 1.5$ models the high virial ratio exaggerates differences in velocity between subgroups of stars. This is because the velocities are scaled by a larger factor than their virial or subvirial counterparts (the opposite extreme would be if the initial velocities were `cold', i.e.\,\,$\alpha_{\rm vir} = 0$).

Fig.~\ref{vdA_med} shows that whilst some of the supervirial simulations have higher velocity dispersions than the virial or sub-virial simulations, in many cases the values overlap, making it essential that the velocity dispersion is compared to a virial mass velocity dispersion estimate, $\sigma_{\rm vir}$, given by Eqn.~\ref{virial_mass}, to distinguish between initial conditions.

\begin{figure*}
  \begin{center}
\setlength{\subfigcapskip}{10pt}
\vspace*{-0.3cm}
\hspace*{-1.cm}\subfigure[$\alpha_{\rm vir} = 0.3, D = 1.6$]{\label{vdC_med-a}\rotatebox{270}{\includegraphics[scale=0.27]{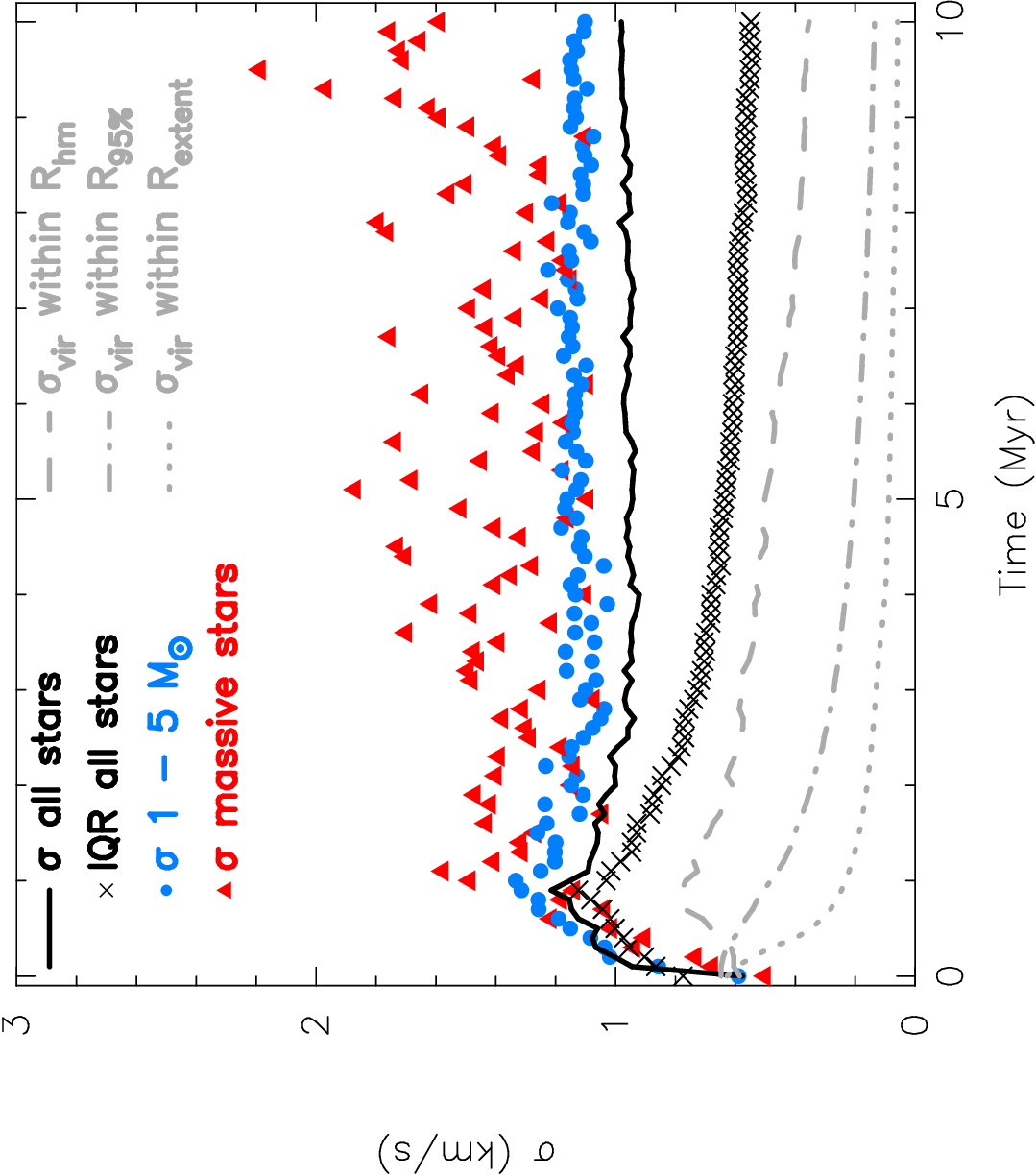}}}
\hspace*{0.3cm} 
\subfigure[$\alpha_{\rm vir} = 0.3, D = 2.0$]{\label{vdC_med-b}\rotatebox{270}{\includegraphics[scale=0.27]{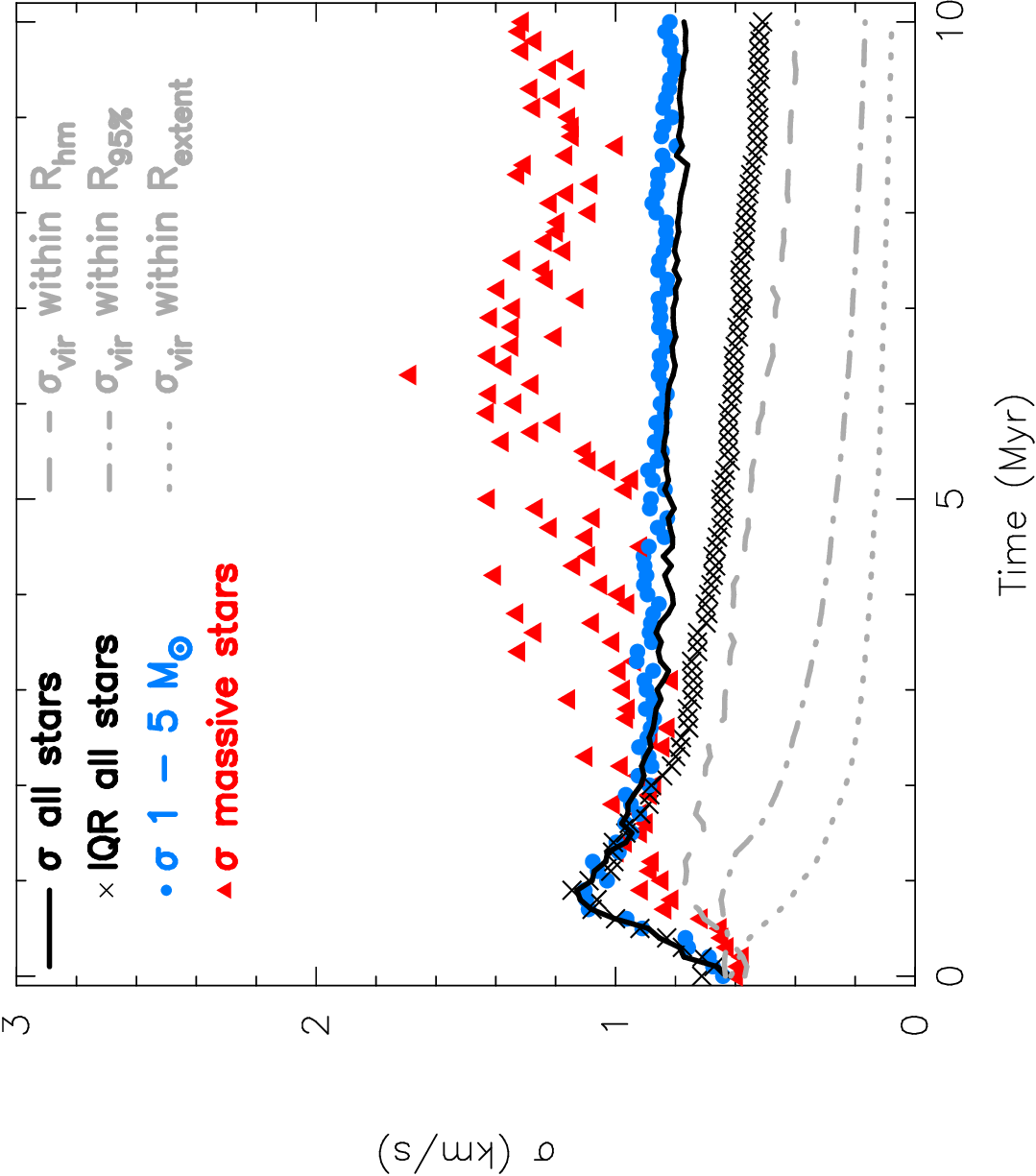}}}
\hspace*{0.3cm} 
\subfigure[$\alpha_{\rm vir} = 0.3, D = 3.0$]{\label{vdC_med-c}\rotatebox{270}{\includegraphics[scale=0.27]{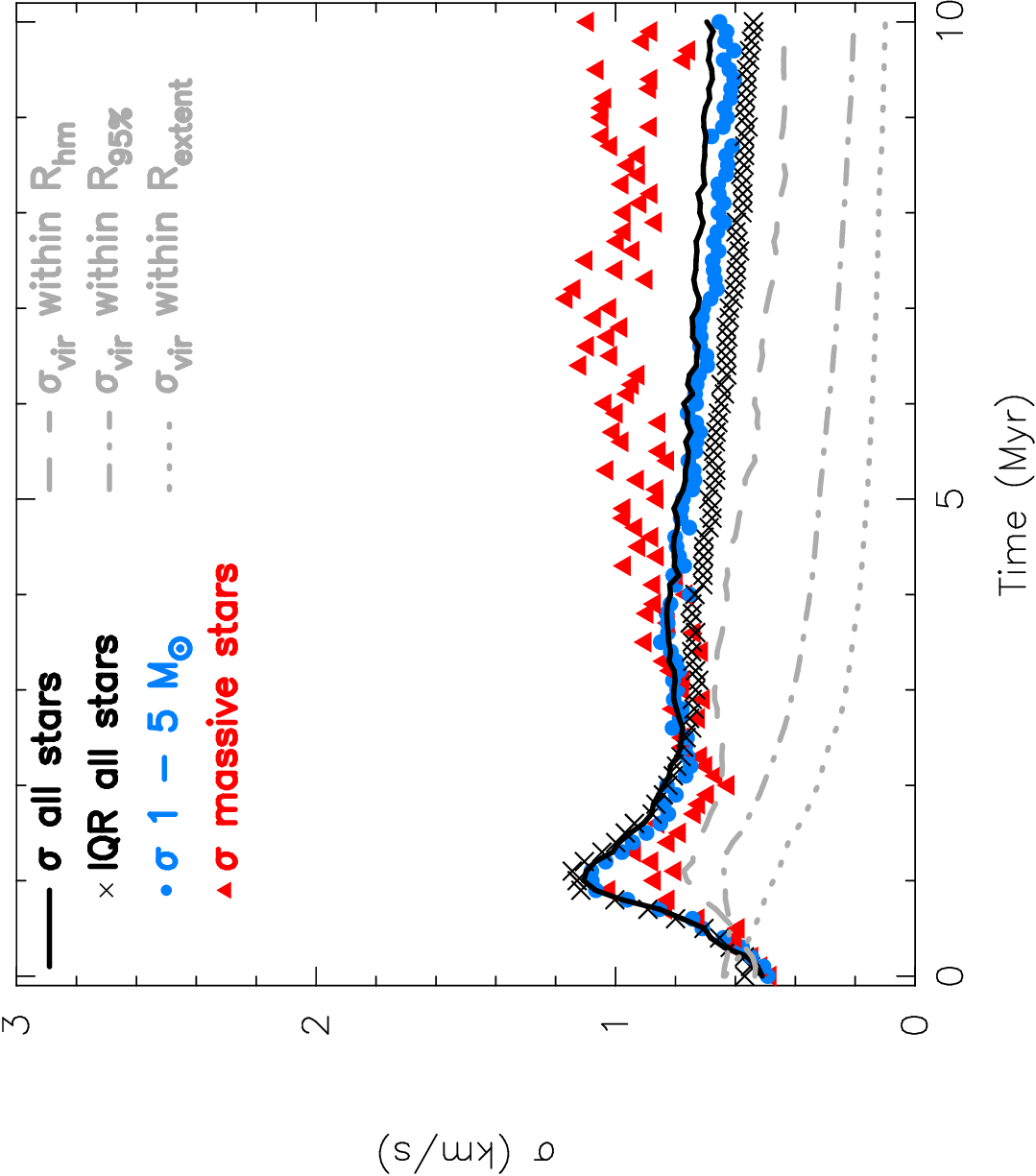}}}
%\vspace*{-0.3cm}
\hspace*{-1.cm}
\subfigure[$\alpha_{\rm vir} = 0.5, D = 1.6$]{\label{vdC_med-d}\rotatebox{270}{\includegraphics[scale=0.27]{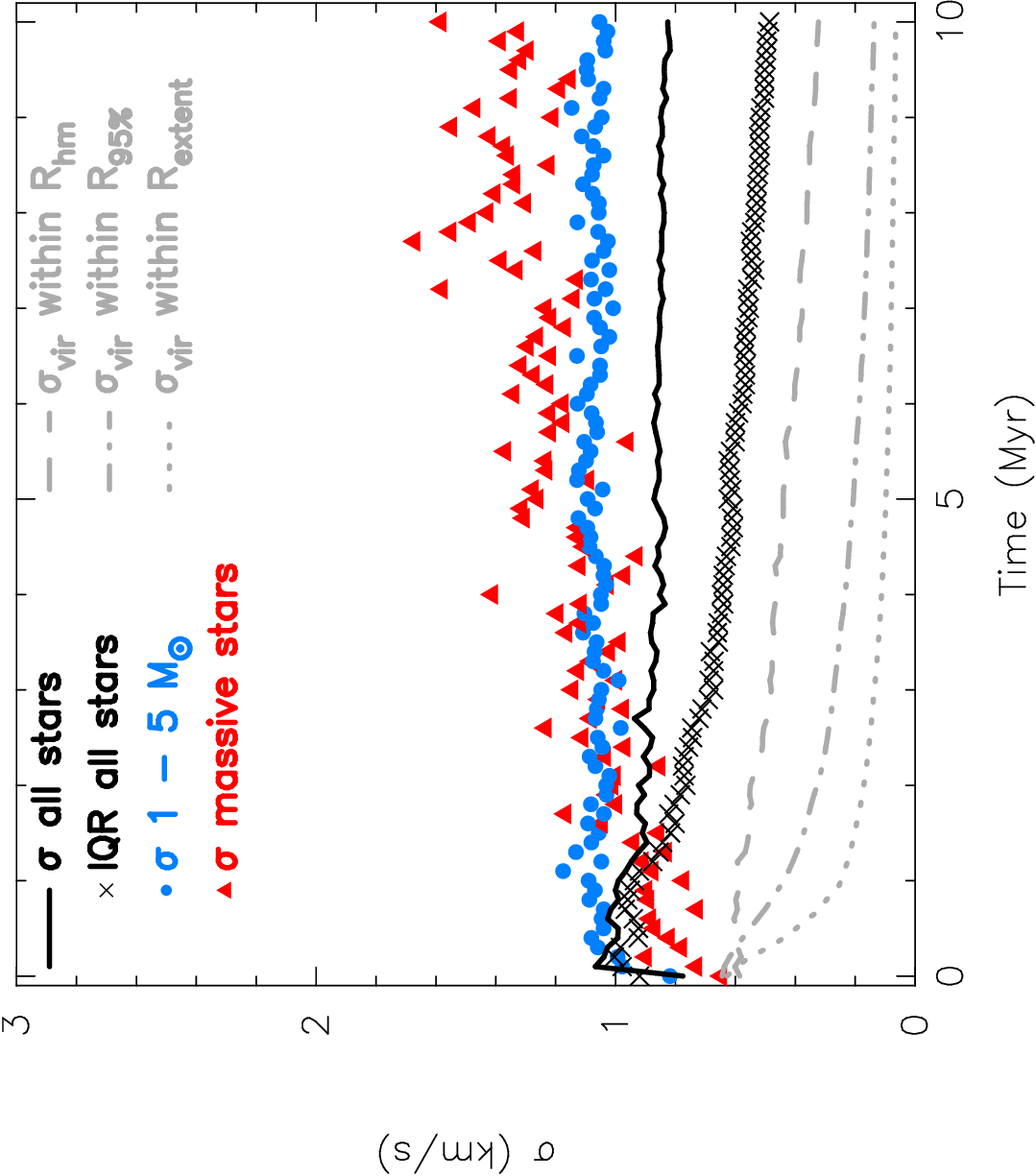}}}
\hspace*{0.3cm} 
\subfigure[$\alpha_{\rm vir} = 0.5, D = 2.0$]{\label{vdC_med-e}\rotatebox{270}{\includegraphics[scale=0.27]{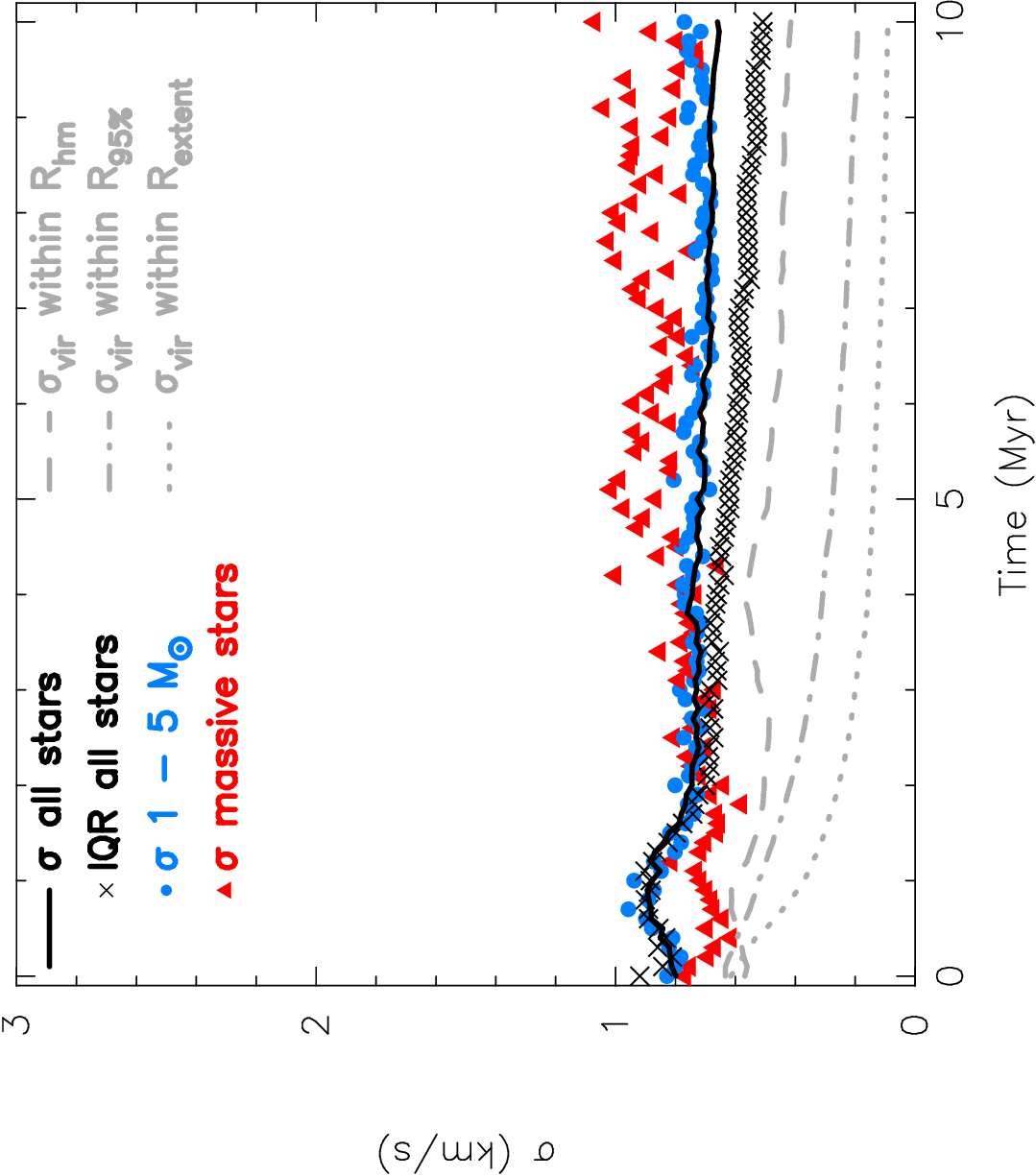}}}
\hspace*{0.3cm} 
\subfigure[$\alpha_{\rm vir} = 0.5, D = 3.0$]{\label{vdC_med-f}\rotatebox{270}{\includegraphics[scale=0.27]{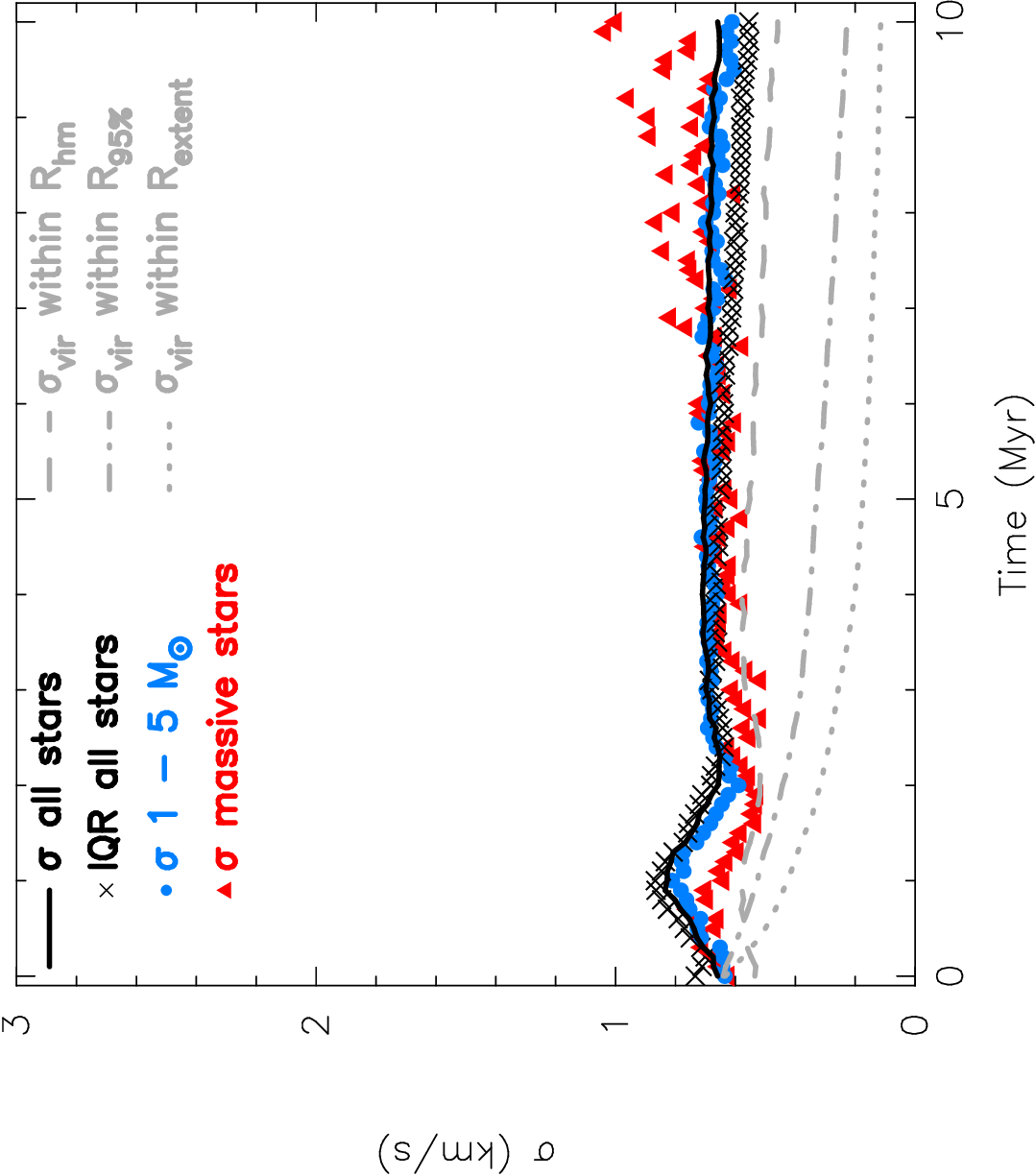}}}
\hspace*{-1.cm}
\subfigure[$\alpha_{\rm vir} = 1.5, D = 1.6$]{\label{vdC_med-g}\rotatebox{270}{\includegraphics[scale=0.27]{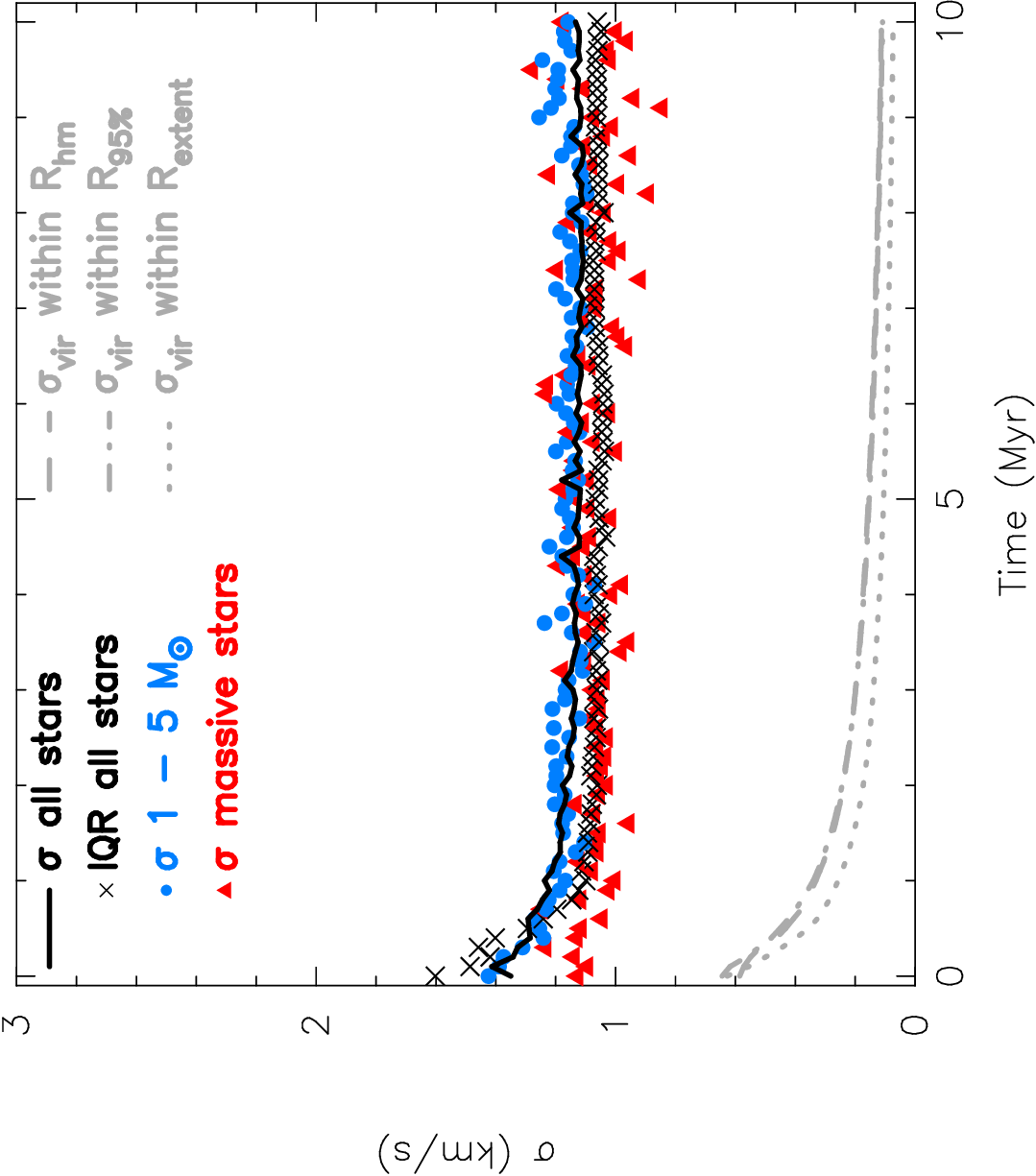}}}
\hspace*{0.3cm} 
\subfigure[$\alpha_{\rm vir} = 1.5, D = 2.0$]{\label{vdC_med-h}\rotatebox{270}{\includegraphics[scale=0.27]{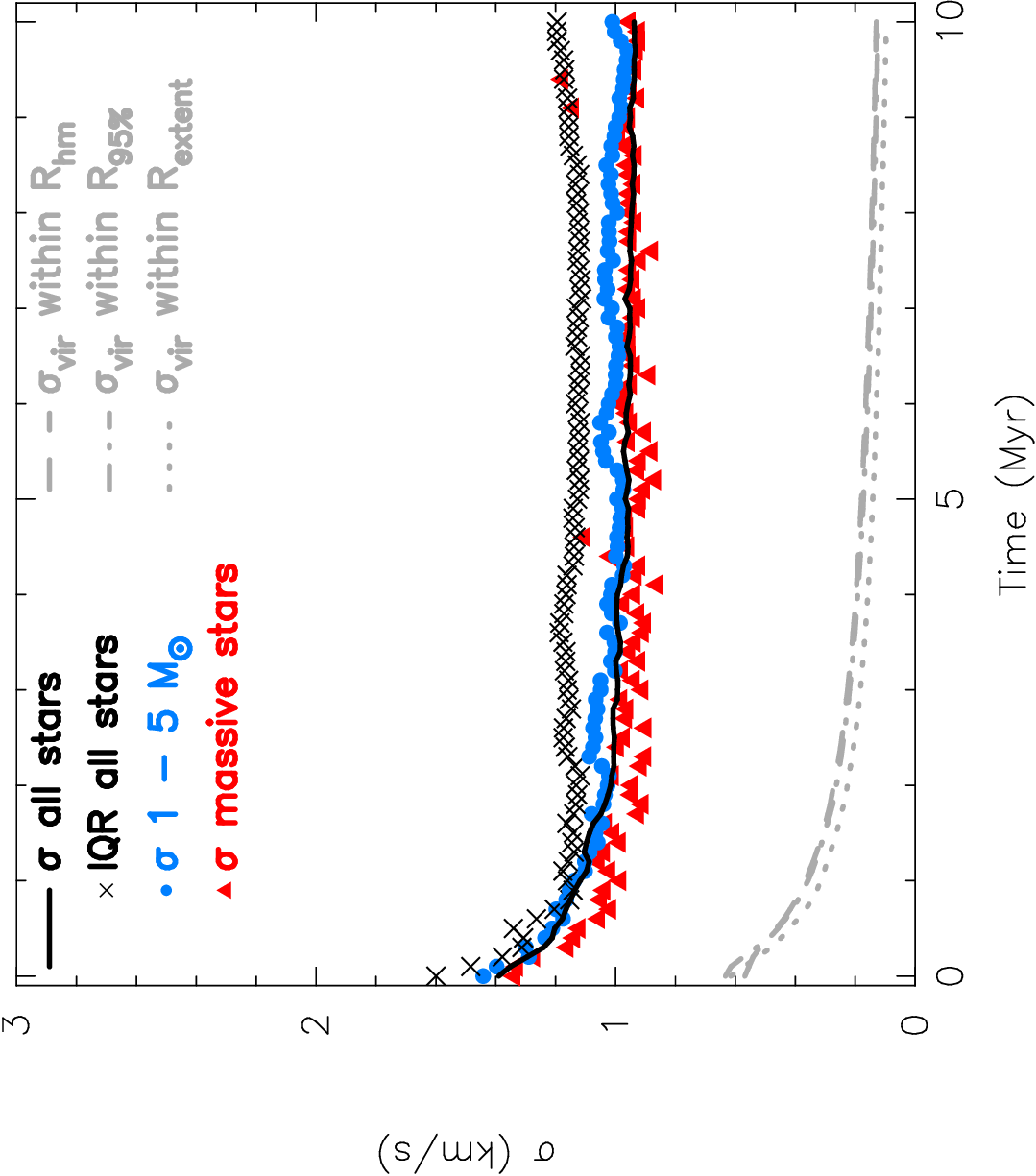}}}
\hspace*{0.3cm} 
\subfigure[$\alpha_{\rm vir} = 1.5, D = 3.0$]{\label{vdC_med-i}\rotatebox{270}{\includegraphics[scale=0.27]{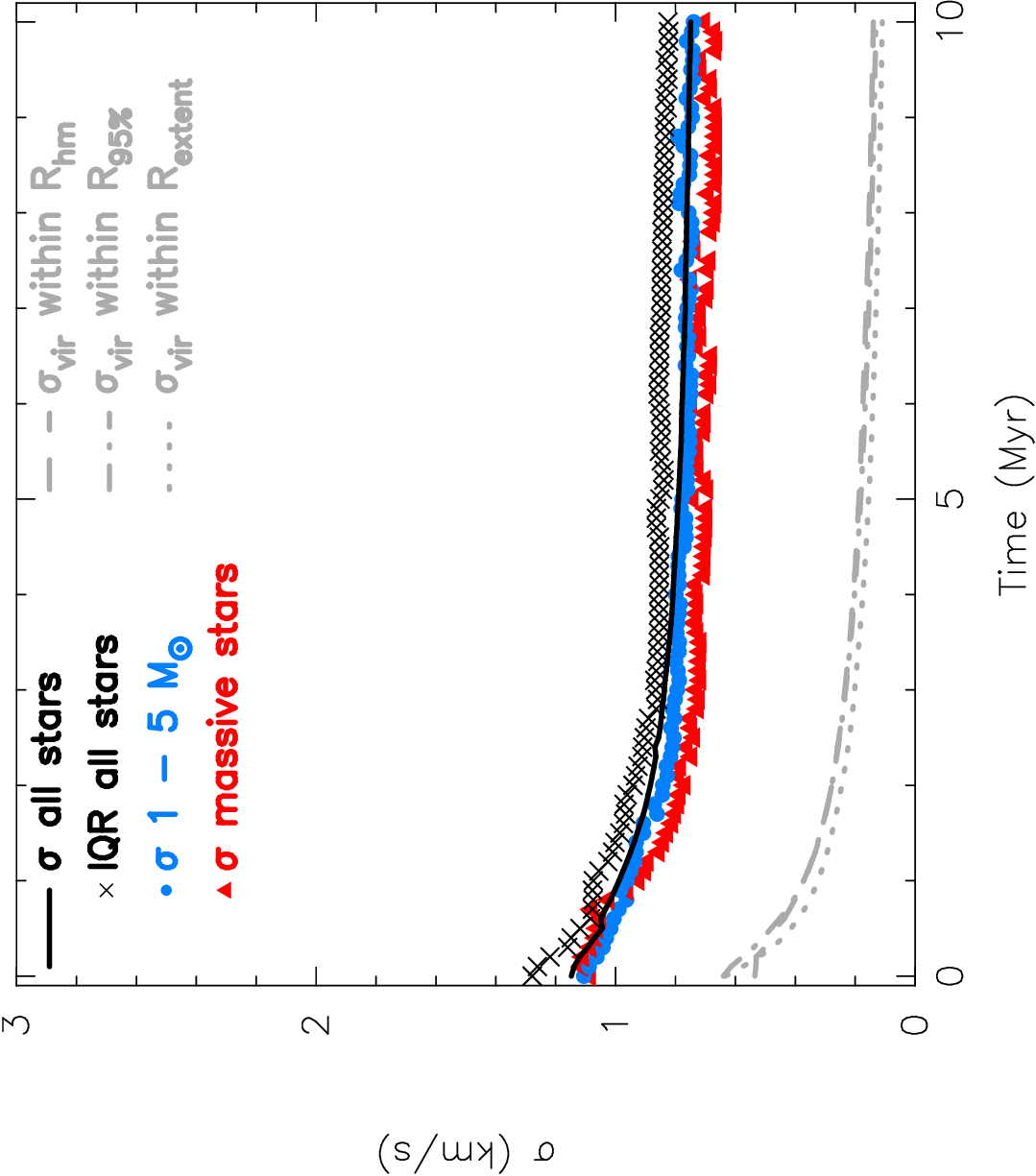}}}
\caption[bf]{Evolution of the median velocity dispersion for all simulations. The median value (from twenty simulations) of the velocity dispersion from each simulation is shown by the solid black line. The median value for stars with masses 1 -- 5\,M$_\odot$ is shown by the blue circles, and the median value for the 10 most massive stars is shown by the filled red triangles. We also show the median value for the IQR dispersion for all stars by the black crosses. We show the median maximum velocity dispersion for the region to still be in virial equilibrium by the broken grey lines; the dashed line is for objects within the half-mass radius, the dot-dashed line is for all stars within 95\,per cent of the extent of the region, and the dotted line is for the full region in the simulations. }
\label{vdC_med}
  \end{center}
\end{figure*}

\subsubsection{Evolution of the relative velocity dispersion}

We show the evolution of the median velocity dispersion from all 20 simulations in Fig.~\ref{vdC_med}. The solid lines show the statistical dispersion about the mean -- this is the median value from the 20 simulations. The dotted grey lines indicate $\sigma_{\rm vir}$ provided by Equation~\ref{virial_mass} for the entire extent of the region (i.e.\,\,all stars in the simulations, including those that have been dynamically ejected). The dot-dashed lines are the virial velocity dispersion estimate for the mass of the most central 95\,per cent of stars in the region, and the dashed grey lines are the virial velocity dispersion estimate within the half-mass radius. Again, we take the median value from 20 simulations at each timestep. The red triangles indicate the median velocity dispersions at each timestep for the 10 most massive stars, and the blue circles are the median velocity dispersion across 20 simulations for stars with masses 1 -- 5\,M$_\odot$. Finally, the crosses are the median values from 20 simulations for the IQR velocity dispersions for all stars. 

Several trends are apparent from Fig.~\ref{vdC_med}. First, in the regions undergoing some degree of violent relaxation ($\alpha_{\rm vir} = 0.3$, and $\alpha_{\rm vir} = 0.5, D = 1.6 - 2.0$) the most massive stars attain higher velocity dispersions as the simulations progress. \citet{Allison10} and \citet{Allison11} show that massive stars  within a star-forming region undergoing violent relaxation tend to form Trapezium-like systems, which dynamically evolve on shorter timescales than the entire region and `decouple' from the other stars. For this reason, they tend to have increased kinetic energy with respect to low-mass stars throughout the later stages of the simulations. Intermediate mass stars (1 -- 5\,M$_\odot$) also have (on average) higher velocity dispersions than the dispersion of the whole region in simulations that have high substructure and (sub)virial velocities (panels a and d). Secondly, the velocity dispersion defined by the IQR falls significantly below the statistical dispersion about the mean radial velocity in the regions undergoing the most extreme violent relaxation ($\alpha_{\rm vir} = 0.3, D = 1.6 - 2.0$ and $\alpha_{\rm vir} = 0.5, D = 1.6$), suggesting that the velocity distribution is evolving significantly due to dynamical interactions. Third, unless we restrict the virial mass estimate of the velocity dispersion, $\sigma_{\rm vir}$, to within the half-mass radius of the clusters that form, all of the regions evolve to be out of virial equilibrium -- i.e.\,\,supervirial, after 5\,Myr.

This third and final point warrants further investigation. Many studies have compared the observed radial velocity dispersion in star-forming regions and clusters and compared the measured $\sigma$ to the virial mass estimate,  $\sigma_{\rm vir}$, to draw conclusions on the virial `state' of a particular region. We have attempted to mimic an observational analysis here, but we also have access to the full data in the simulations and can therefore measure the virial ratio directly. 

\begin{figure*}
  \begin{center}
\setlength{\subfigcapskip}{10pt}
\vspace*{-0.3cm}
\hspace*{-1.cm}\subfigure[$\alpha_{\rm vir} = 0.3, D = 1.6$]{\label{vir_rat-a}\rotatebox{270}{\includegraphics[scale=0.27]{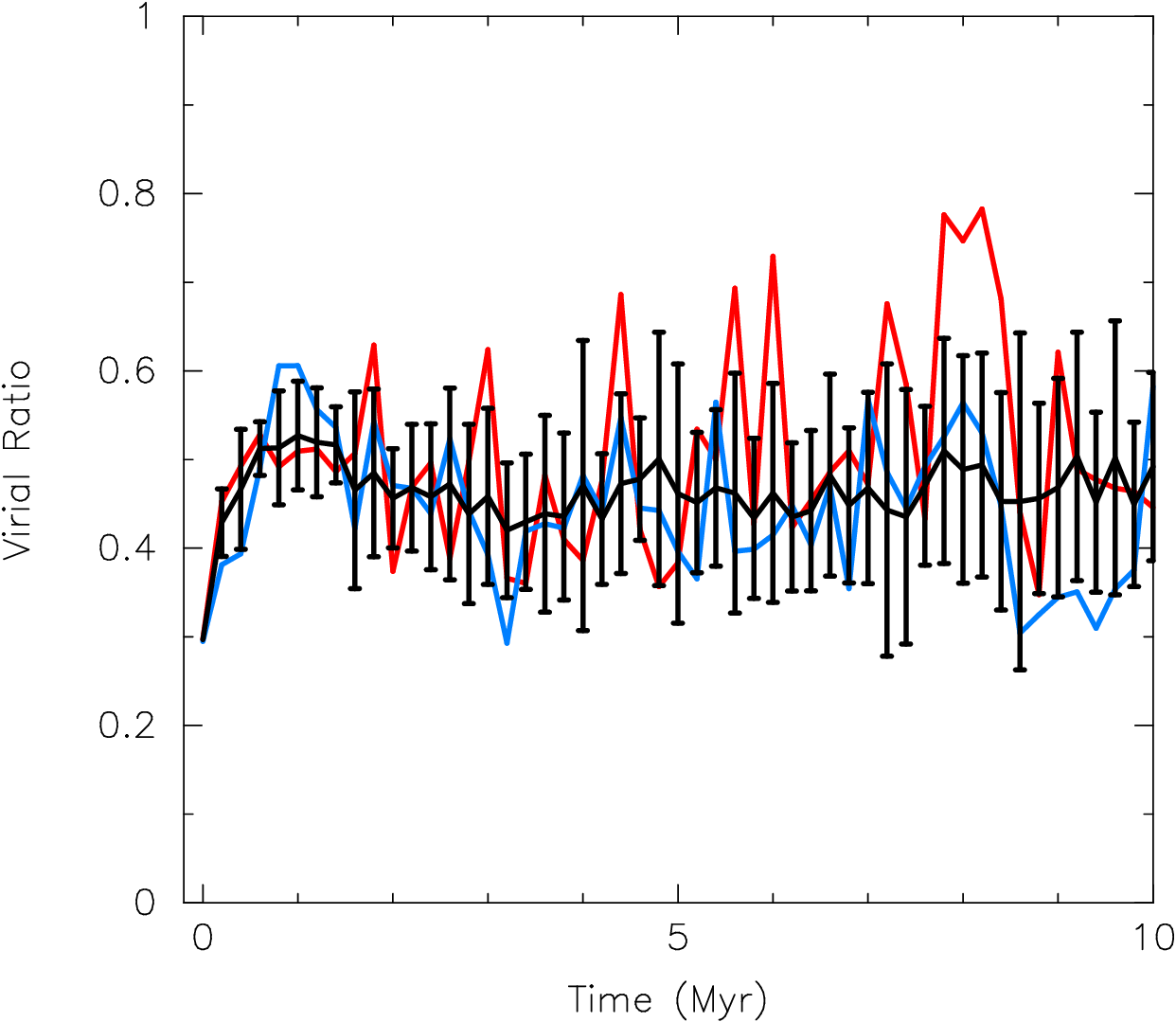}}}
\hspace*{0.3cm} 
\subfigure[$\alpha_{\rm vir} = 0.3, D = 2.0$]{\label{vir_rat-b}\rotatebox{270}{\includegraphics[scale=0.27]{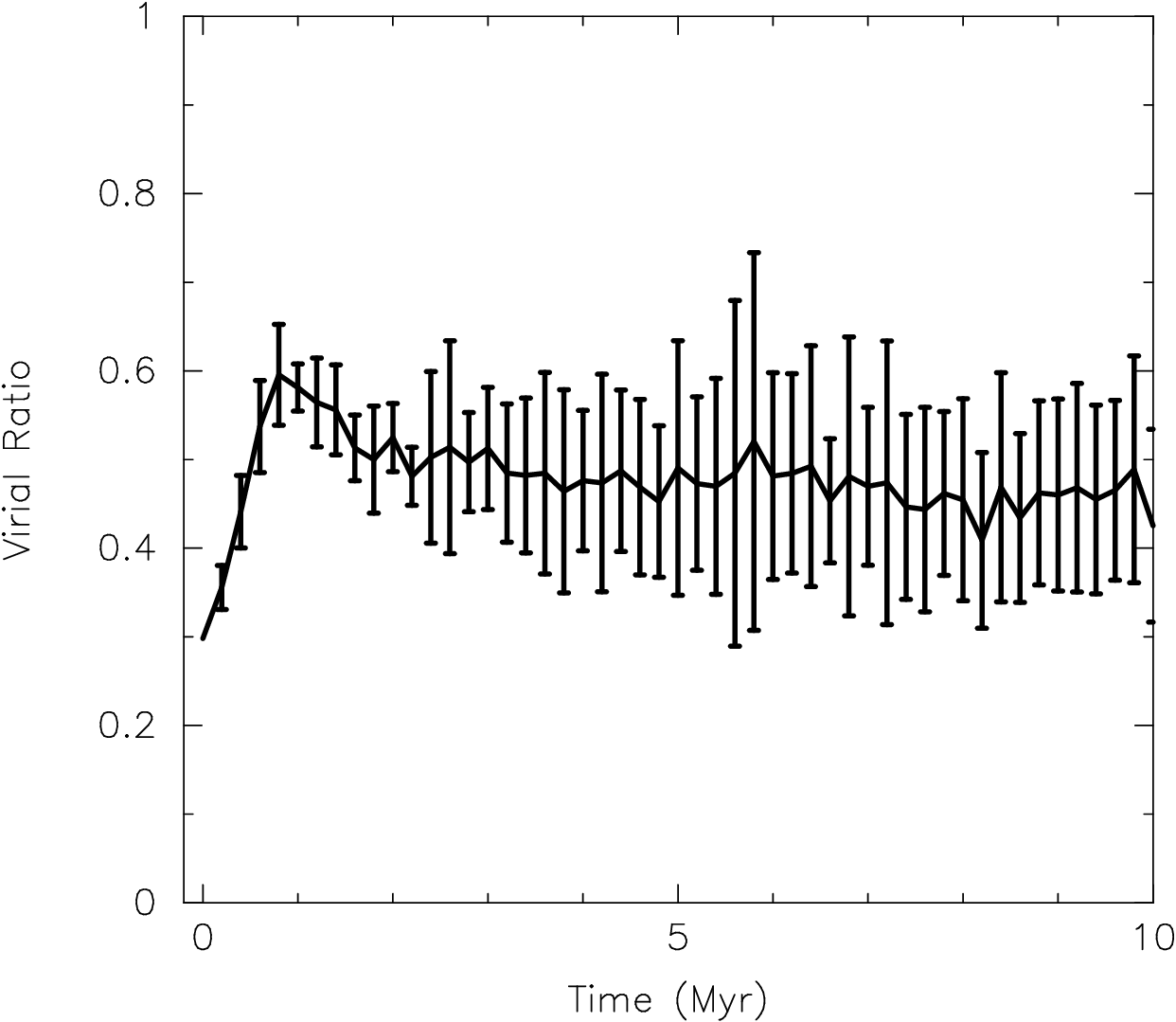}}}
\hspace*{0.3cm} 
\subfigure[$\alpha_{\rm vir} = 0.3, D = 3.0$]{\label{vir_rat-c}\rotatebox{270}{\includegraphics[scale=0.27]{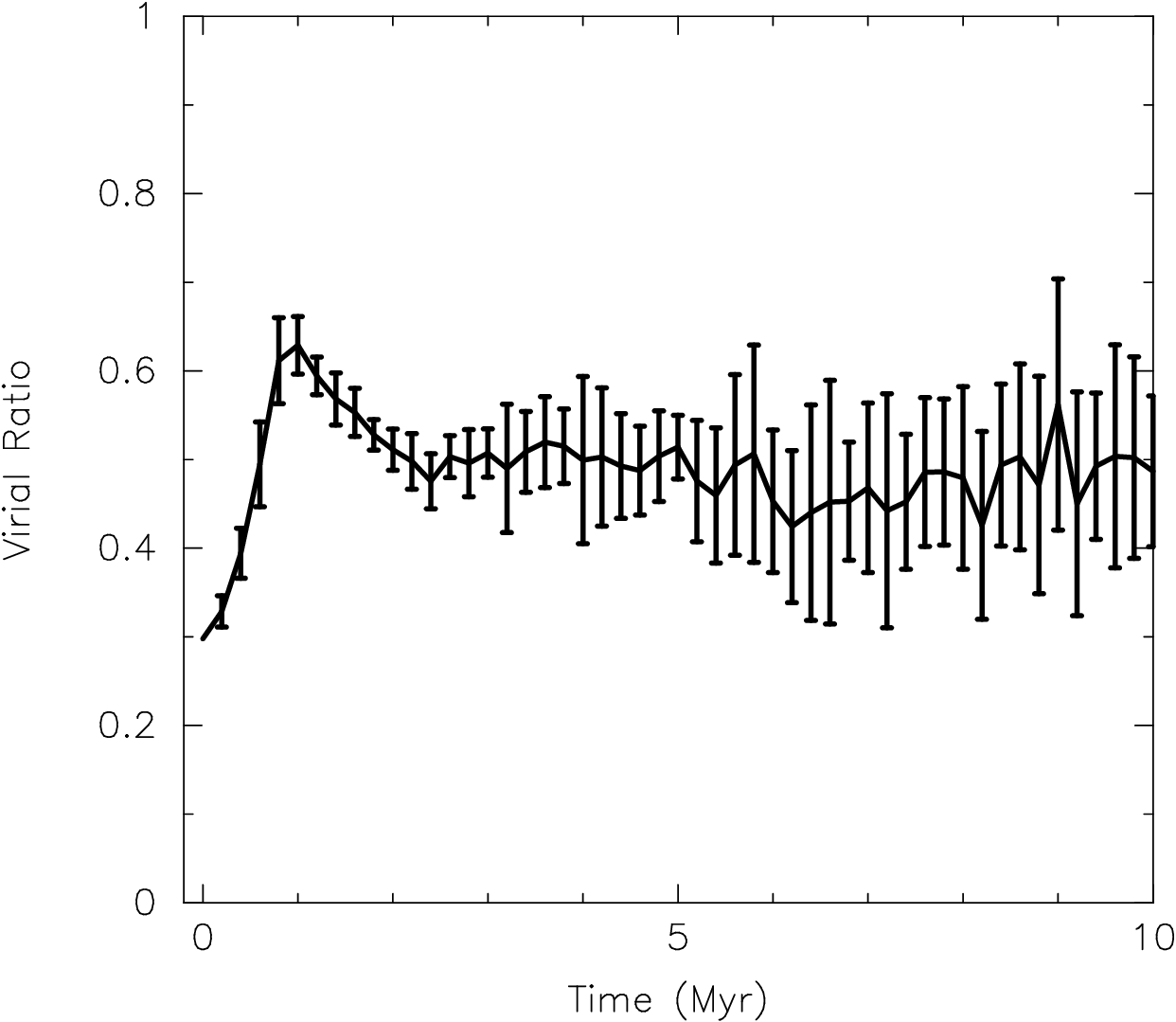}}}
%\vspace*{-0.3cm}
\hspace*{-1.cm}
\subfigure[$\alpha_{\rm vir} = 0.5, D = 1.6$]{\label{vir_rat-d}\rotatebox{270}{\includegraphics[scale=0.27]{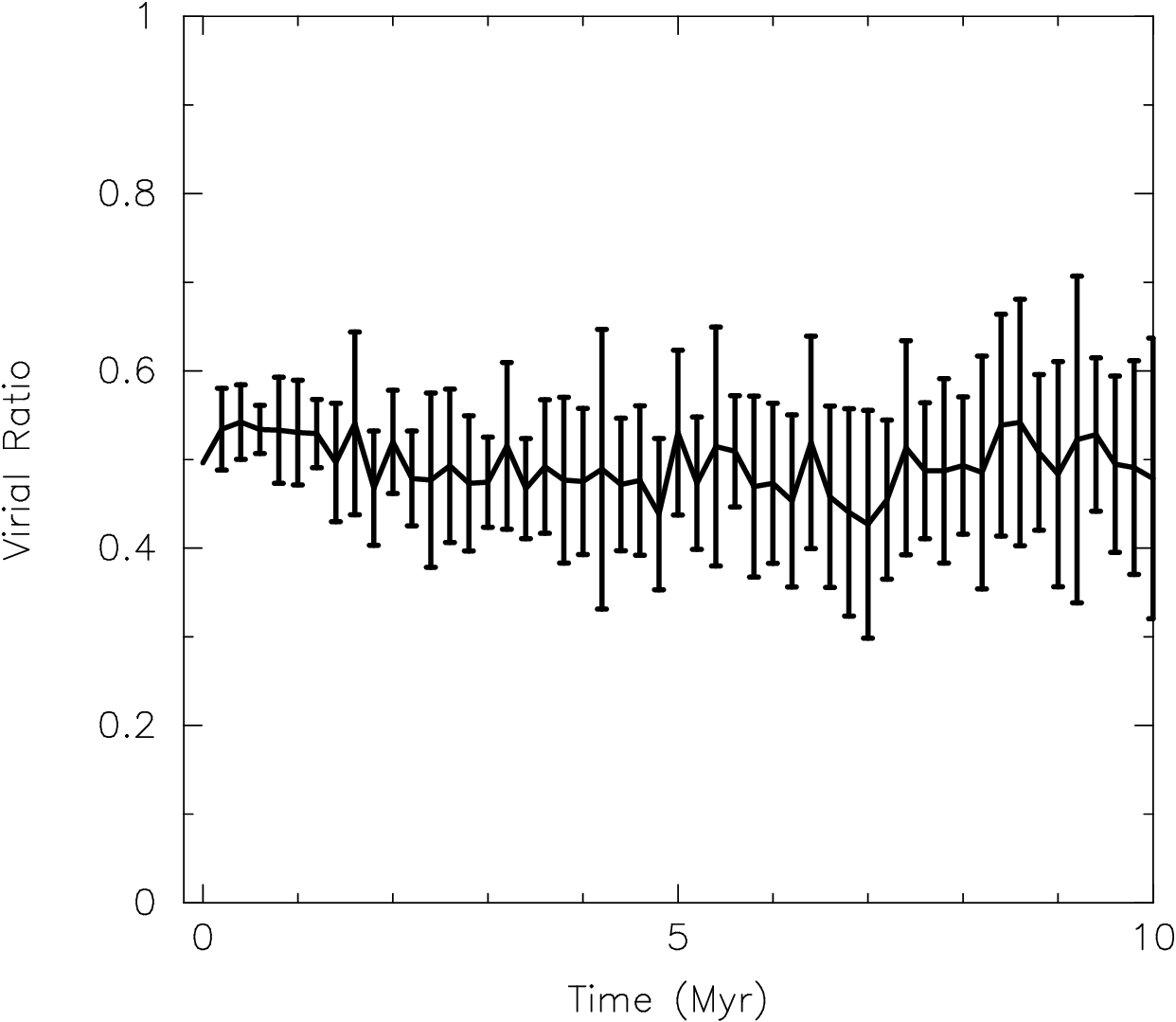}}}
\hspace*{0.3cm} 
\subfigure[$\alpha_{\rm vir} = 0.5, D = 2.0$]{\label{vir_rat-e}\rotatebox{270}{\includegraphics[scale=0.27]{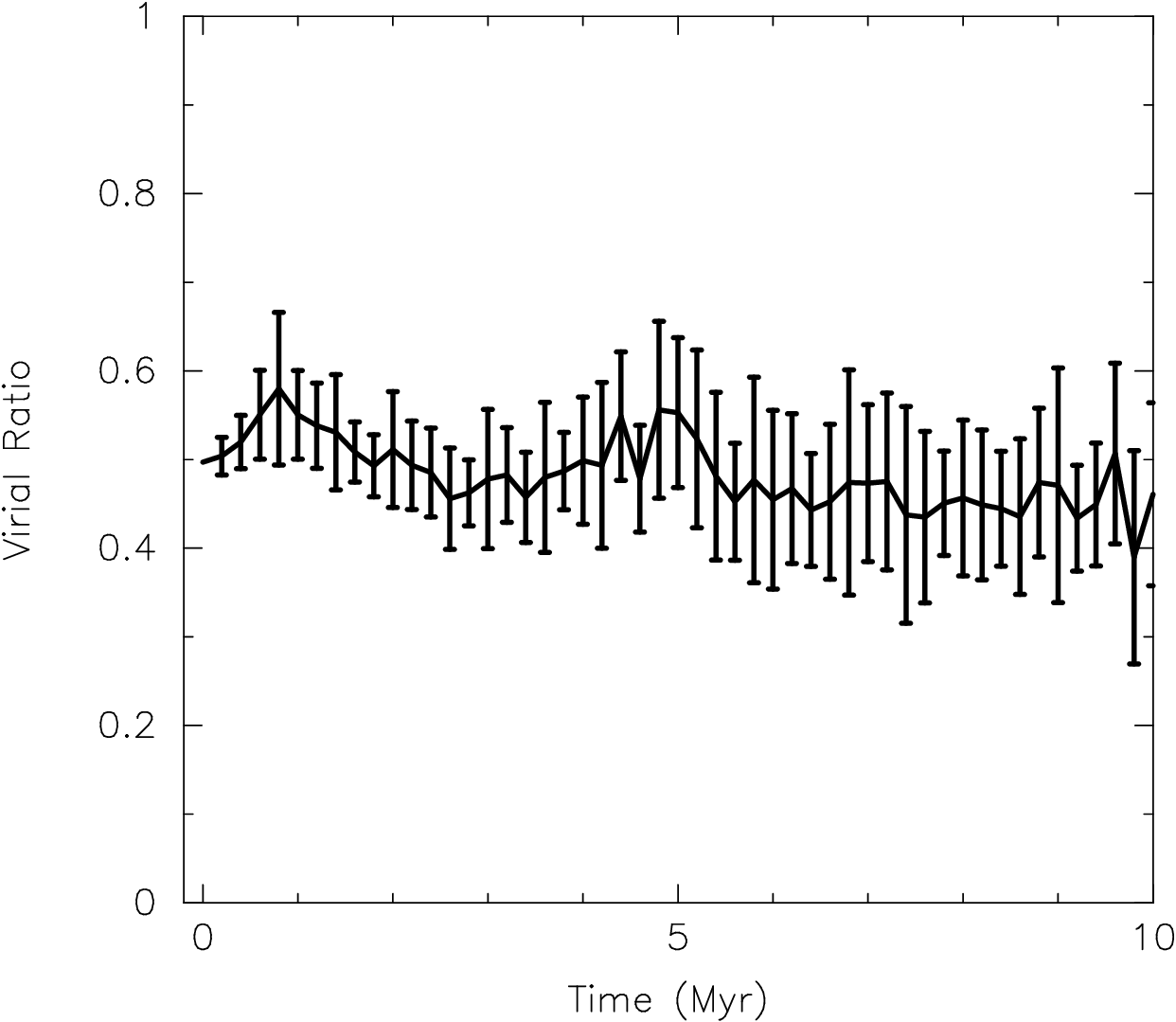}}}
\hspace*{0.3cm} 
\subfigure[$\alpha_{\rm vir} = 0.5, D = 3.0$]{\label{vir_rat-f}\rotatebox{270}{\includegraphics[scale=0.27]{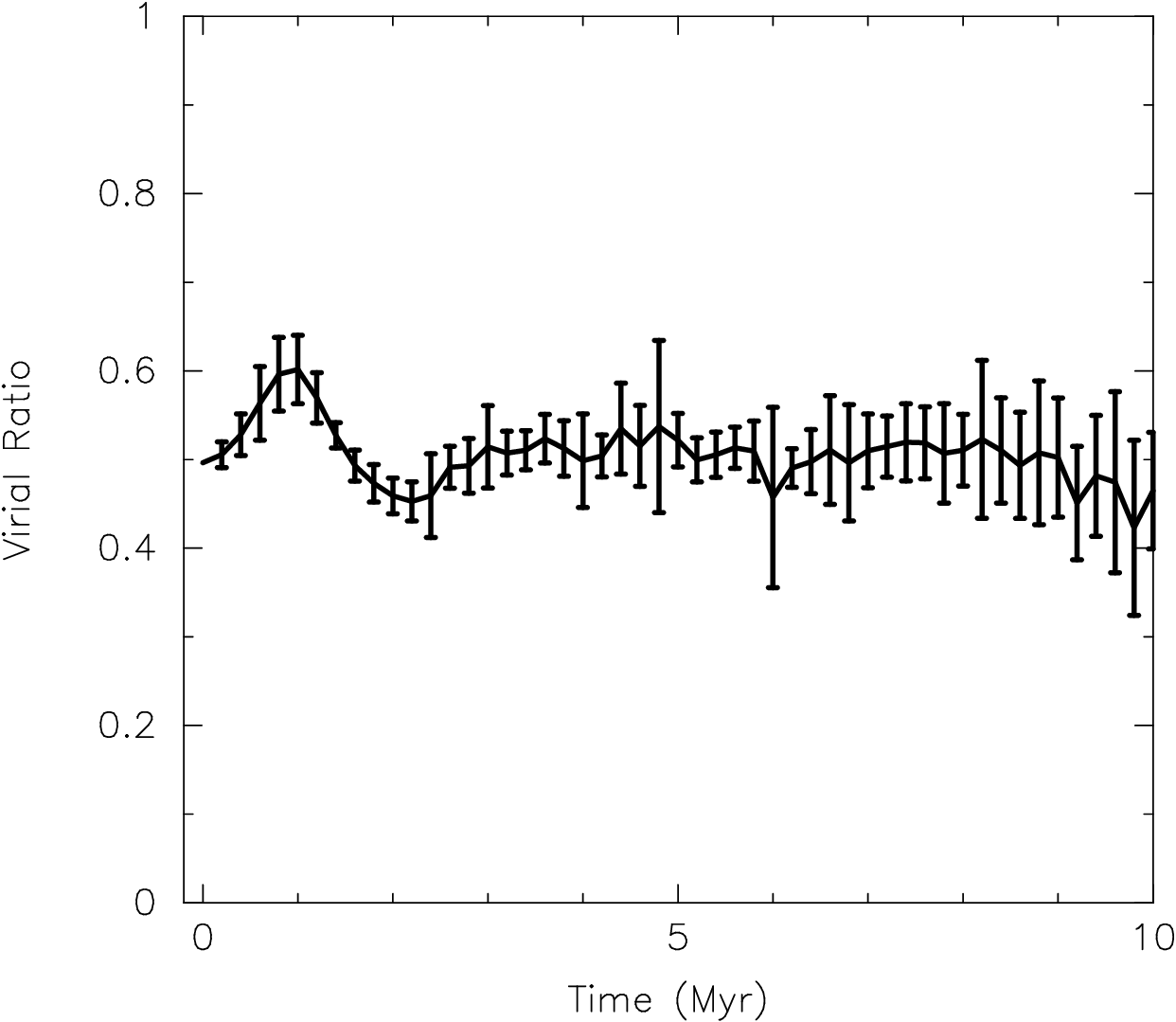}}}
\hspace*{-1.cm}
\subfigure[$\alpha_{\rm vir} = 1.5, D = 1.6$]{\label{vir_rat-g}\rotatebox{270}{\includegraphics[scale=0.27]{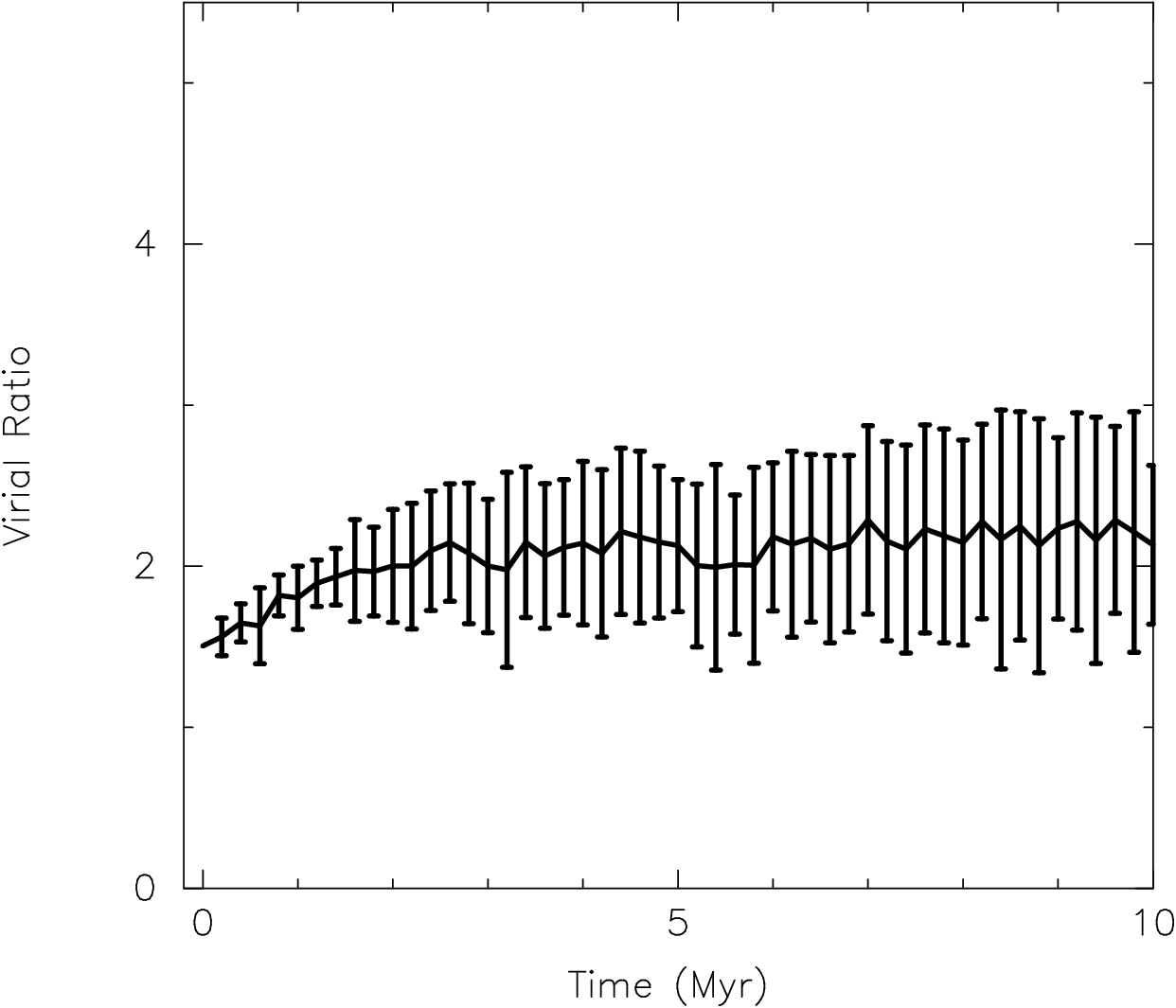}}}
\hspace*{0.3cm} 
\subfigure[$\alpha_{\rm vir} = 1.5, D = 2.0$]{\label{vir_rat-h}\rotatebox{270}{\includegraphics[scale=0.27]{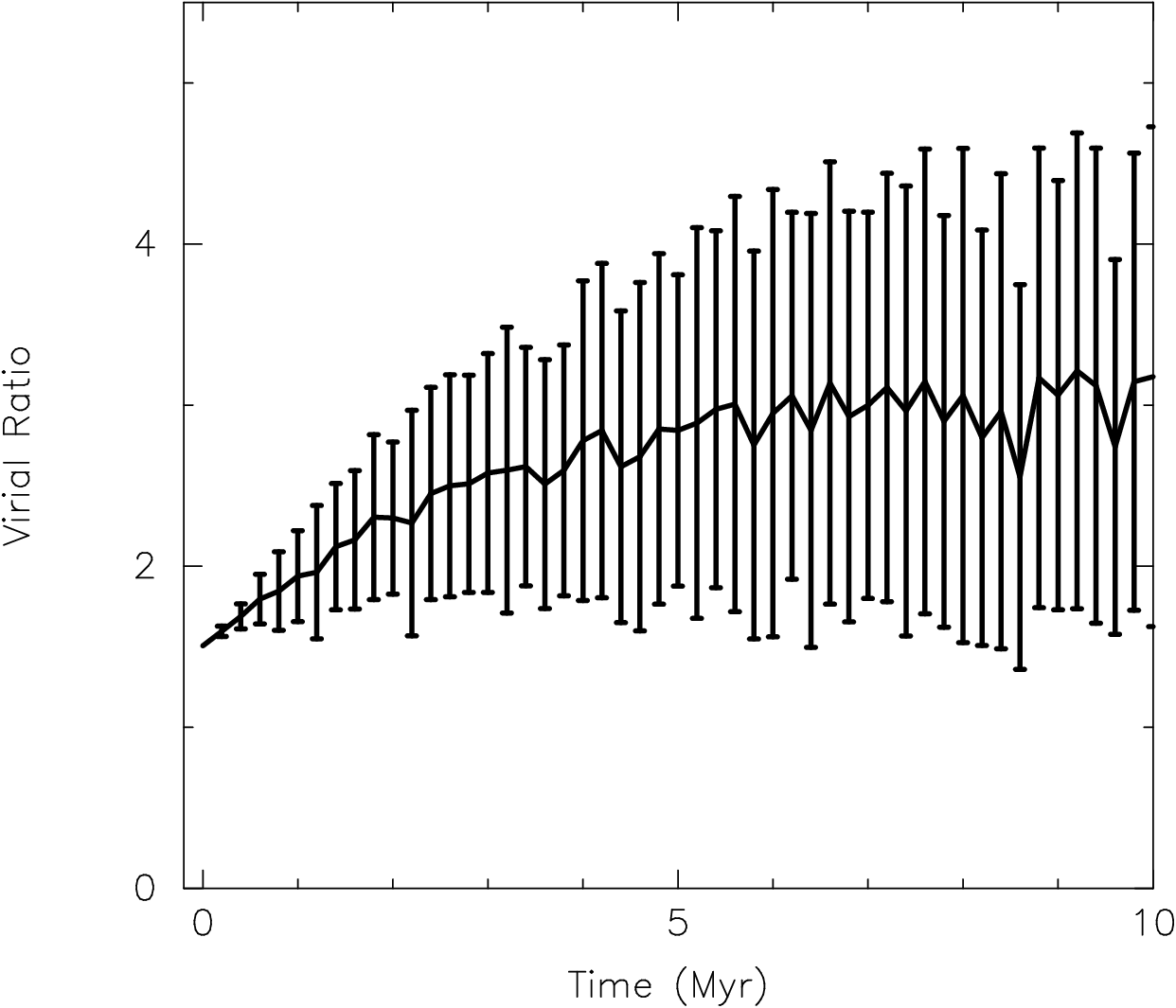}}}
\hspace*{0.3cm} 
\subfigure[$\alpha_{\rm vir} = 1.5, D = 3.0$]{\label{vir_rat-i}\rotatebox{270}{\includegraphics[scale=0.27]{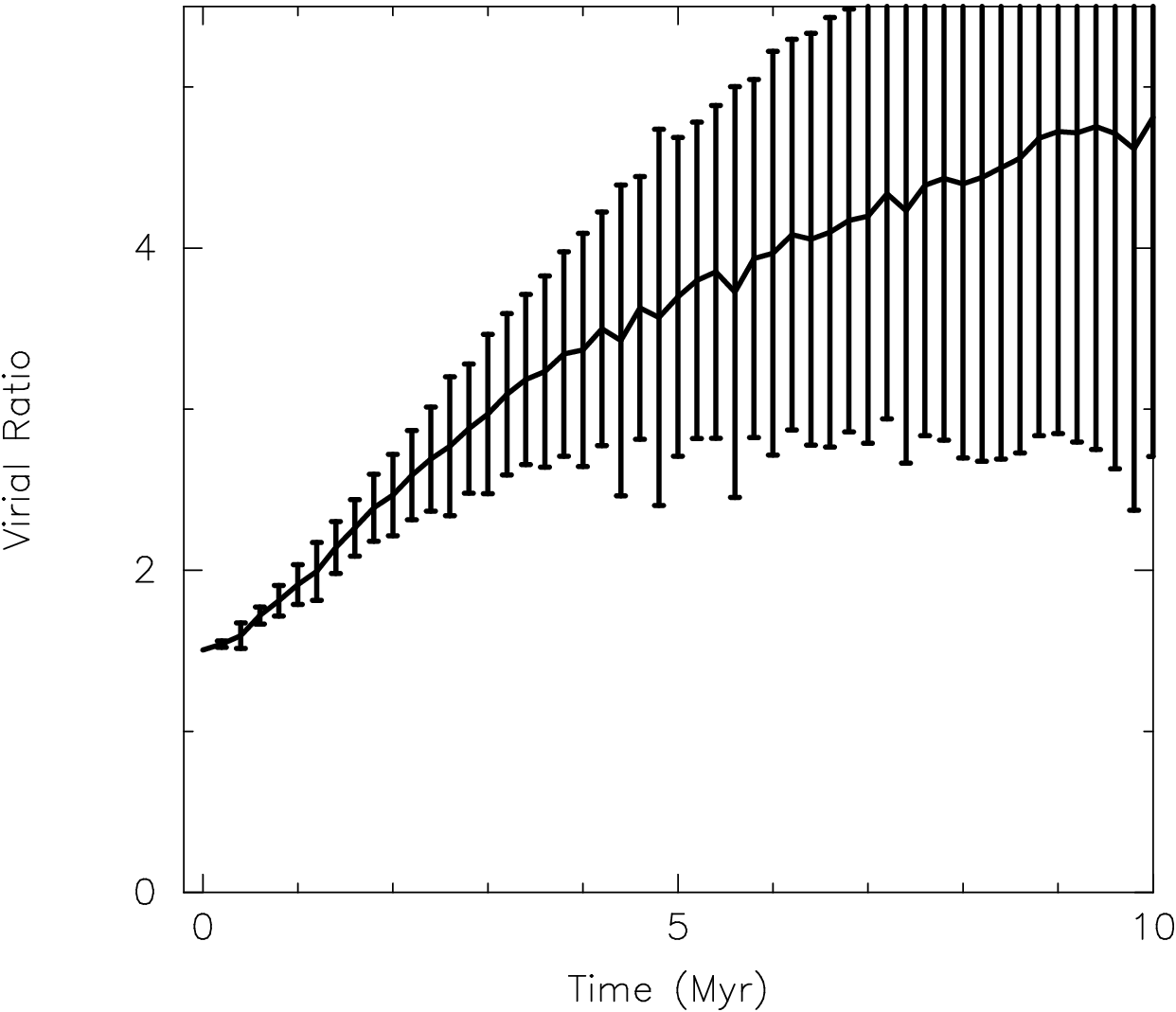}}}
\caption[bf]{Evolution of the average virial ratio (with one sigma uncertainties) from 10 realisations of each set of initial conditions in the $N = 1500$ simulations. In panel (a) we also show the evolution of the virial ratio in two individual simulations (the red and blue lines) to demonstrate the level of fluctuation that occurs. Note the difference in scale on the y-axis in panels (g -- i) -- the supervirial simulations.}
\label{vir_rat}
  \end{center}
\end{figure*}

\subsubsection{Evolution of the virial ratio}

Once dynamical evolution takes place, the evolution of the virial ratio is rather noisy \citep[see also][]{Moeckel10,Smith11}. In Fig.~\ref{vir_rat} we show the evolution of the average virial ratio from 10 different realisations of the same initial conditions at a given time, with 1-sigma uncertainties. In Fig.~\ref{vir_rat-a} we also show the evolution of the individual virial ratio from two of these simulations by the red and blue lines, to demonstrate how noisy its evolution really is. However, taking the average suggests that the regions that begin with cool velocities ($\alpha_{\rm vir} = 0.3$) are likely to undergo a slightly supervirial phase within the first 3\,Myr, where they may reach $\alpha_{\rm vir} = 0.6$, before relaxing and reaching virial equilibrium (albeit with some fluctuation -- note the large uncertainties in panels (a -- c)).

The simulations that start in virial equilbrium  ($\alpha_{\rm vir} = 0.5$; panels d -- f) fluctuate, but remain close to virial equilibrium throughout. The information provided by the radial velocity dispersions (Fig.~\ref{vdC_med}) seems to contradict this picture, even when considering the average behaviour of 20 simulations. In part, this is likely due to the arbitrary way in which the virial mass velocity dispersion is defined, but is most probably because the one-dimensional velocity dispersion is being used as a proxy for the full seven-dimensional calculation of the virial ratio, which requires the mass, position vector and velocity vector for every star to be known. Even with this information, the virial ratio fluctuates due to the violent dynamical evolution within the system.

Finally, the simulations that start supervirial ($\alpha_{\rm vir} = 1.5$) remain supervirial, and their virial ratio increases (panels g -- i). The simulations that have $D = 3.0$, i.e. smooth, spherical initial conditions display the largest increase in virial ratio. This is due to the absence of substructure, so there is little velocity correlation on local scales and the regions therefore expand at a faster rate. These supervirial regions are the only simulations that are consistent with the virial state estimate from the velocity dispersion analysis.

\begin{figure*}
  \begin{center}
\setlength{\subfigcapskip}{10pt}
\vspace*{-0.3cm}
\hspace*{-1.cm}\subfigure[$\alpha_{\rm vir} = 0.3, D = 1.6$]{\label{vdC_med_bound-a}\rotatebox{270}{\includegraphics[scale=0.27]{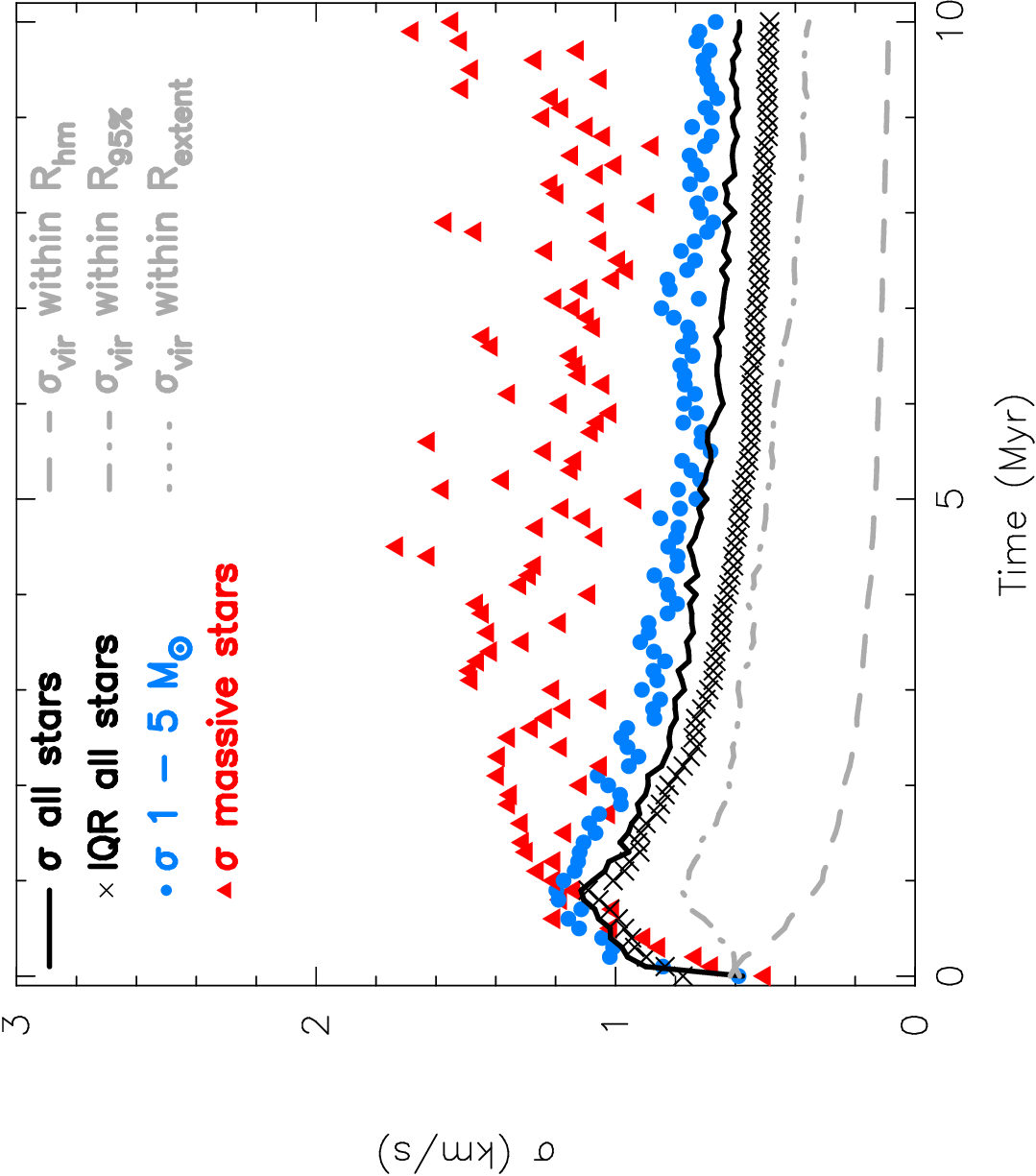}}}
\hspace*{0.3cm} 
\subfigure[$\alpha_{\rm vir} = 0.3, D = 2.0$]{\label{vdC_med_bound-b}\rotatebox{270}{\includegraphics[scale=0.27]{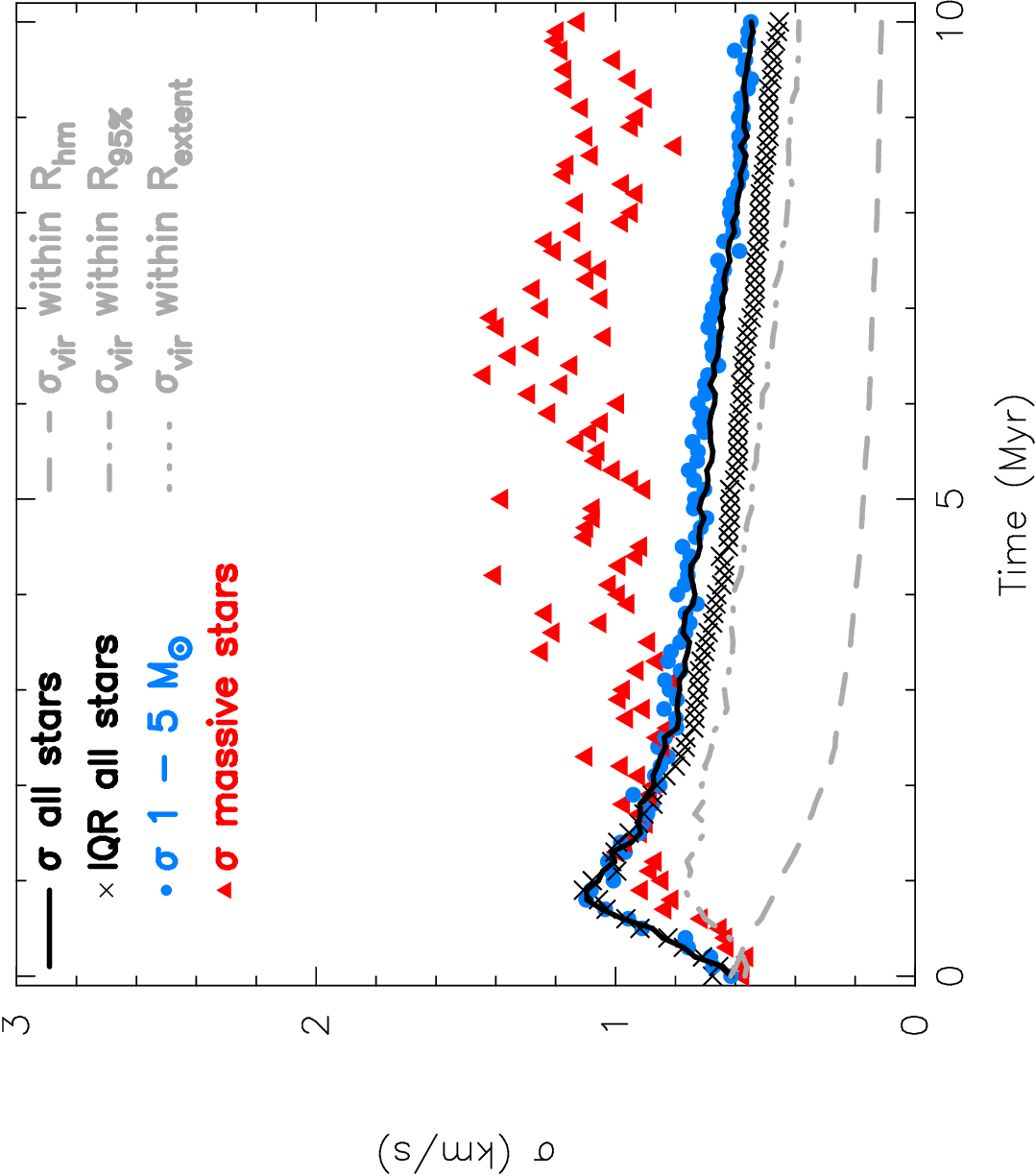}}}
\hspace*{0.3cm} 
\subfigure[$\alpha_{\rm vir} = 0.3, D = 3.0$]{\label{vdC_med_bound-c}\rotatebox{270}{\includegraphics[scale=0.27]{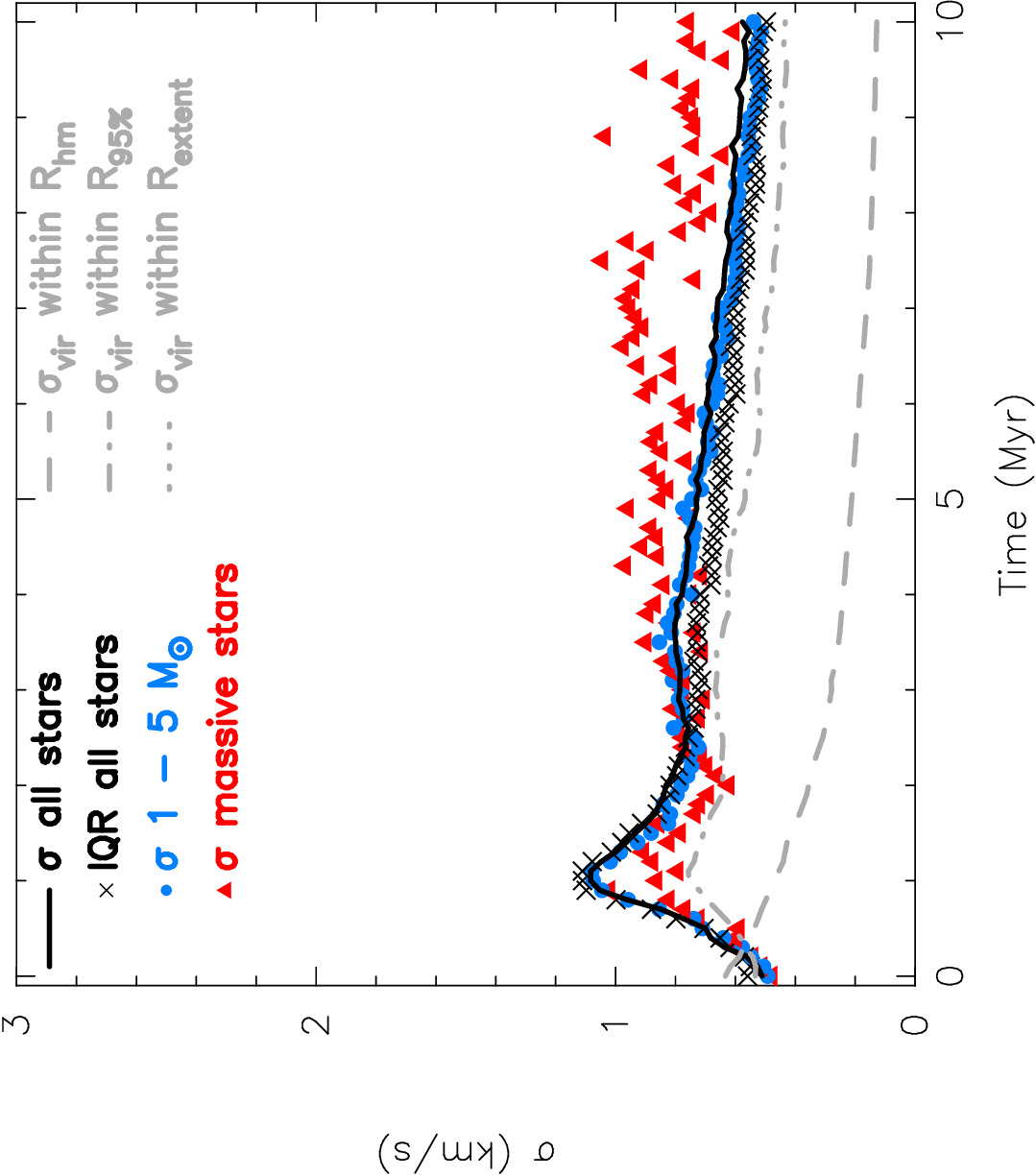}}}
%\vspace*{-0.3cm}
\caption[bf]{Evolution of the median velocity dispersion values with time for bound stars only for the three sets of subvirial initial conditions. The median value (from twenty simulations) of the velocity dispersion for all stars is shown by the solid black line; the median value for stars with masses 1 -- 5\,M$_\odot$ is shown by the blue circles, and the median value for the 10 most massive stars is shown by the filled red triangles. The median velocity dispersion calculated using the IQR is shown by the black crosses. We show the median maximum velocity dispersion for the region to still be in virial equilibrium by the broken grey lines; the dashed line assumes the furthest bound star comprises the cluster radius and the dot-dashed line is the estimate using the half-mass radii of all bound stars.}
\label{vdC_med_bound}
  \end{center}
\end{figure*}

When determining the virial ratio, the sum of kinetic and potential energy for every star is used. It is possible that the effects of ejected stars have more influence on the radial velocity dispersion than the virial ratio. In Fig.~\ref{vdC_med_bound} we show the median velocity dispersions for stars that remain bound to the region, where a bound star is an object that has negative total energy ($T_i + V_i < 0$, where $T_i$ and $V_i $ are the kinetic and potential energies of a star, respectively). The potential energy is given by
\begin{equation}
V_i = - \sum\limits_{i \not= j} \frac{Gm_im_j}{r_{ij}},
\end{equation} 
where $m_i$ and $m_j$ are the masses of two stars and $r_{ij}$ is the distance between them. The kinetic energy of a star, $T_i$ is given thus:
\begin{equation}
T_i = \frac{1}{2}m_i|{\bf v}_i - {\bf v}_{\rm cl}|^2, 
\end{equation}
where ${\bf v}_i$ and ${\bf v}_{\rm cl}$ are the velocity vectors of the star and the centre of mass of the region, respectively.

In Fig.~\ref{vdC_med_bound} we only show the subvirial simulations, but the effects are also similar for the initially virial simulations (the regions that start supervirial are unbound by definition, and their velocity dispersions evolve accordingly). As in Fig.~\ref{vdC_med}, the solid line is the median statistical dispersion, the black crosses are the IQR (both for all bound stars). The median value for bound stars with masses 1 -- 5\,M$_\odot$ is shown by the blue circles, and the median value for the 10 most massive bound stars is shown by the filled red triangles. 

The IQR dispersion still drops below the statistical dispersion due to dynamical evolution, but not to the same extent as for the full data. We also plot the evolution of the virial mass dispersion estimate, $\sigma_{\rm vir}$, using both the extent of the furthest bound star as the radius the dashed line, and the half-mass radius of the bound stars (dot-dashed line). As in the analysis which includes every star, the virial mass estimate still falls below the measured velocity dispersions, suggesting that the regions are supervirial when in reality they have attained virial equilibrium.

\subsubsection{Kinematic evolution of low density systems}

\begin{figure*}
  \begin{center}
\setlength{\subfigcapskip}{10pt}
\vspace*{-0.3cm}
\hspace*{-1.cm}\subfigure[$\alpha_{\rm vir} = 0.3, D = 1.6$]{\label{vdC_medOH-a}\rotatebox{270}{\includegraphics[scale=0.27]{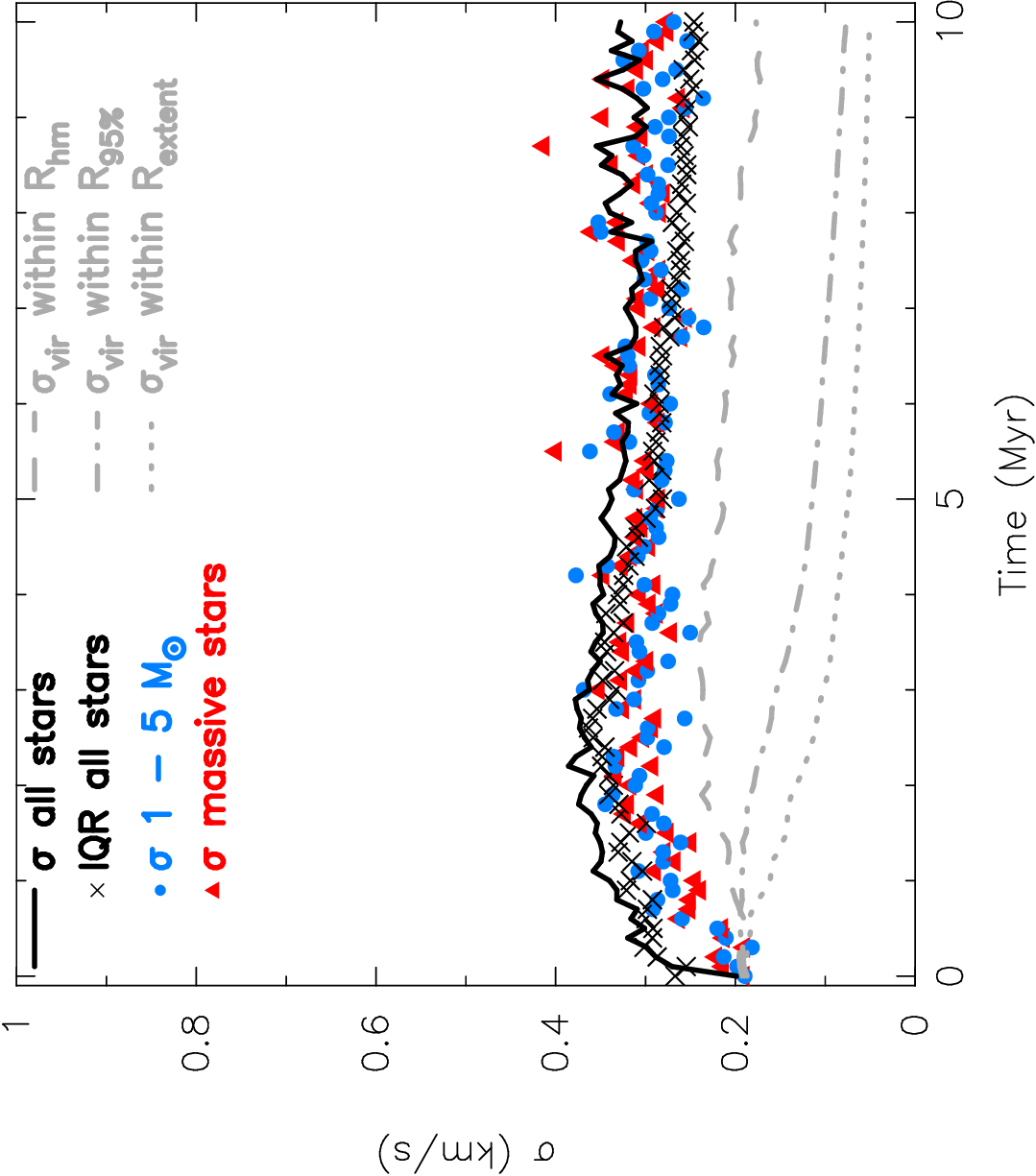}}}
\hspace*{0.3cm} 
\subfigure[$\alpha_{\rm vir} = 0.3, D = 2.0$]{\label{vdC_medOH-b}\rotatebox{270}{\includegraphics[scale=0.27]{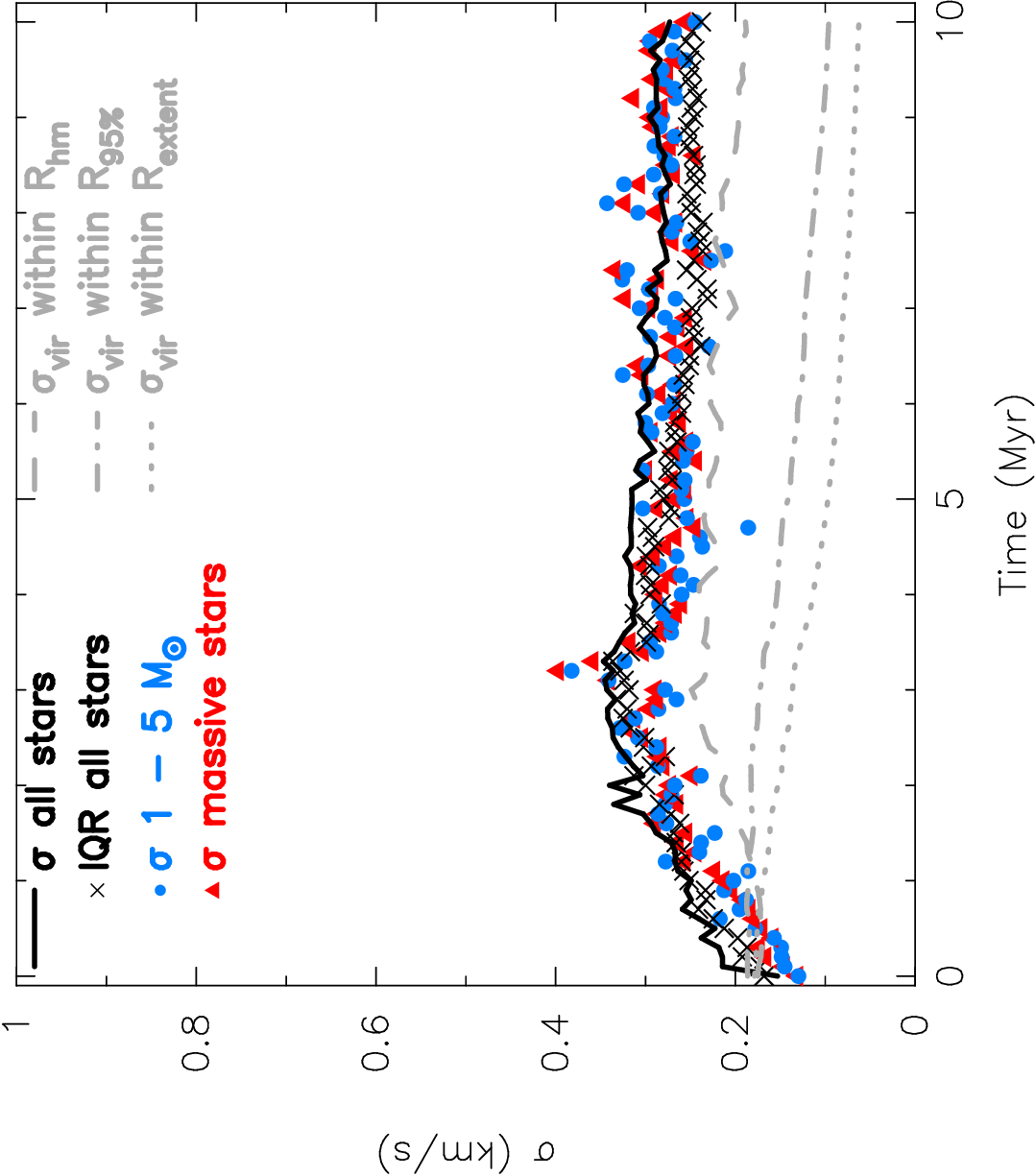}}}
\hspace*{0.3cm} 
\subfigure[$\alpha_{\rm vir} = 0.3, D = 3.0$]{\label{vdC_medOH-c}\rotatebox{270}{\includegraphics[scale=0.27]{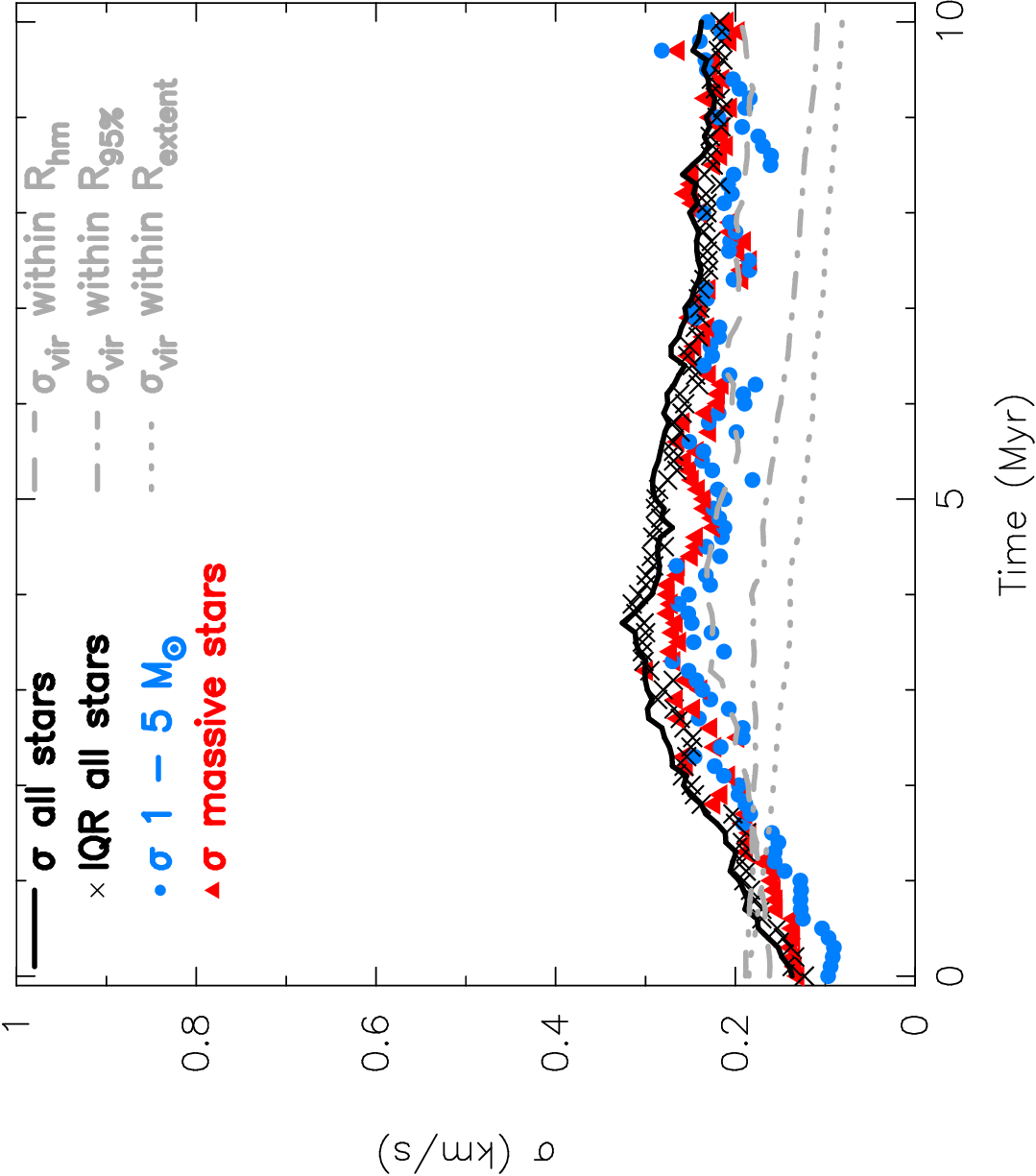}}}
%\vspace*{-0.3cm}
\hspace*{-1.cm}
\subfigure[$\alpha_{\rm vir} = 0.5, D = 1.6$]{\label{vdC_medOH-d}\rotatebox{270}{\includegraphics[scale=0.27]{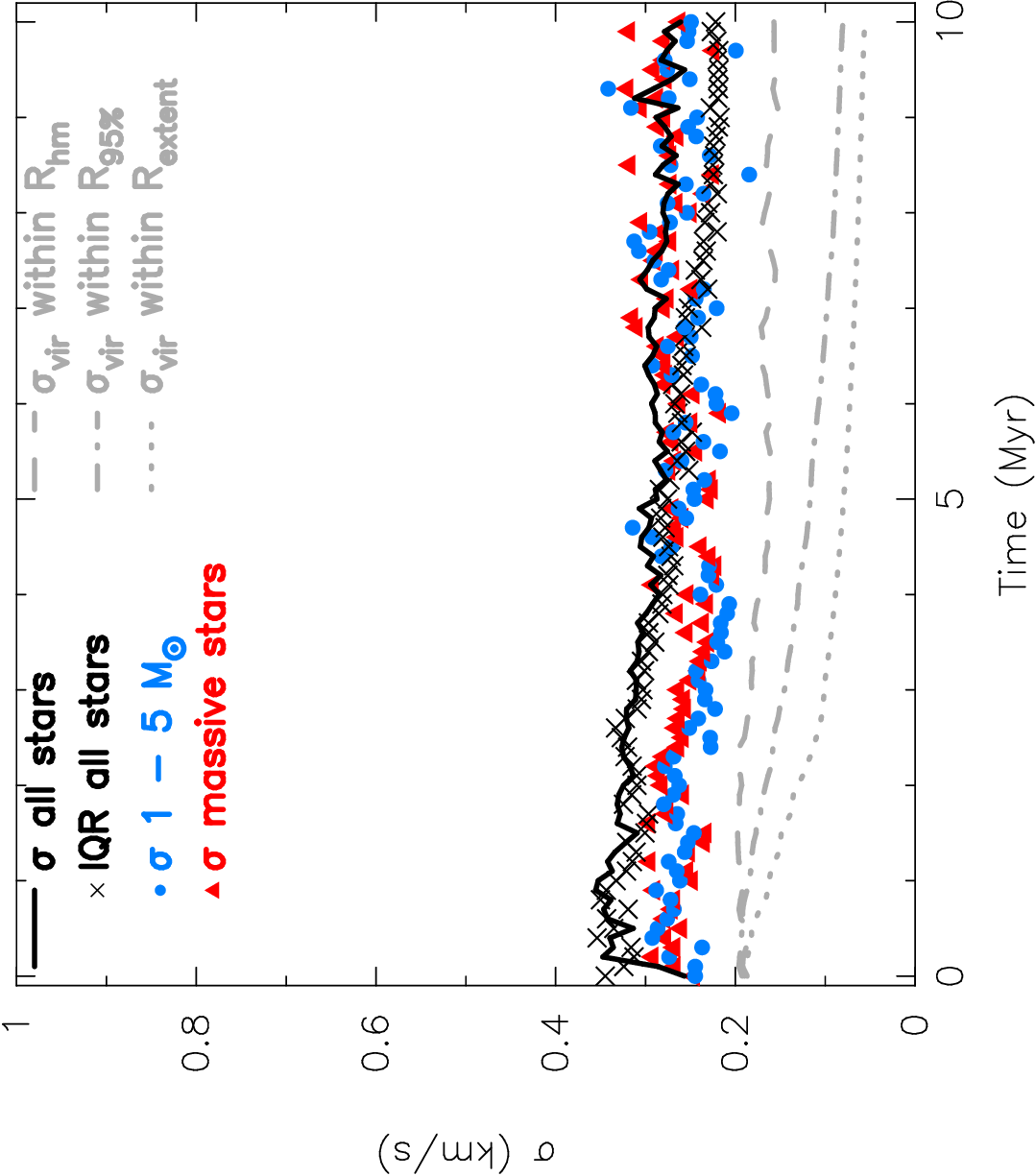}}}
\hspace*{0.3cm} 
\subfigure[$\alpha_{\rm vir} = 0.5, D = 2.0$]{\label{vdC_medOH-e}\rotatebox{270}{\includegraphics[scale=0.27]{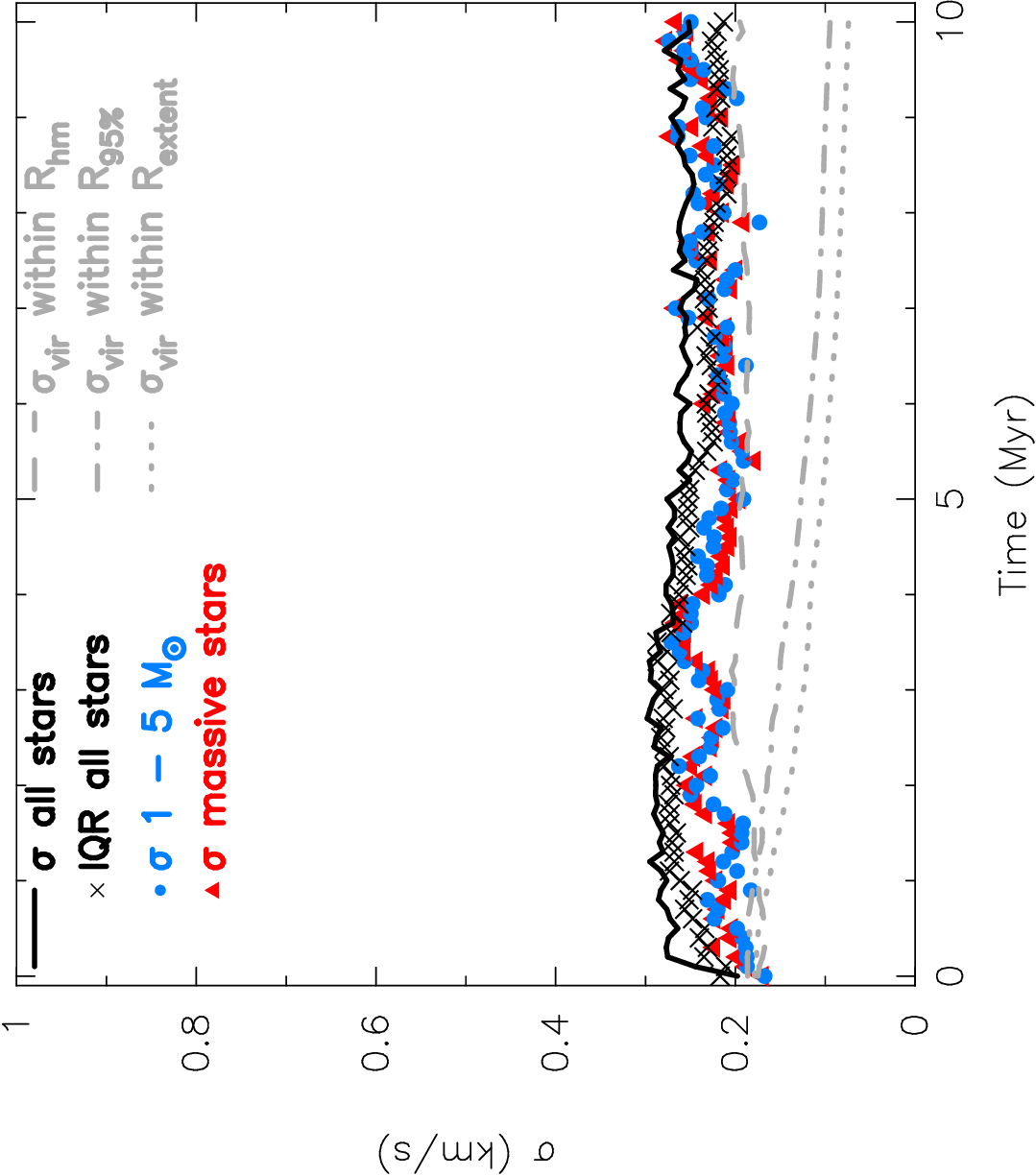}}}
\hspace*{0.3cm} 
\subfigure[$\alpha_{\rm vir} = 0.5, D = 3.0$]{\label{vdC_medOH-f}\rotatebox{270}{\includegraphics[scale=0.27]{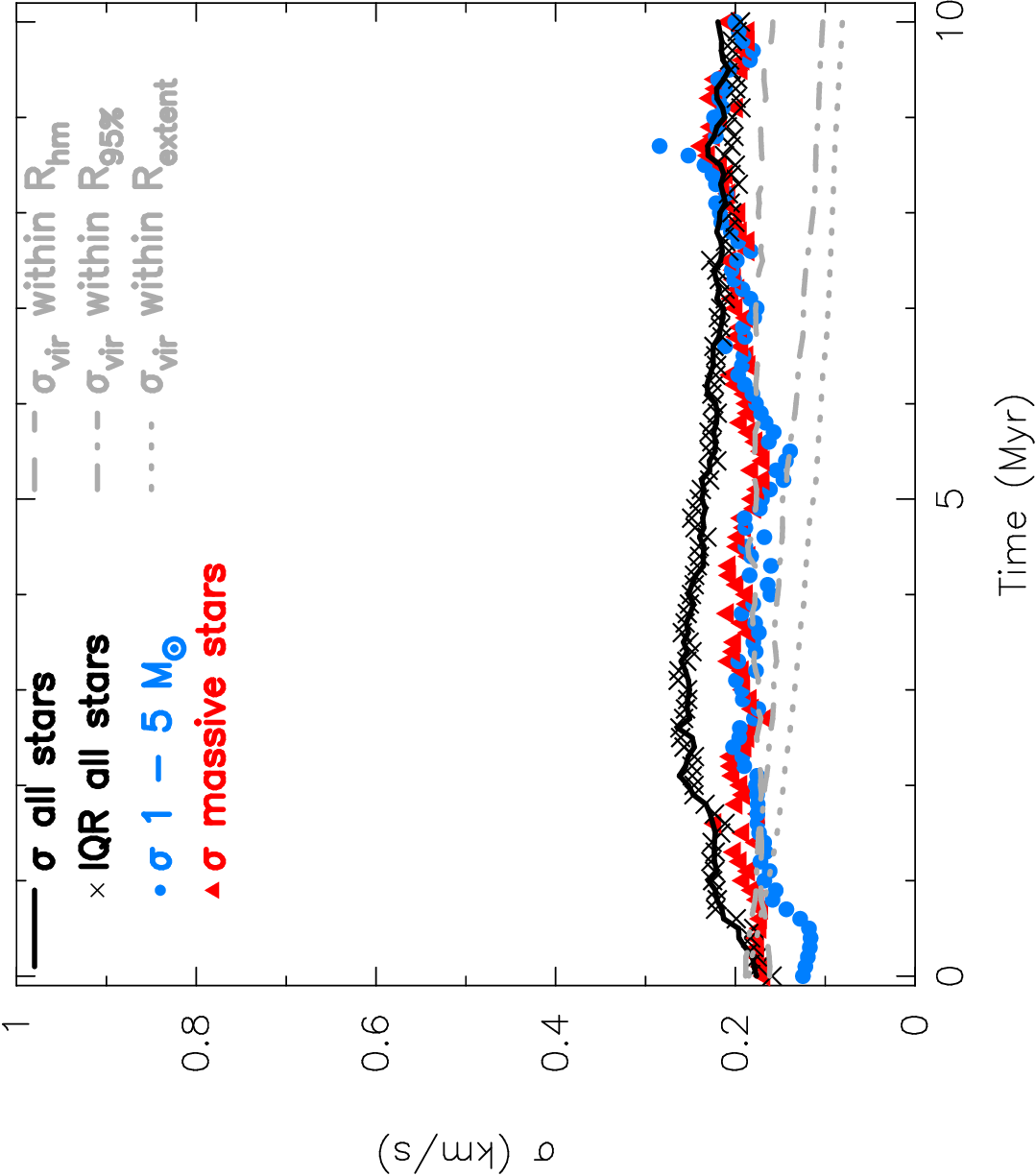}}}
\hspace*{-1.cm}
\subfigure[$\alpha_{\rm vir} = 1.5, D = 1.6$]{\label{vdC_medOH-g}\rotatebox{270}{\includegraphics[scale=0.27]{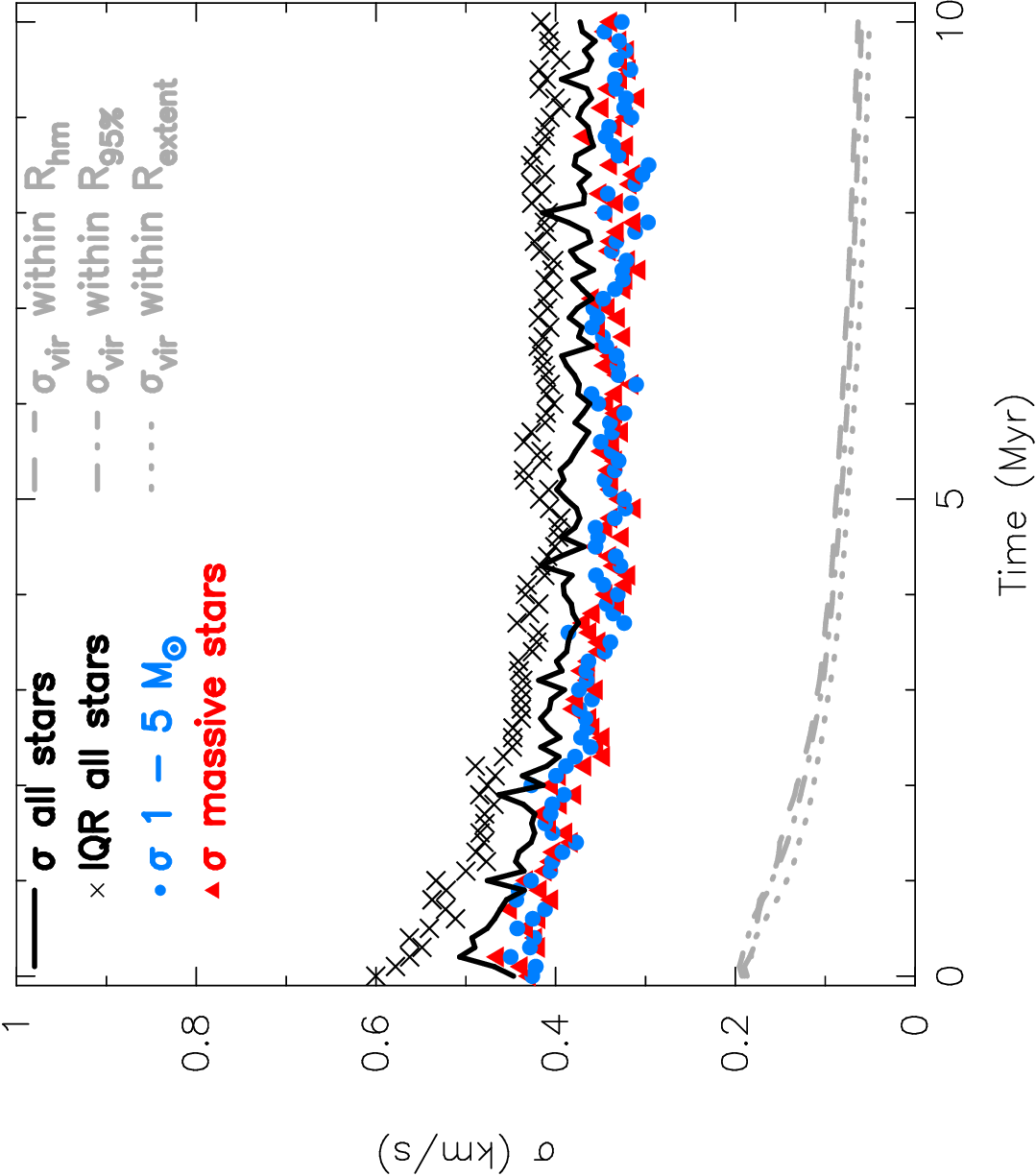}}}
\hspace*{0.3cm} 
\subfigure[$\alpha_{\rm vir} = 1.5, D = 2.0$]{\label{vdC_medOH-h}\rotatebox{270}{\includegraphics[scale=0.27]{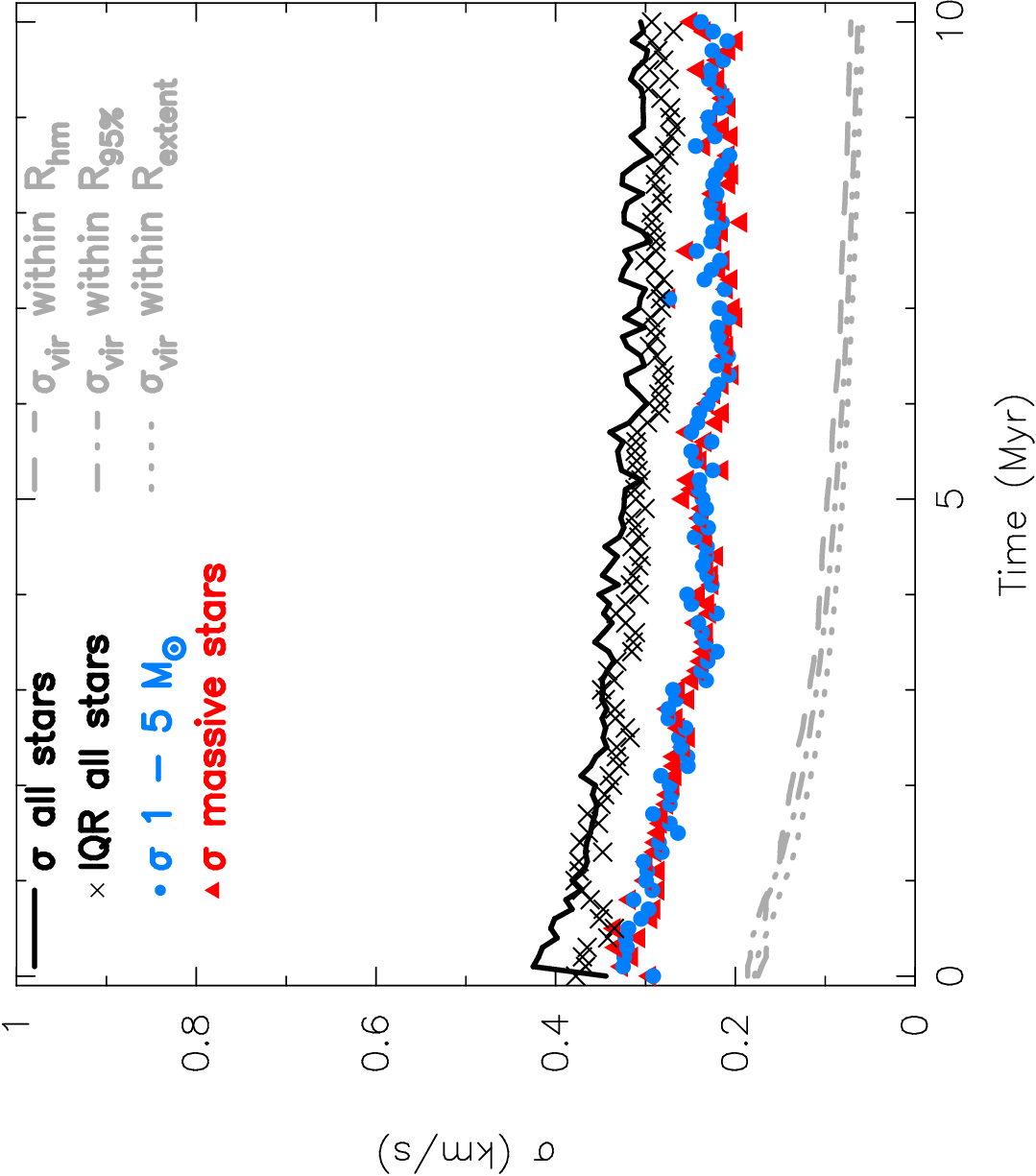}}}
\hspace*{0.3cm} 
\subfigure[$\alpha_{\rm vir} = 1.5, D = 3.0$]{\label{vdC_medOH-i}\rotatebox{270}{\includegraphics[scale=0.27]{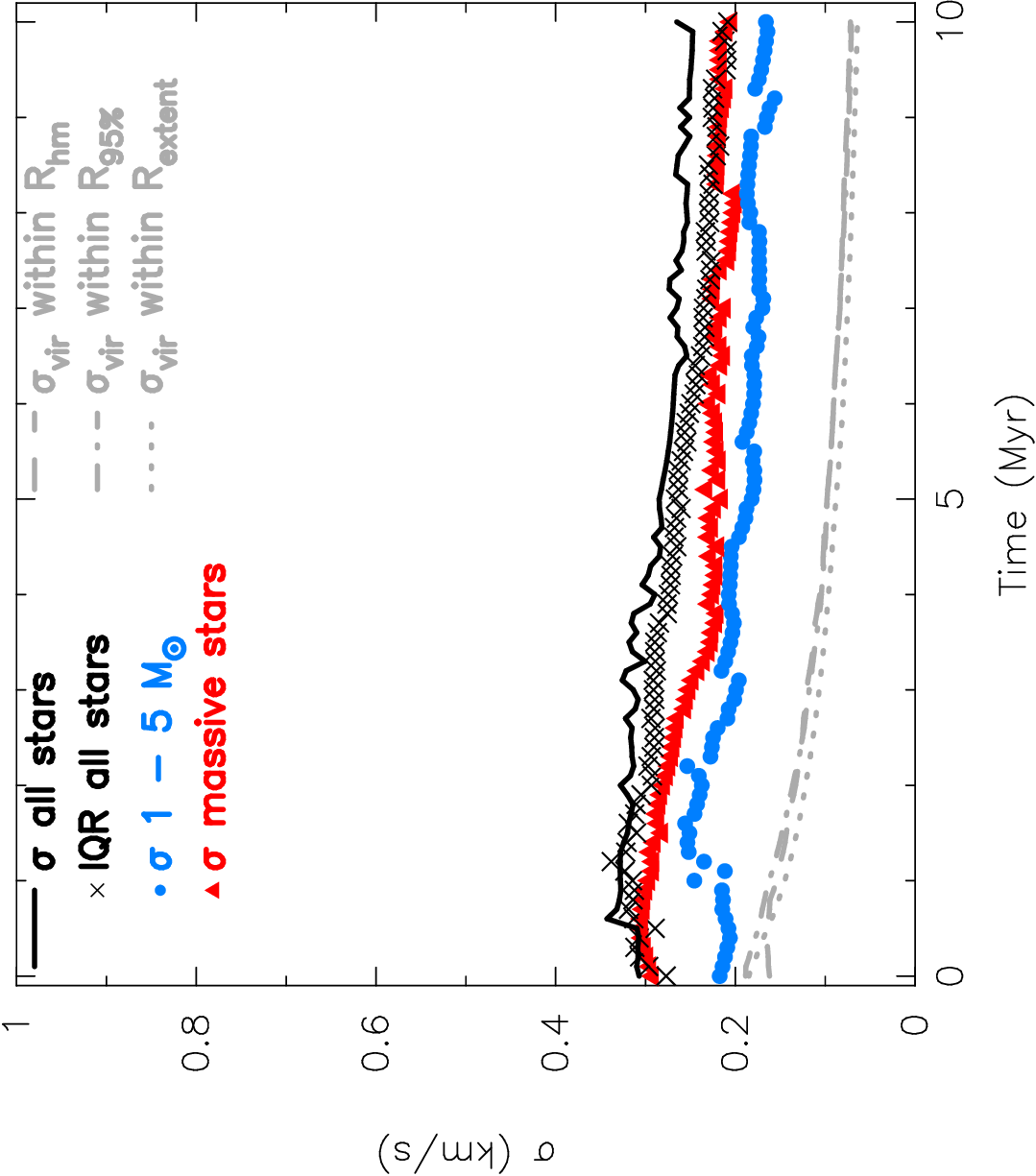}}}
\caption[bf]{As Fig.~\ref{vdC_med} but for simulations containing 150 stars each. Evolution of the median velocity dispersion values with time for all simulations. The median value (from twenty simulations) of the velocity dispersion for all stars is shown by the solid black line. The median value for stars with masses 1 -- 5\,M$_\odot$ is shown by the blue circles, and the median value for the 10 most massive stars is shown by the filled red triangles. We also show the median value for the IQR dispersion for all stars by the black crosses. We show the median maximum velocity dispersion for the region to still be in virial equilibrium by the broken grey lines; the dashed line is for objects within the half-mass radius, the dot-dashed line is for all stars within 95\,per cent of the extent of the region, and the dotted line is for the full region in the simulations.  }
\label{vdC_medOH}
  \end{center}
\end{figure*}

So far, we have only considered our simulations with $N = 1500$ stars, which have typical initial densities between $10^3 - 10^4$\,M$_\odot$\,pc$^{-3}$. These simulations provide a high statistical significance when determining the velocity dispersion or virial ratio, but they are also dynamically active, and as we have seen, both of these quantities (either directly or indirectly measured) can change on short timescales. Many nearby star-forming regions however have fewer stars (typically 100s, rather thans 1000s), such as IC\,348, NGC\,1333, $\rho$~Oph, Cham~I. In Fig.~\ref{vdC_medOH} we present the evolution of the median velocity dispersions from 20 simulations (as Fig.~\ref{vdC_med}) for our simulations which contain only 150 stars, and have a median density of $\sim 10^2$\,M$_\odot$\,pc$^{-3}$.

Due to the lower number of stars, the initial velocity dispersions are much lower than in the $N = 1500$ simulations (typically 0.2\,km\,s$^{-1}$ for the subvirial regions, compared to 0.5\,km\,s$^{-1}$ for the higher-N simulations). The subvirial ($\alpha_{\rm vir} = 0.3$) simulations still undergo a cool collapse phase, but over much longer timescales than the comparable $N = 1500$ regions (the free-fall time is 5\,Myr instead of 0.5\,Myr, and the corresponding \emph{local} crossing times are 1\,Myr for the low-density simulations and 0.1\,Myr for the higher density runs). Unlike in the higher number, higher density simulations, the most massive stars in the low density cool-collapse regions do not reach faster velocity dispersions, and if anything are marginally slower. Furthermore, the velocity dispersion as defined by the IQR never dips significantly below the statistical velocity dispersion about the mean radial velocity.

According to the maximum velocity dispersion for the region to still be in virial equilibrium, as defined by the virial mass and radius, the $\alpha_{\rm vir} = 0.3 - 0.5; N = 150$ regions remain (sub)virial until 5 -- 10\,Myr, i.e.\,\,much longer than the $N = 1500$ regions. However, as we saw for the $N = 1500$ regions, this appears to be a poor indicator of the true virial ratio. Indeed, the evolution of the virial ratio is virtually identical to the $N = 1500$ counterparts per each set of initial conditions, and we do not show those plots here. The regions that are supervirial to begin with have velocity dispersions that exceed the virial mass estimate at times beyond 4\,Myr (compare the black solid line to the grey broken lines in Figs.~\ref{vdC_medOH-g}~--~\ref{vdC_medOH-i}). 

In all of these simulations, the velocity dispersion of the 10 most massive stars (the red triangles) is very similar to the 1 -- 5\,M$_\odot$ stars (the blue circles) due to stochastic sampling of the IMF; the 10 most massive stars are little over a solar mass. However, in the supervirial, smooth initial conditions (panel i) the two have slightly different velocity dispersions initially (a difference of less than 1\,km\,s$^{-1}$). Due to the paucity of dynamical interactions in these low-density expanding regions, this difference remains for the duration of the simulation. 

\subsection{Spatial structure versus velocity}

\begin{figure*}
  \begin{center}
\setlength{\subfigcapskip}{10pt}
\vspace*{-0.3cm}
\hspace*{-1.cm}\subfigure[$\alpha_{\rm vir} = 0.3, D = 1.6$]{\label{Q_vd-a}\rotatebox{270}{\includegraphics[scale=0.35]{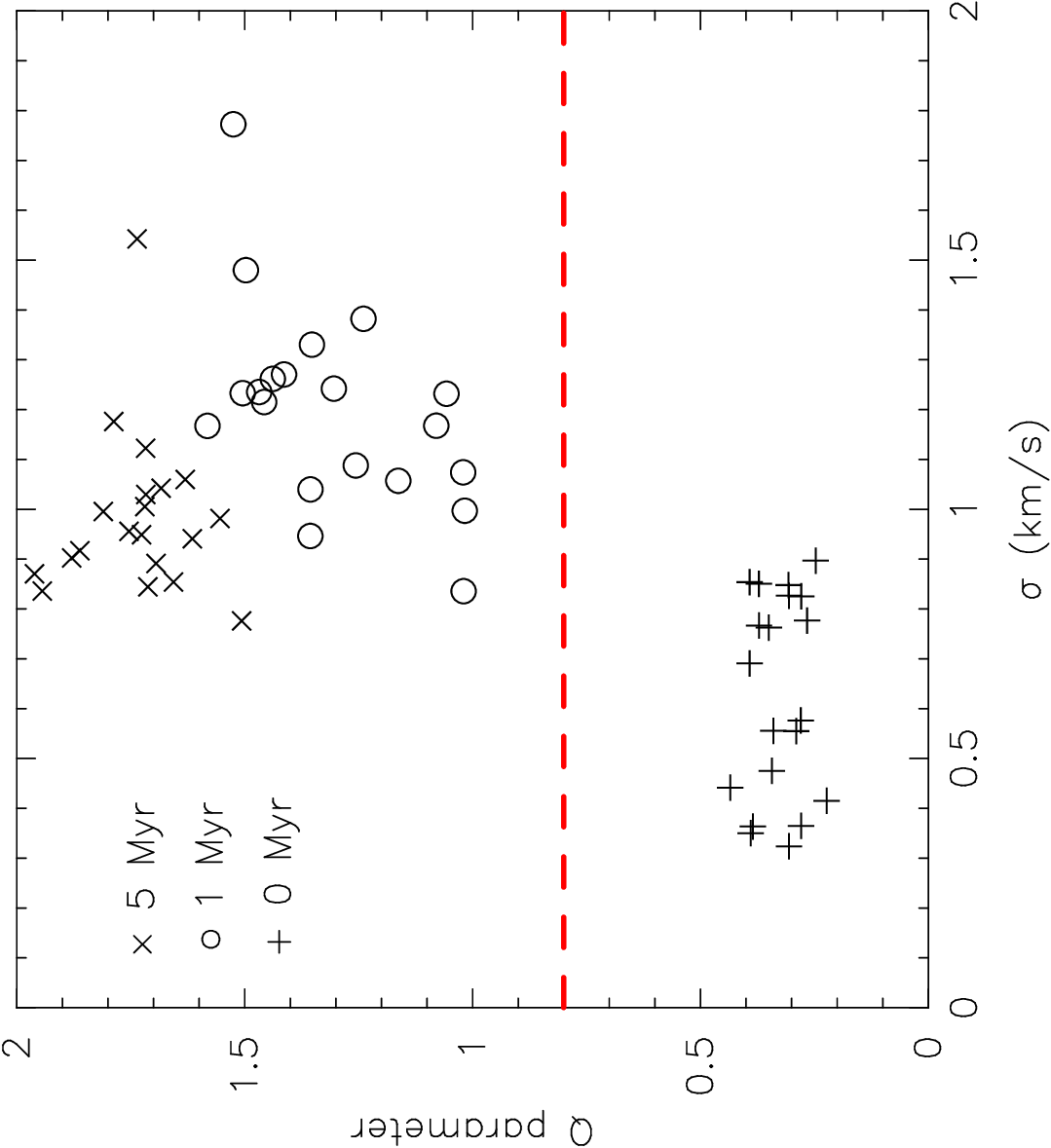}}}
\hspace*{0.5cm} 
\subfigure[$\alpha_{\rm vir} = 1.5, D = 1.6$]{\label{Q_vd-g}\rotatebox{270}{\includegraphics[scale=0.35]{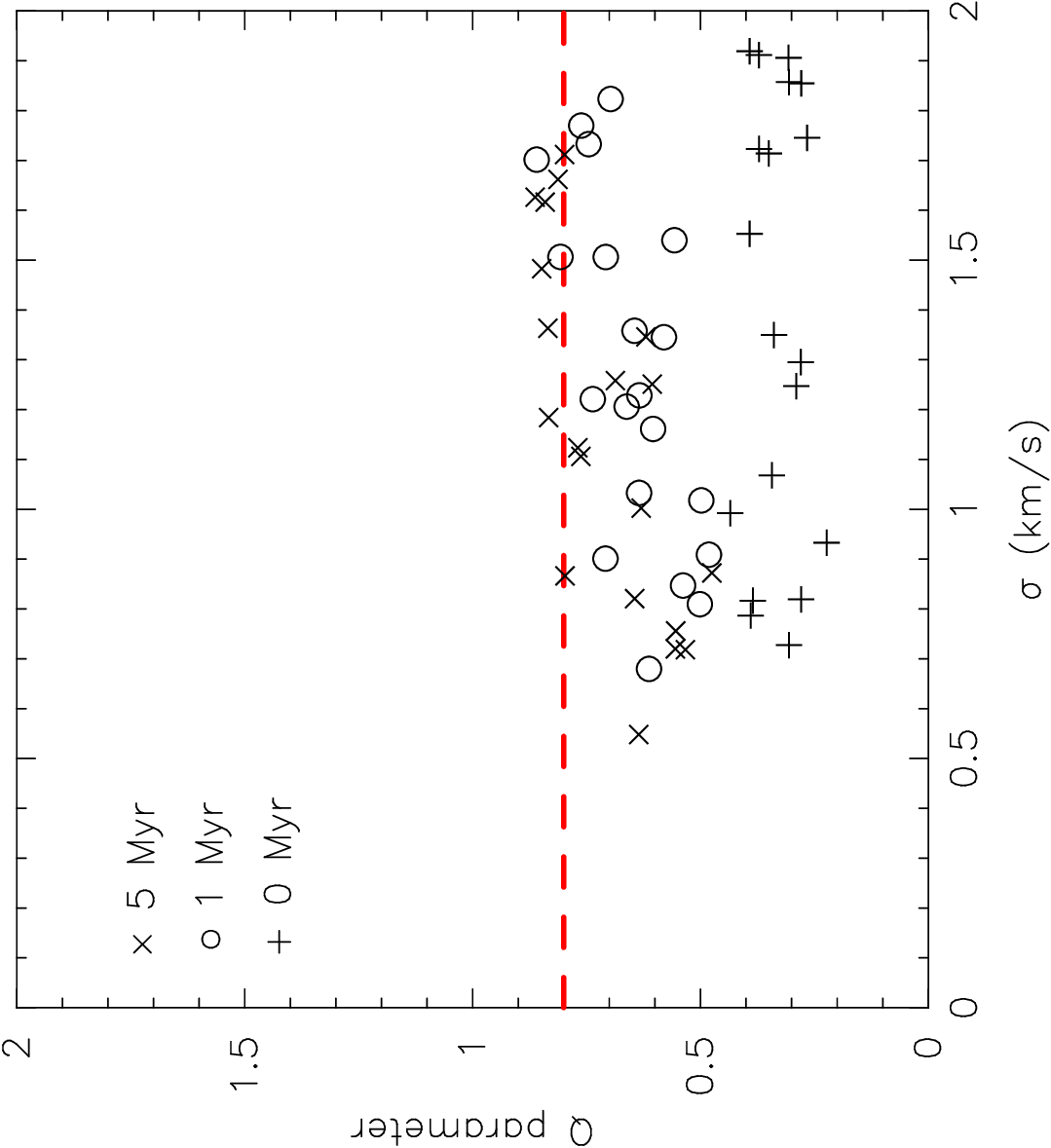}}}
\caption[bf]{Evolution of structure as measured by the $\mathcal{Q}$-parameter versus the velocity dispersion for all stars for substructured, subvirial initial conditions (panel a) and substructured, supervirial initial conditions (panel b) in the simulations containing $N = 1500$ stars. For each set of initial conditions we show the values from 20 realisations of the same simulation, and plot values at 0\,Myr (before dynamical evolution) -- the plus signs; 1\,Myr -- the open circles; and 5\,Myr -- the crosses. The horizontal dashed line indicates the boundary between substructured ($\mathcal{Q}<0.8$) and smooth, centrally concentrated ($\mathcal{Q}>0.8$) spatial distributions.}
\label{Q_vd}
  \end{center}
\end{figure*}

\begin{figure*}
  \begin{center}
\setlength{\subfigcapskip}{10pt}
\vspace*{-0.3cm}
\hspace*{-1.cm}\subfigure[$\alpha_{\rm vir} = 0.3, D = 1.6$]{\label{vd_IQR_med-a}\rotatebox{270}{\includegraphics[scale=0.27]{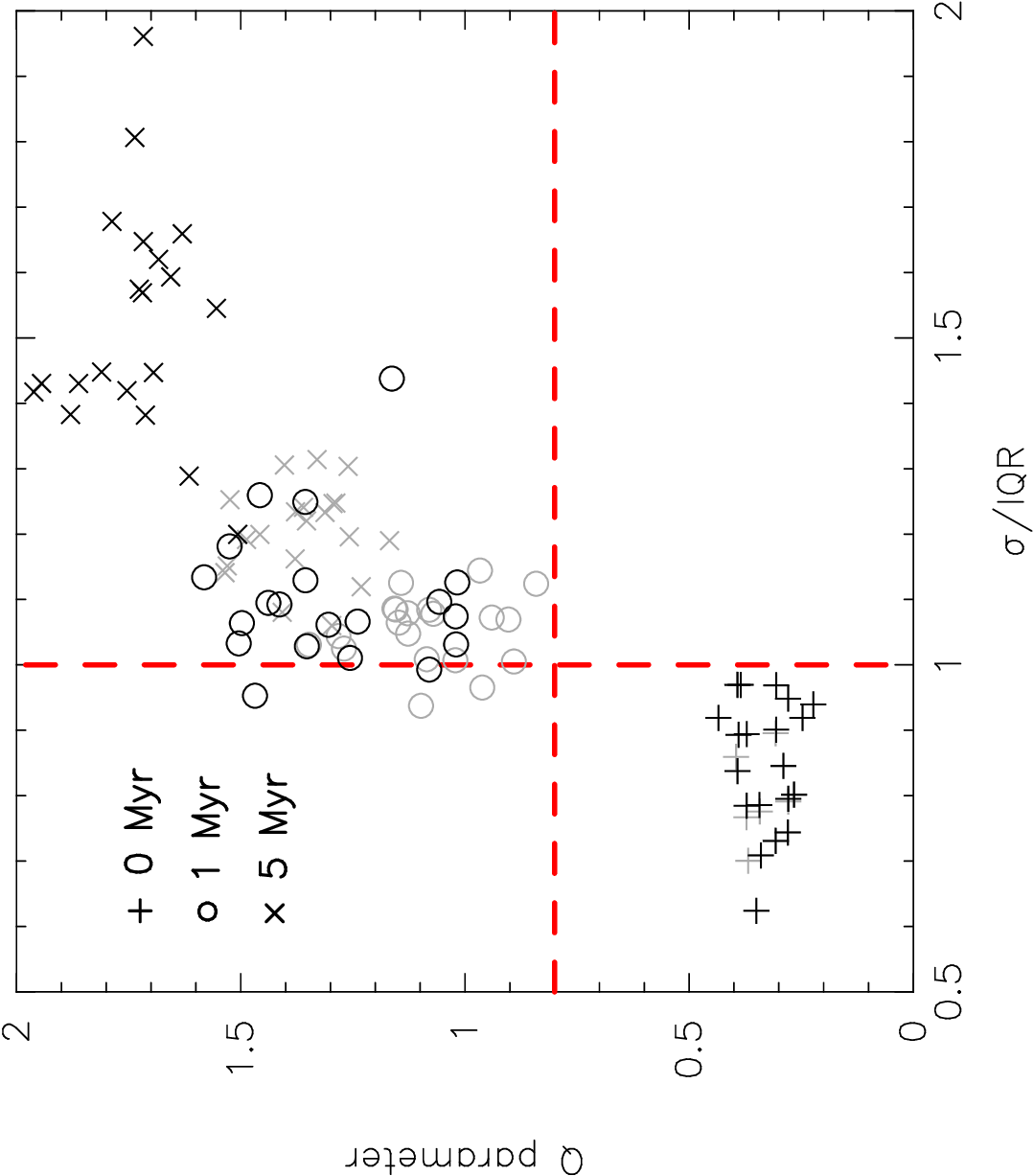}}}
\hspace*{0.3cm} 
\subfigure[$\alpha_{\rm vir} = 0.3, D = 2.0$]{\label{vd_IQR_med-b}\rotatebox{270}{\includegraphics[scale=0.27]{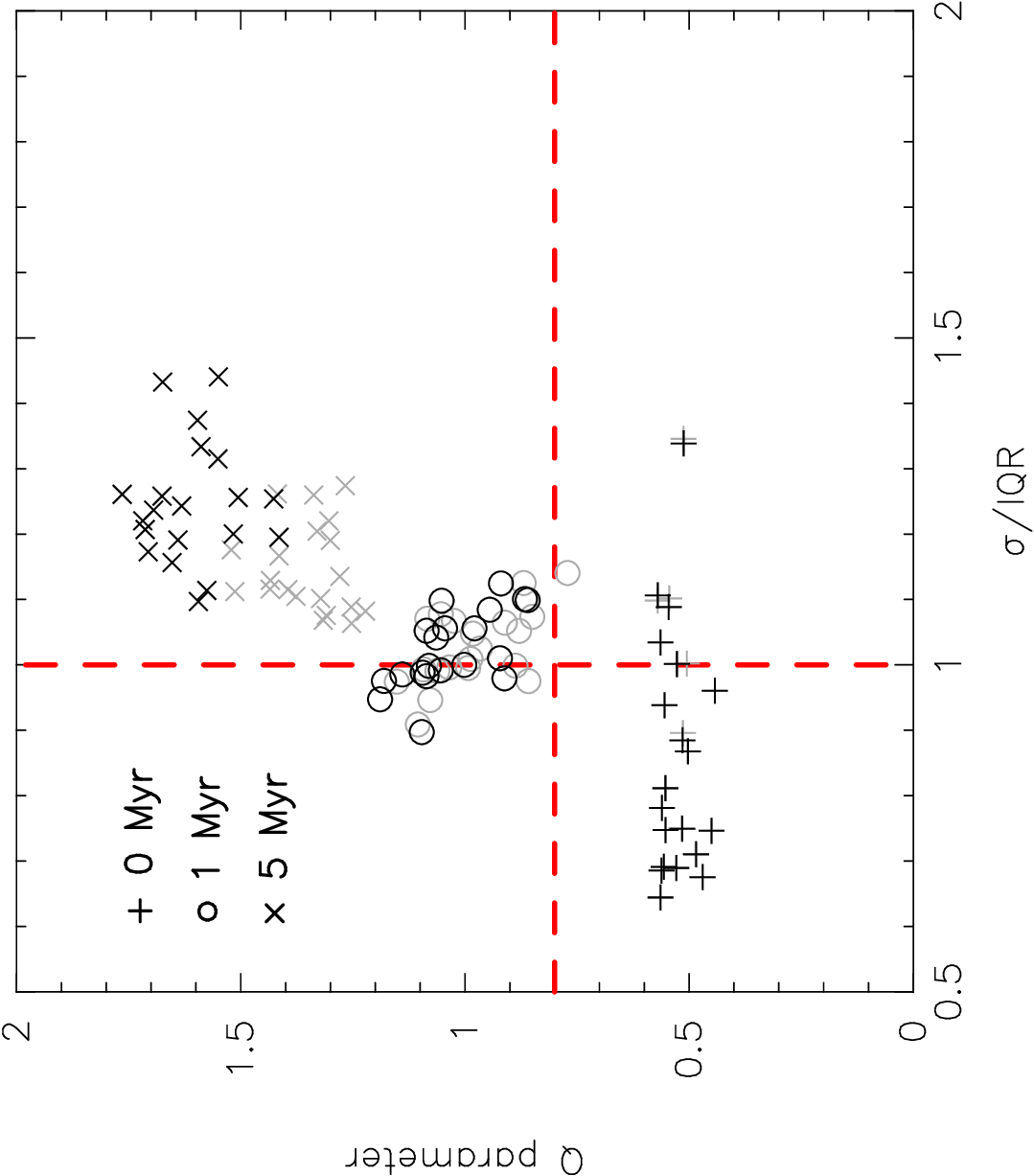}}}
\hspace*{0.3cm} 
\subfigure[$\alpha_{\rm vir} = 0.3, D = 3.0$]{\label{vd_IQR_med-c}\rotatebox{270}{\includegraphics[scale=0.27]{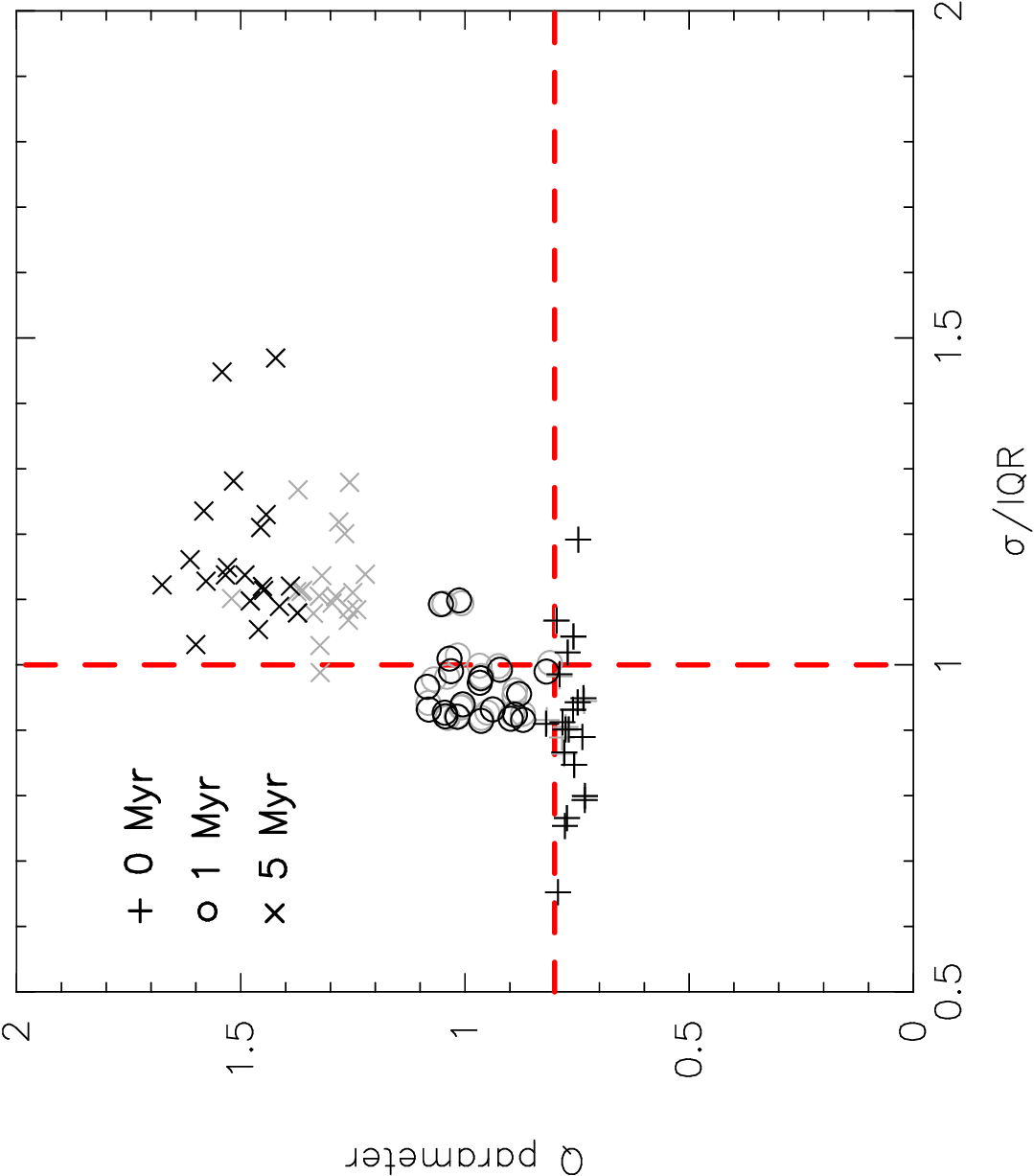}}}
%\vspace*{-0.3cm}
\hspace*{-1.cm}
\subfigure[$\alpha_{\rm vir} = 0.5, D = 1.6$]{\label{vd_IQR_med-d}\rotatebox{270}{\includegraphics[scale=0.27]{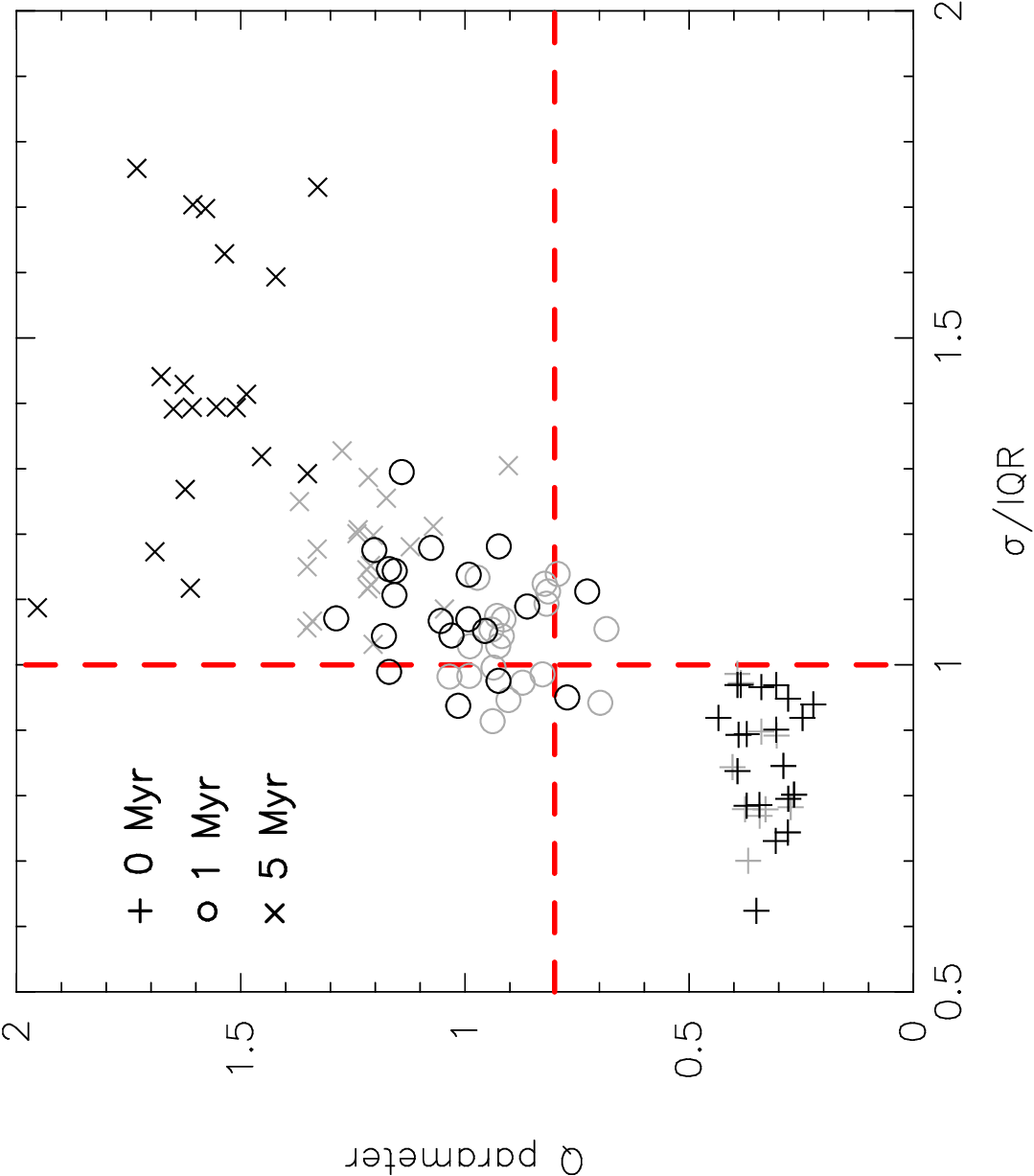}}}
\hspace*{0.3cm} 
\subfigure[$\alpha_{\rm vir} = 0.5, D = 2.0$]{\label{vd_IQR_med-e}\rotatebox{270}{\includegraphics[scale=0.27]{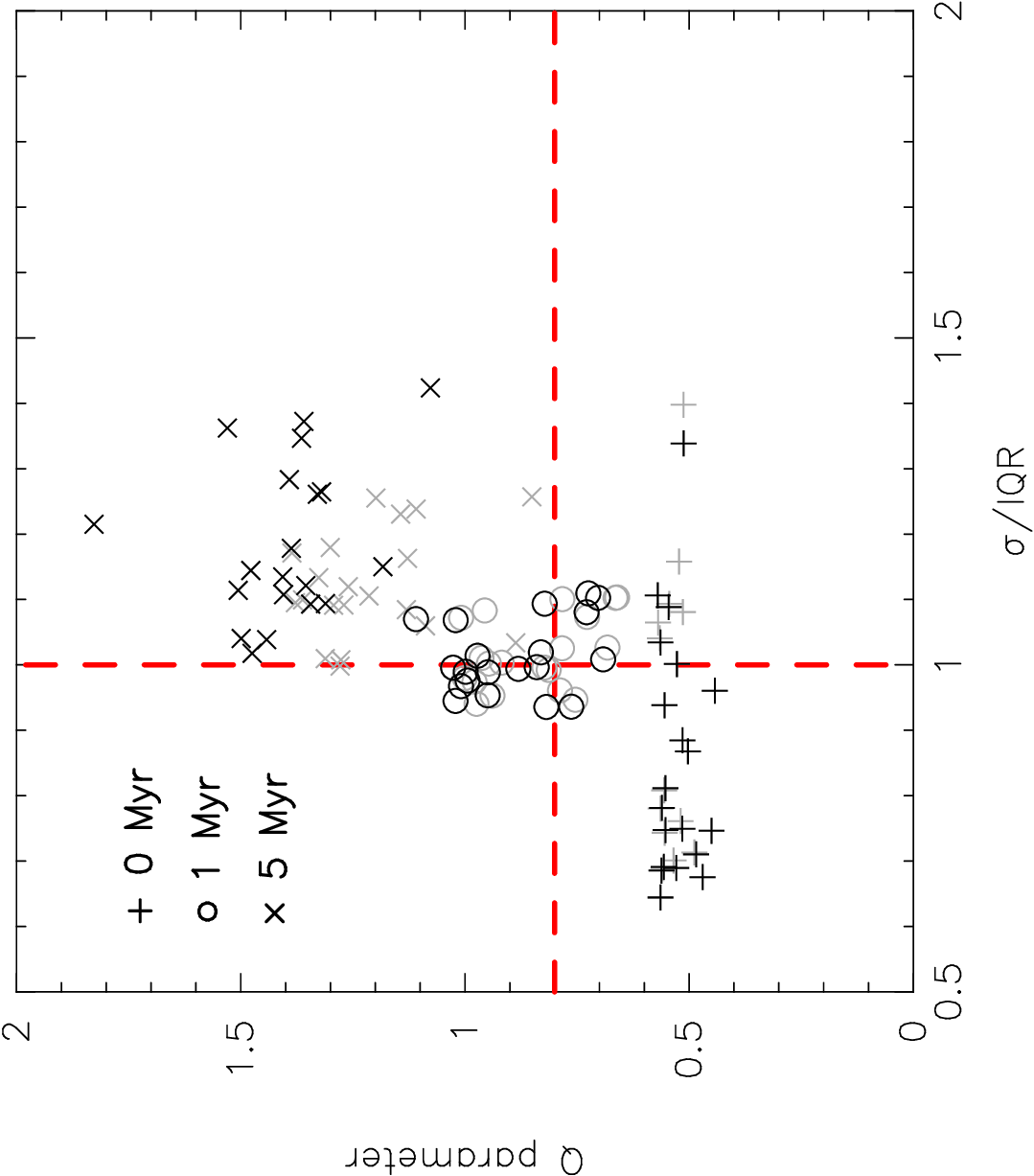}}}
\hspace*{0.3cm} 
\subfigure[$\alpha_{\rm vir} = 0.5, D = 3.0$]{\label{vd_IQR_med-f}\rotatebox{270}{\includegraphics[scale=0.27]{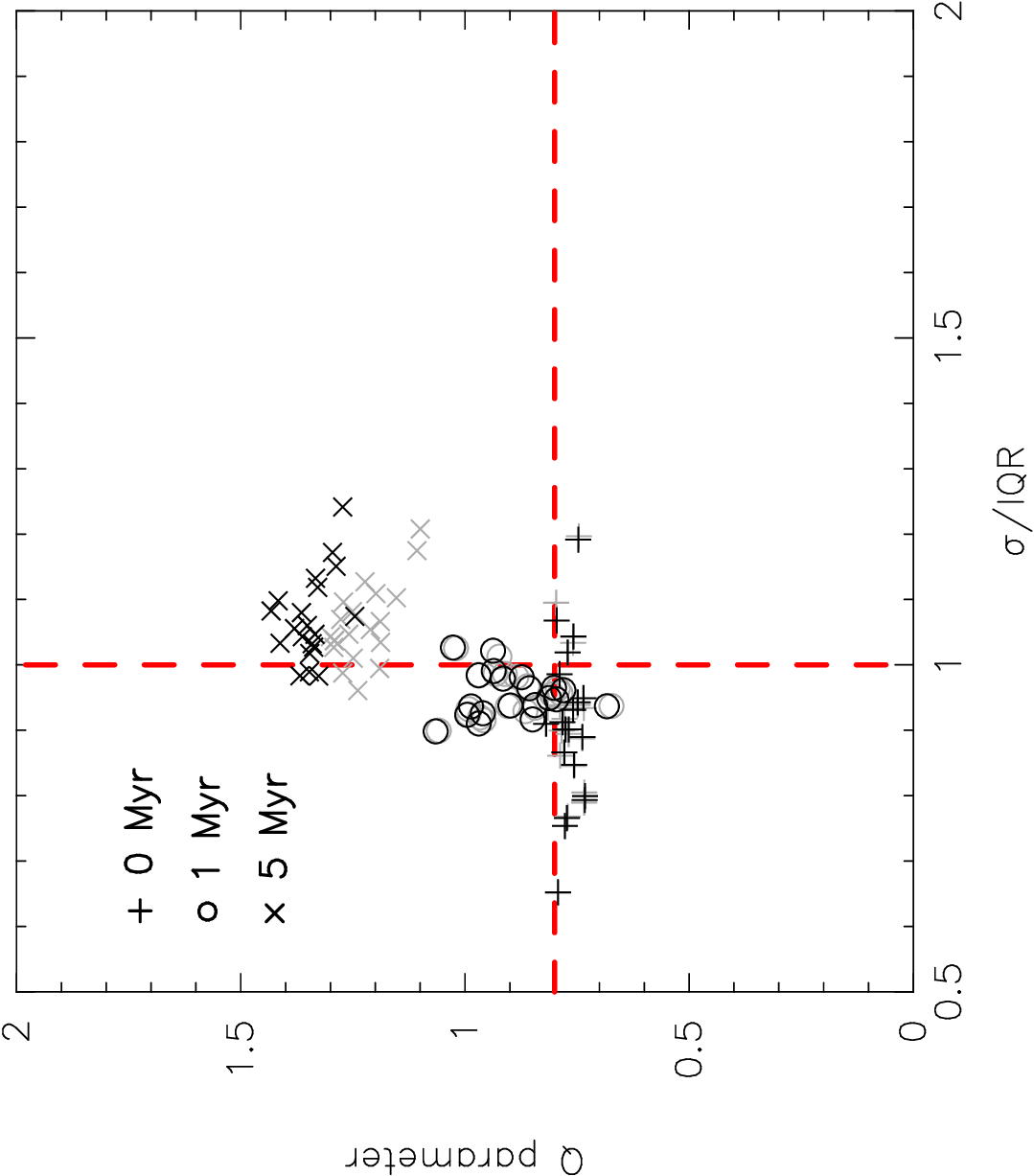}}}
\hspace*{-1.cm}
\subfigure[$\alpha_{\rm vir} = 1.5, D = 1.6$]{\label{vd_IQR_med-g}\rotatebox{270}{\includegraphics[scale=0.27]{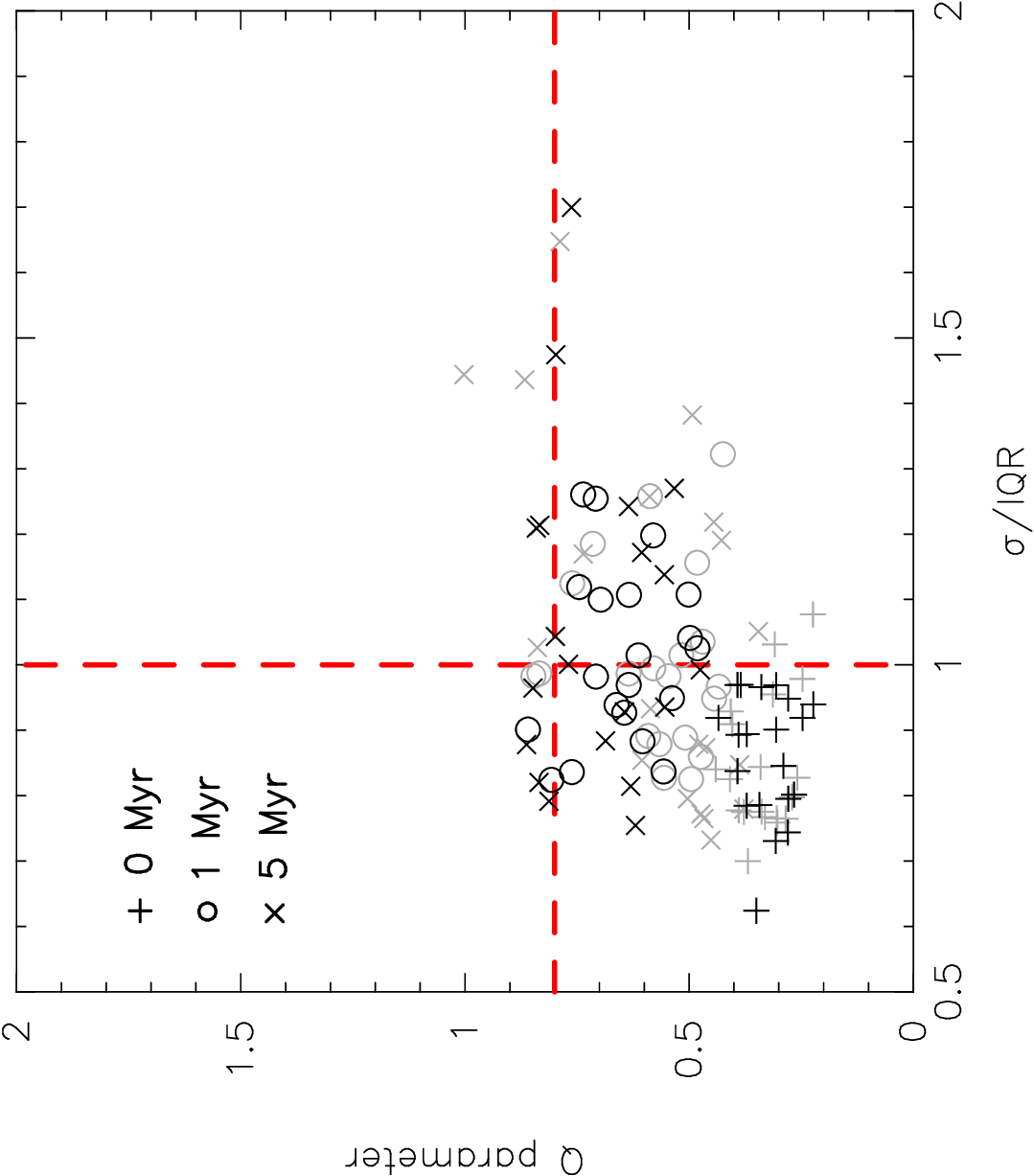}}}
\hspace*{0.3cm} 
\subfigure[$\alpha_{\rm vir} = 1.5, D = 2.0$]{\label{vd_IQR_med-h}\rotatebox{270}{\includegraphics[scale=0.27]{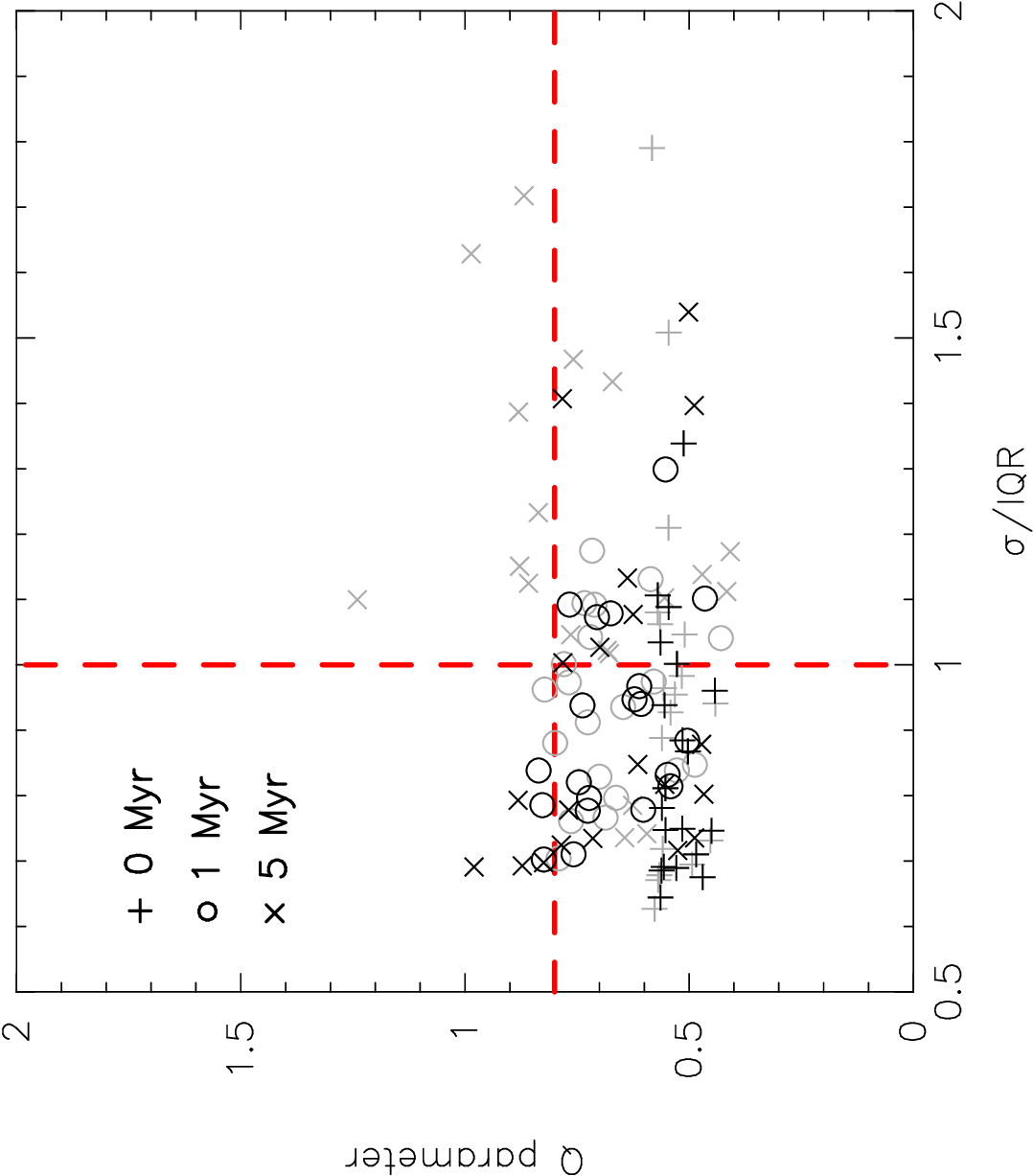}}}
\hspace*{0.3cm} 
\subfigure[$\alpha_{\rm vir} = 1.5, D = 3.0$]{\label{vd_IQR_med-i}\rotatebox{270}{\includegraphics[scale=0.27]{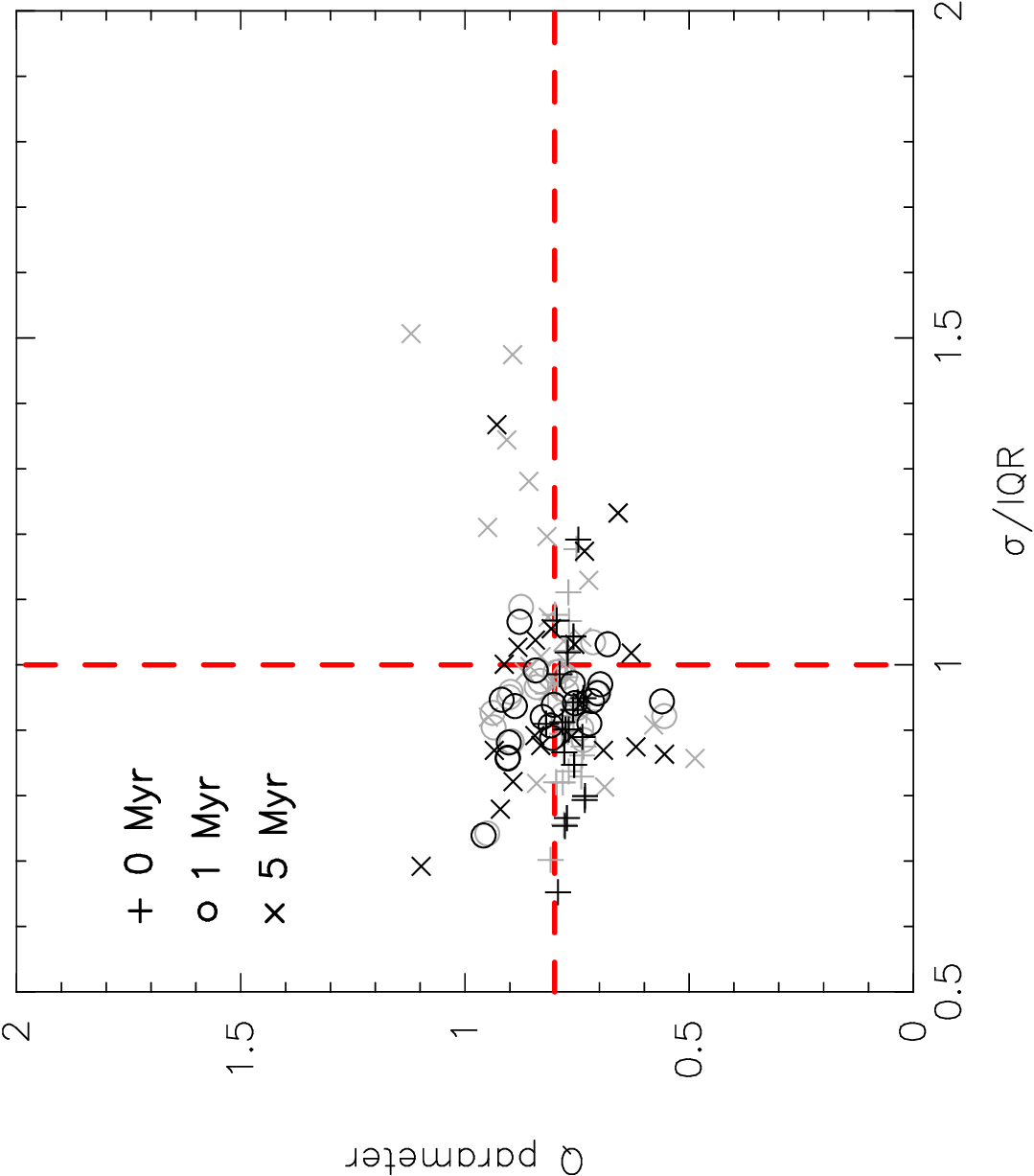}}}
\caption[bf]{Evolution of the  $\mathcal{Q}$-parameter versus $\sigma$/IQR for all initial conditions in the simulations containing $N = 1500$ stars. For each set of initial conditions we show the values from 20 realisations of the same simulation, and plot values at 0\,Myr (before dynamical evolution) -- the plus signs; 1\,Myr -- the open circles; and 5\,Myr -- the crosses. The black symbols are the values for all stars in the simulations, whereas the grey symbols are for bound stars only. The horizontal dashed line indicates the boundary between substructured ($\mathcal{Q}<0.8$) and smooth, centrally concentrated ($\mathcal{Q}>0.8$) spatial distributions. The vertical dashed line indicates  $\sigma$/IQR = 1.  }
\label{Q_vd_IQR}
  \end{center}
\end{figure*}

\begin{figure*}
  \begin{center}
\setlength{\subfigcapskip}{10pt}
\vspace*{-0.3cm}
\hspace*{-1.cm}\subfigure[$\alpha_{\rm vir} = 0.3, D = 1.6$]{\label{OHvd_IQR_med-a}\rotatebox{270}{\includegraphics[scale=0.27]{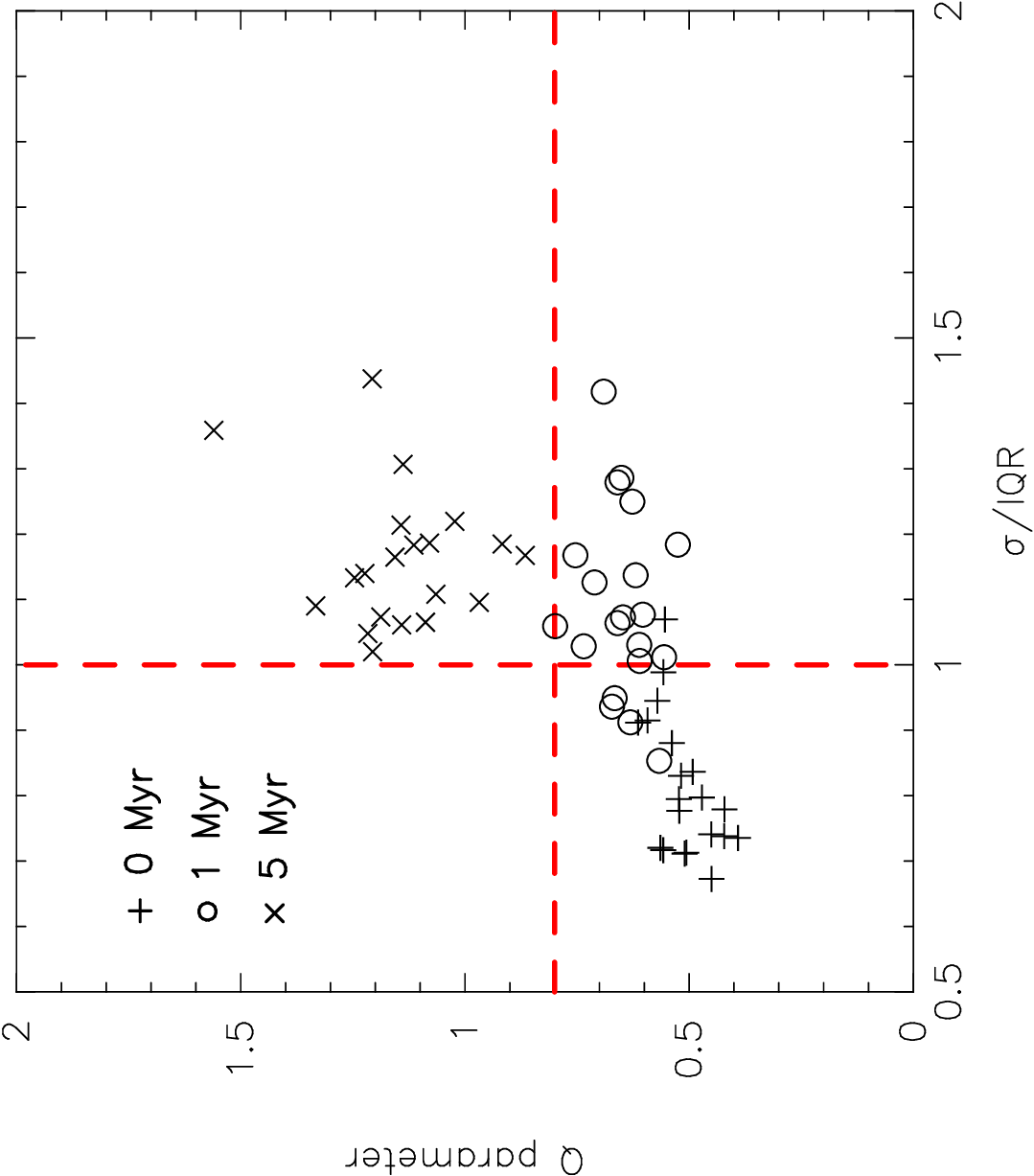}}}
\hspace*{0.3cm} 
\subfigure[$\alpha_{\rm vir} = 0.3, D = 2.0$]{\label{OHvd_IQR_med-b}\rotatebox{270}{\includegraphics[scale=0.27]{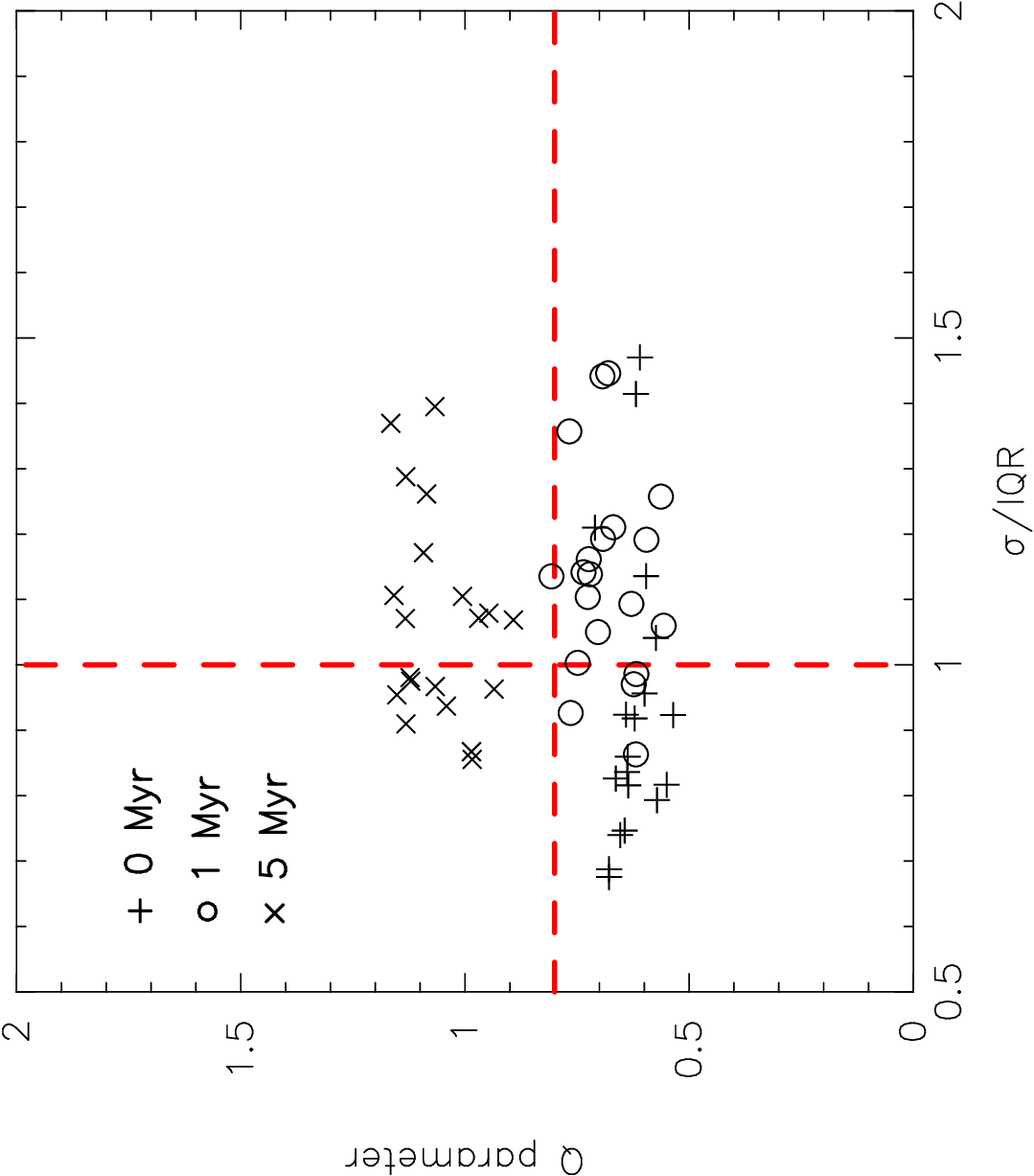}}}
\hspace*{0.3cm} 
\subfigure[$\alpha_{\rm vir} = 0.3, D = 3.0$]{\label{OHvd_IQR_med-c}\rotatebox{270}{\includegraphics[scale=0.27]{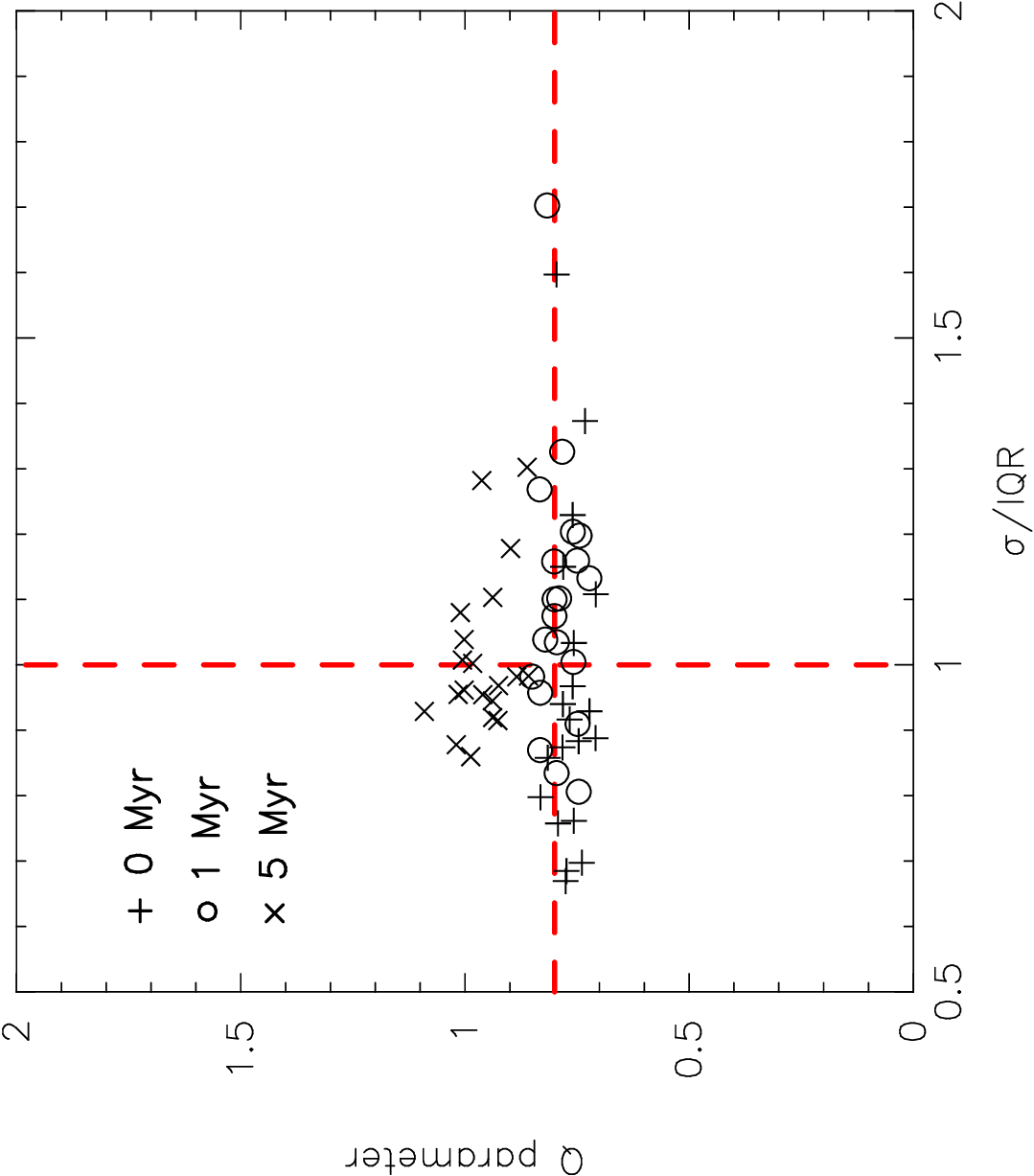}}}
%\vspace*{-0.3cm}
\hspace*{-1.cm}
\subfigure[$\alpha_{\rm vir} = 0.5, D = 1.6$]{\label{OHvd_IQR_med-d}\rotatebox{270}{\includegraphics[scale=0.27]{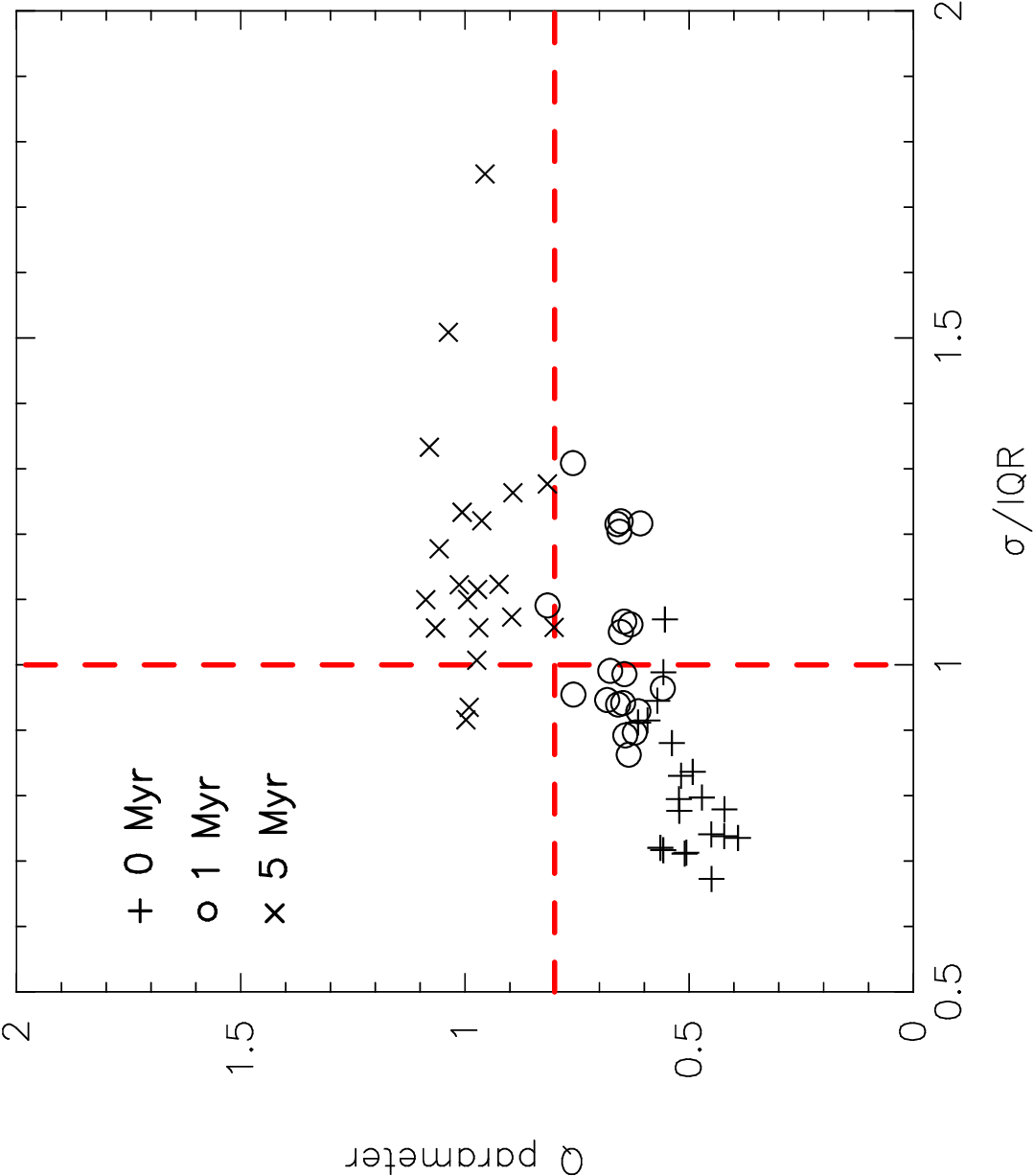}}}
\hspace*{0.3cm} 
\subfigure[$\alpha_{\rm vir} = 0.5, D = 2.0$]{\label{OHvd_IQR_med-e}\rotatebox{270}{\includegraphics[scale=0.27]{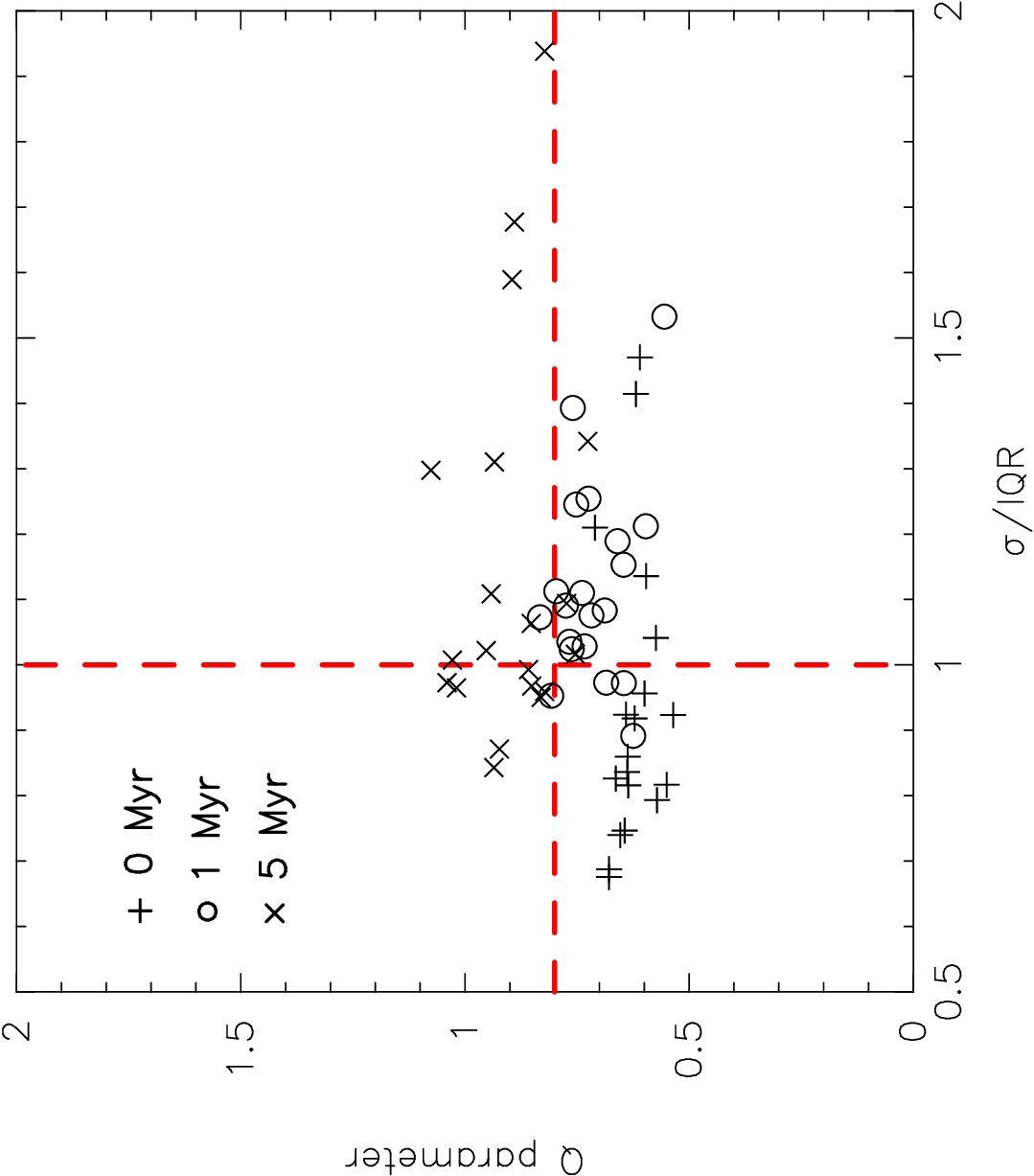}}}
\hspace*{0.3cm} 
\subfigure[$\alpha_{\rm vir} = 0.5, D = 3.0$]{\label{OHvd_IQR_med-f}\rotatebox{270}{\includegraphics[scale=0.27]{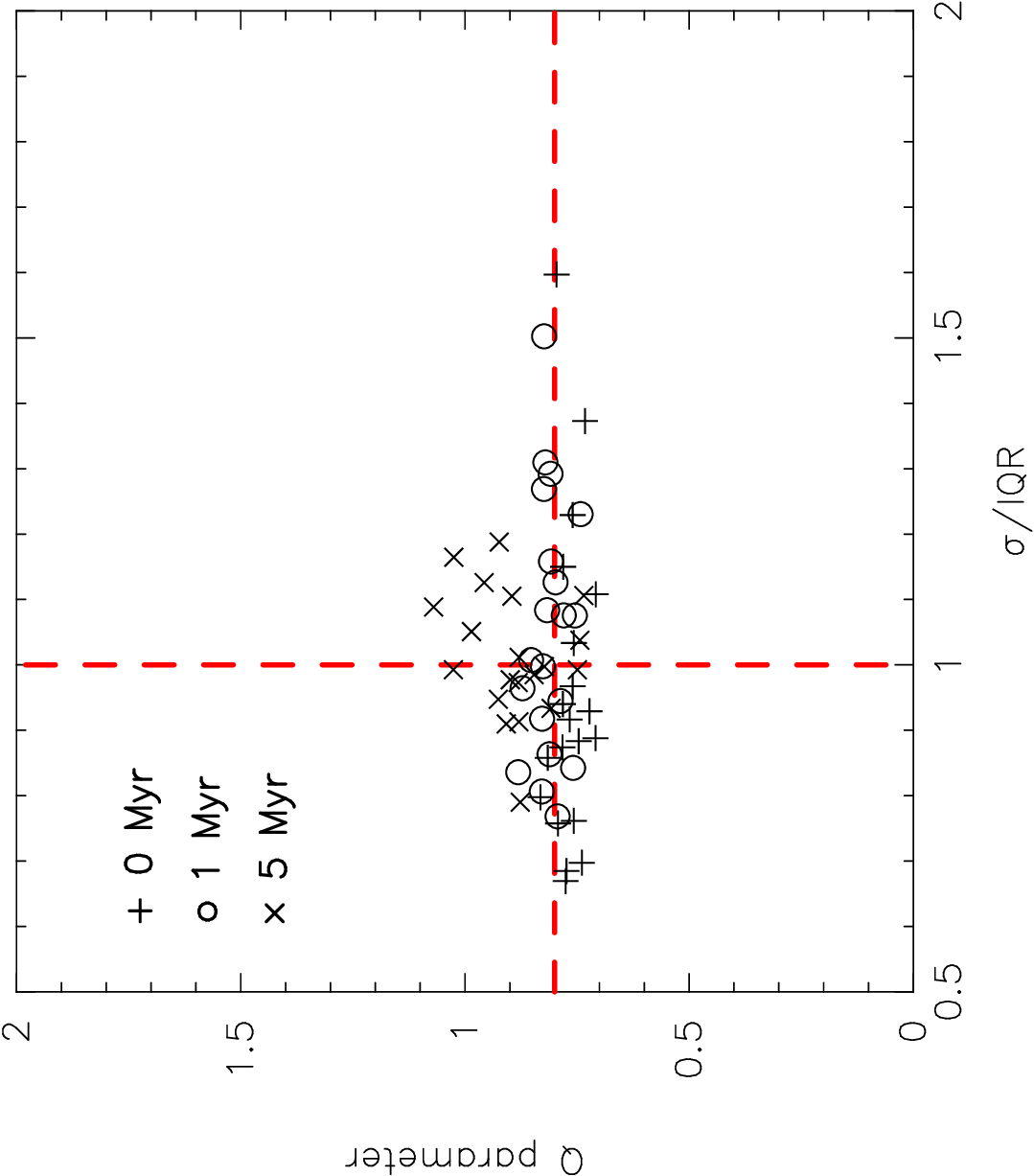}}}
\hspace*{-1.cm}
\subfigure[$\alpha_{\rm vir} = 1.5, D = 1.6$]{\label{OHvd_IQR_med-g}\rotatebox{270}{\includegraphics[scale=0.27]{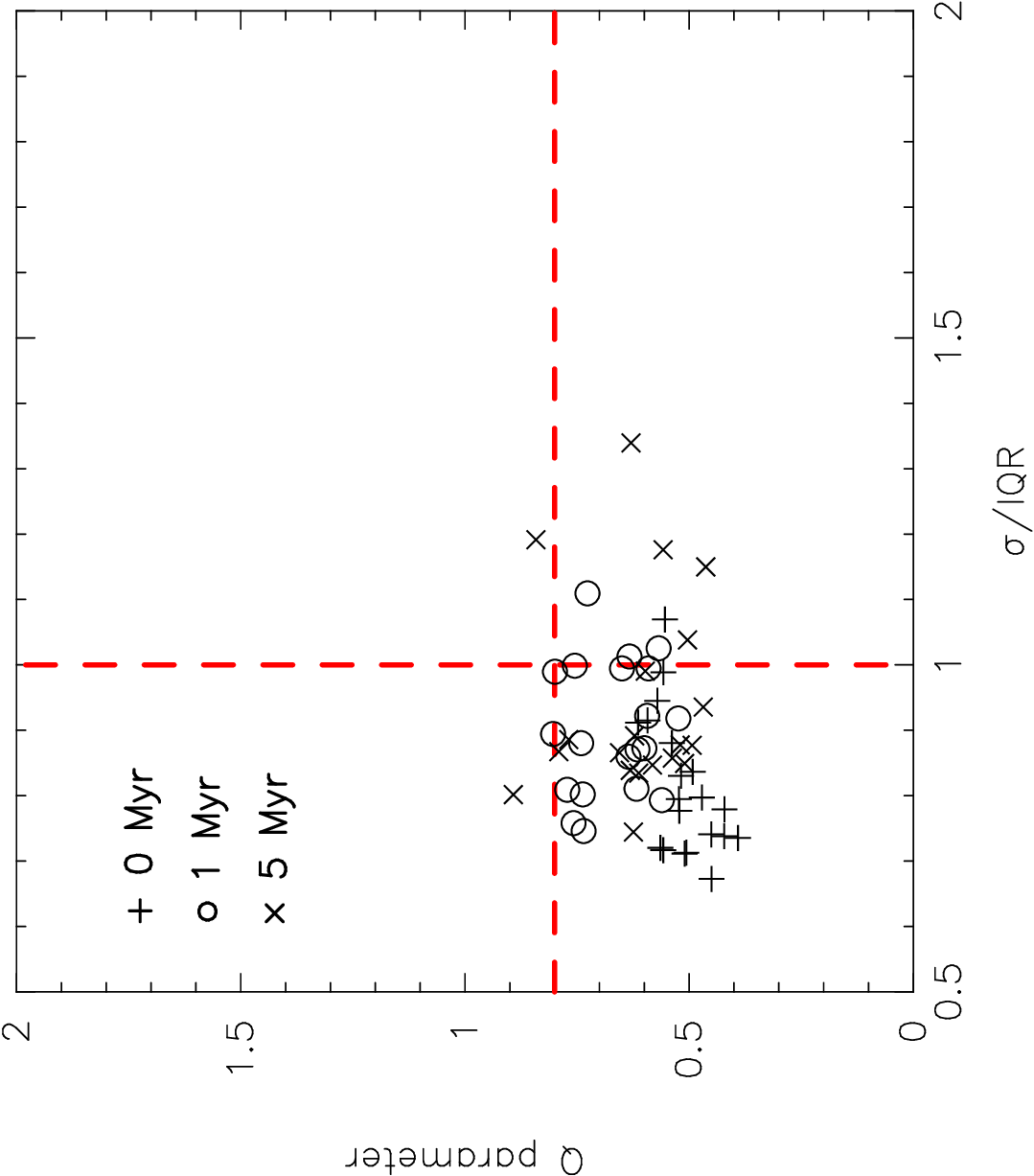}}}
\hspace*{0.3cm} 
\subfigure[$\alpha_{\rm vir} = 1.5, D = 2.0$]{\label{OHvd_IQR_med-h}\rotatebox{270}{\includegraphics[scale=0.27]{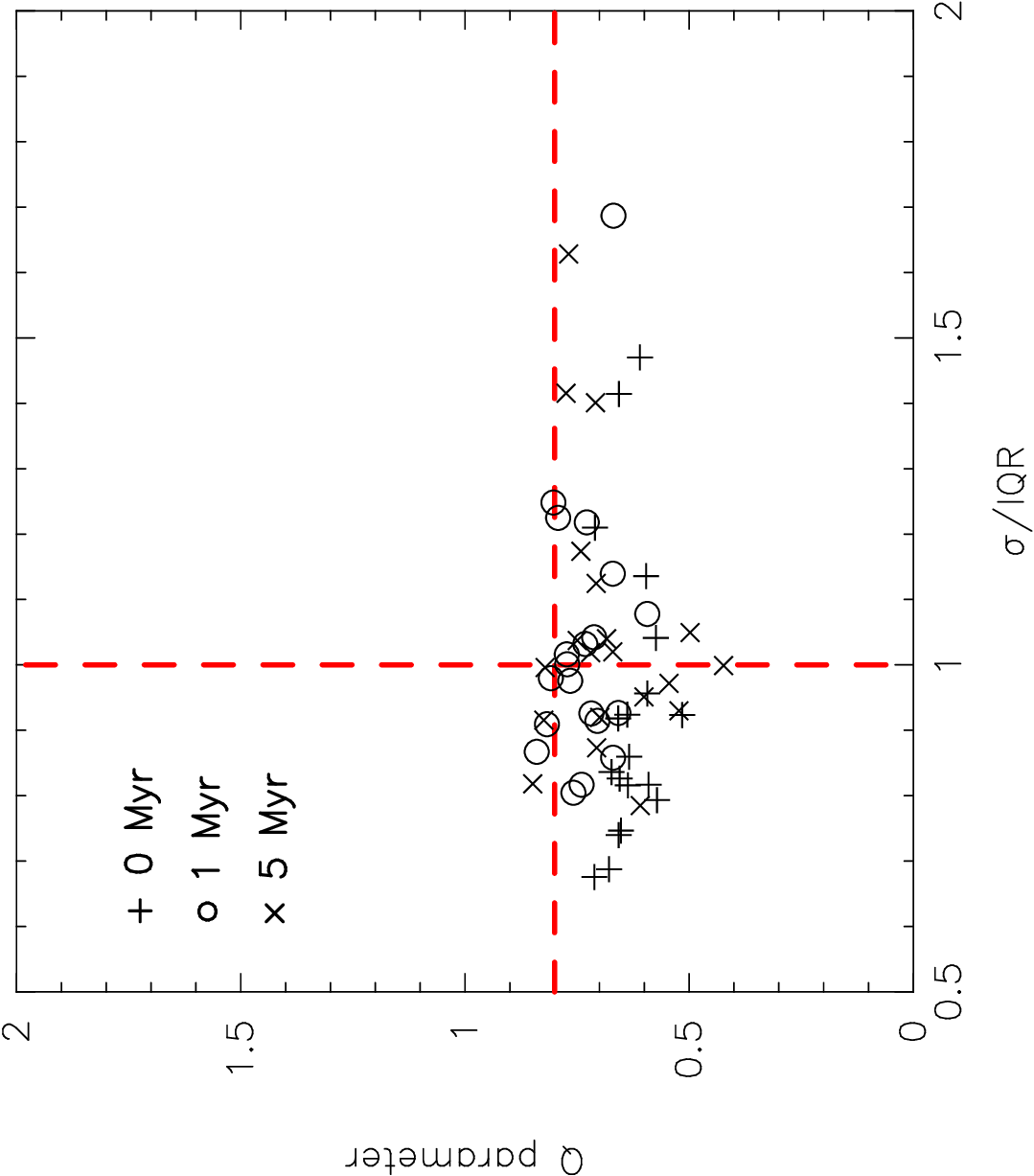}}}
\hspace*{0.3cm} 
\subfigure[$\alpha_{\rm vir} = 1.5, D = 3.0$]{\label{OHvd_IQR_med-i}\rotatebox{270}{\includegraphics[scale=0.27]{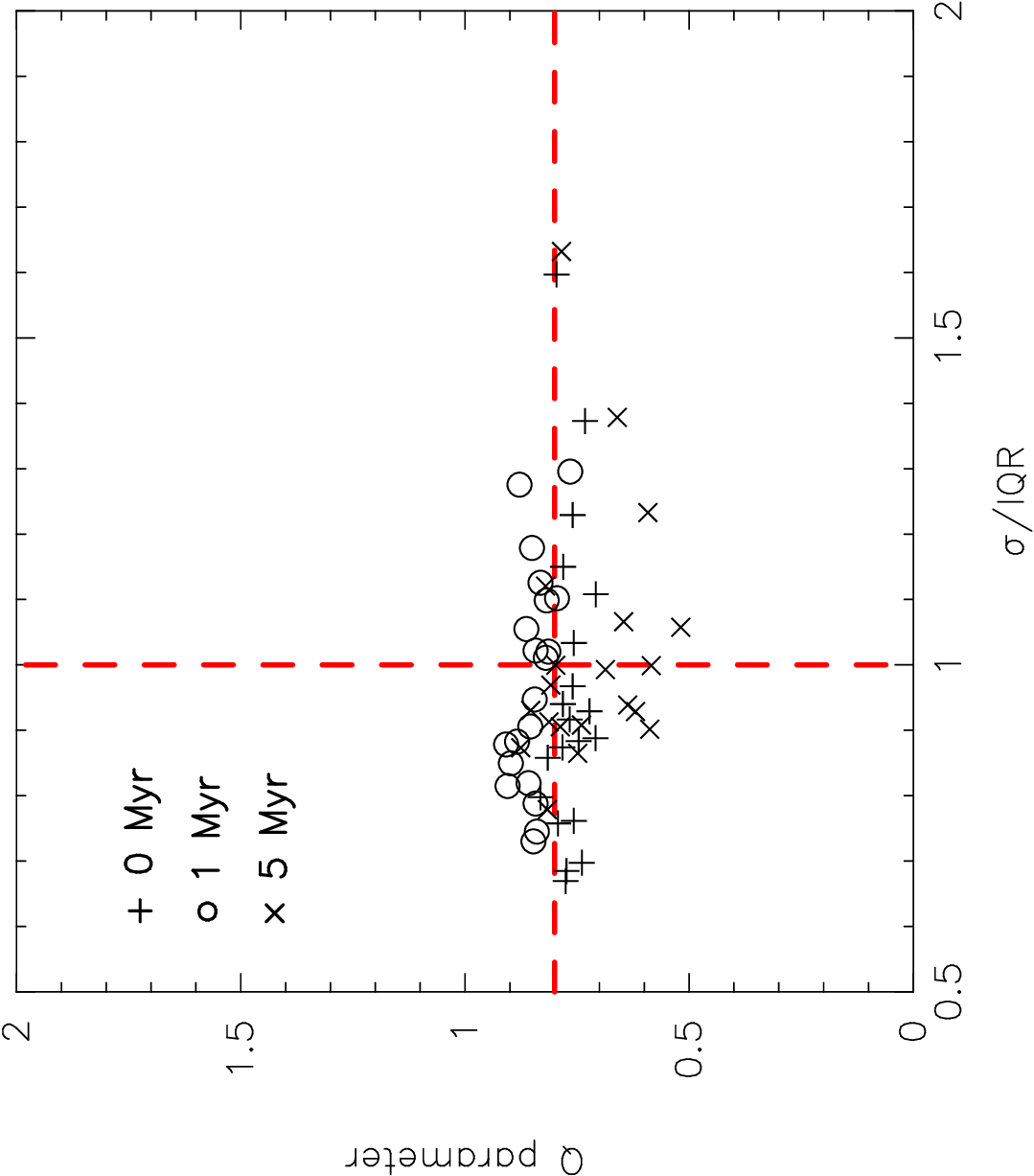}}}
\caption[bf]{Evolution of the  $\mathcal{Q}$-parameter versus $\sigma$/IQR for all initial conditions in the simulations containing $N = 150$ stars. For each set of initial conditions we show the values from 20 realisations of the same simulation, and plot values at 0\,Myr (before dynamical evolution) -- the plus signs; 1\,Myr -- the open circles; and 5\,Myr -- the crosses.  The horizontal dashed line indicates the boundary between substructured ($\mathcal{Q}<0.8$) and smooth, centrally concentrated ($\mathcal{Q}>0.8$) spatial distributions. The vertical dashed line indicates  $\sigma$/IQR = 1.  }
\label{Q_OHvd_IQR}
  \end{center}
\end{figure*}

\begin{figure*}
  \begin{center}
\setlength{\subfigcapskip}{10pt}
\vspace*{-0.3cm}
\hspace*{-1.cm}\subfigure[$\alpha_{\rm vir} = 0.3, D = 1.6$]{\label{Sig_vd_IQR-a}\rotatebox{270}{\includegraphics[scale=0.27]{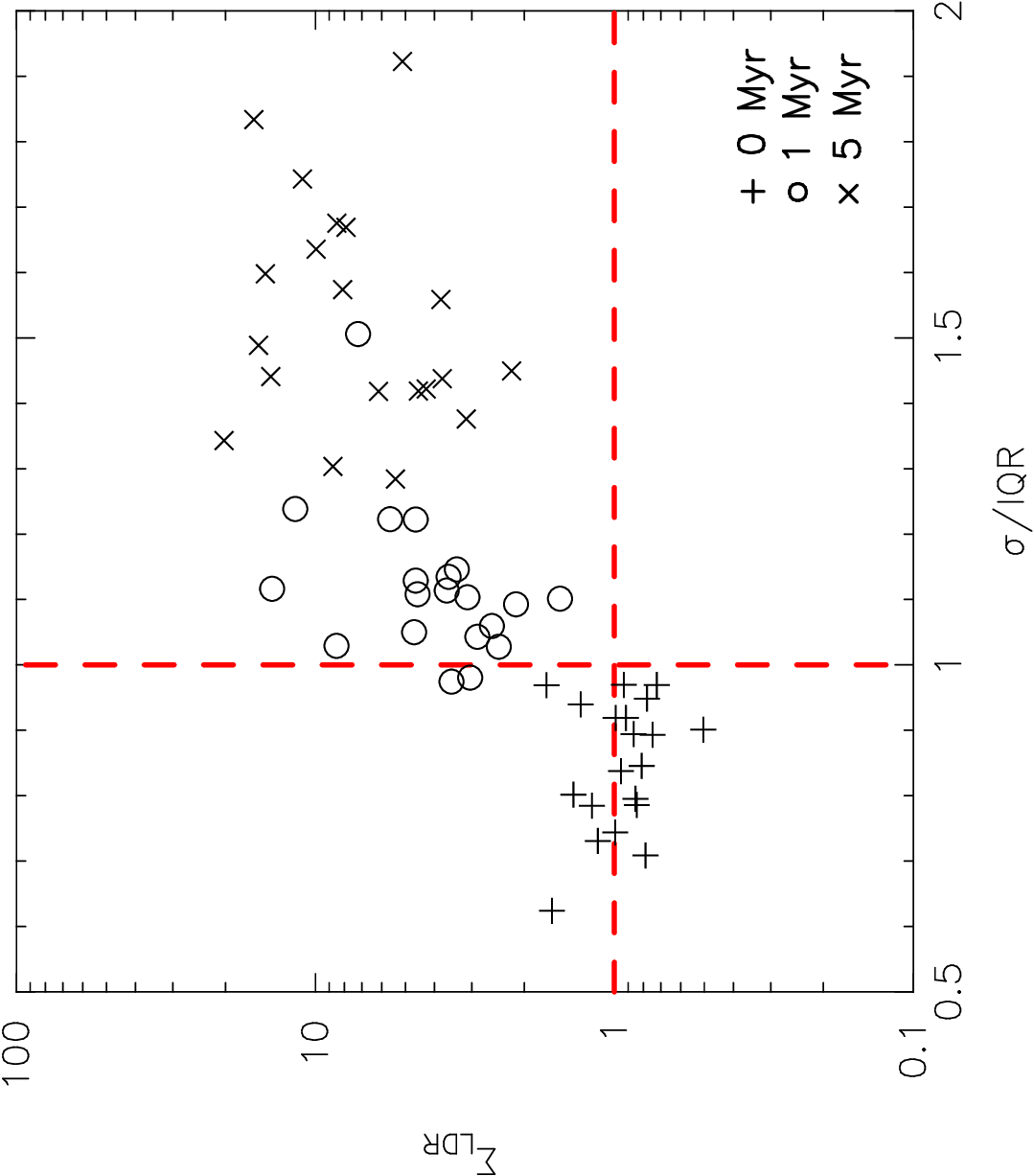}}}
\hspace*{0.3cm} 
\subfigure[$\alpha_{\rm vir} = 0.3, D = 2.0$]{\label{Sig_vd_IQR-b}\rotatebox{270}{\includegraphics[scale=0.27]{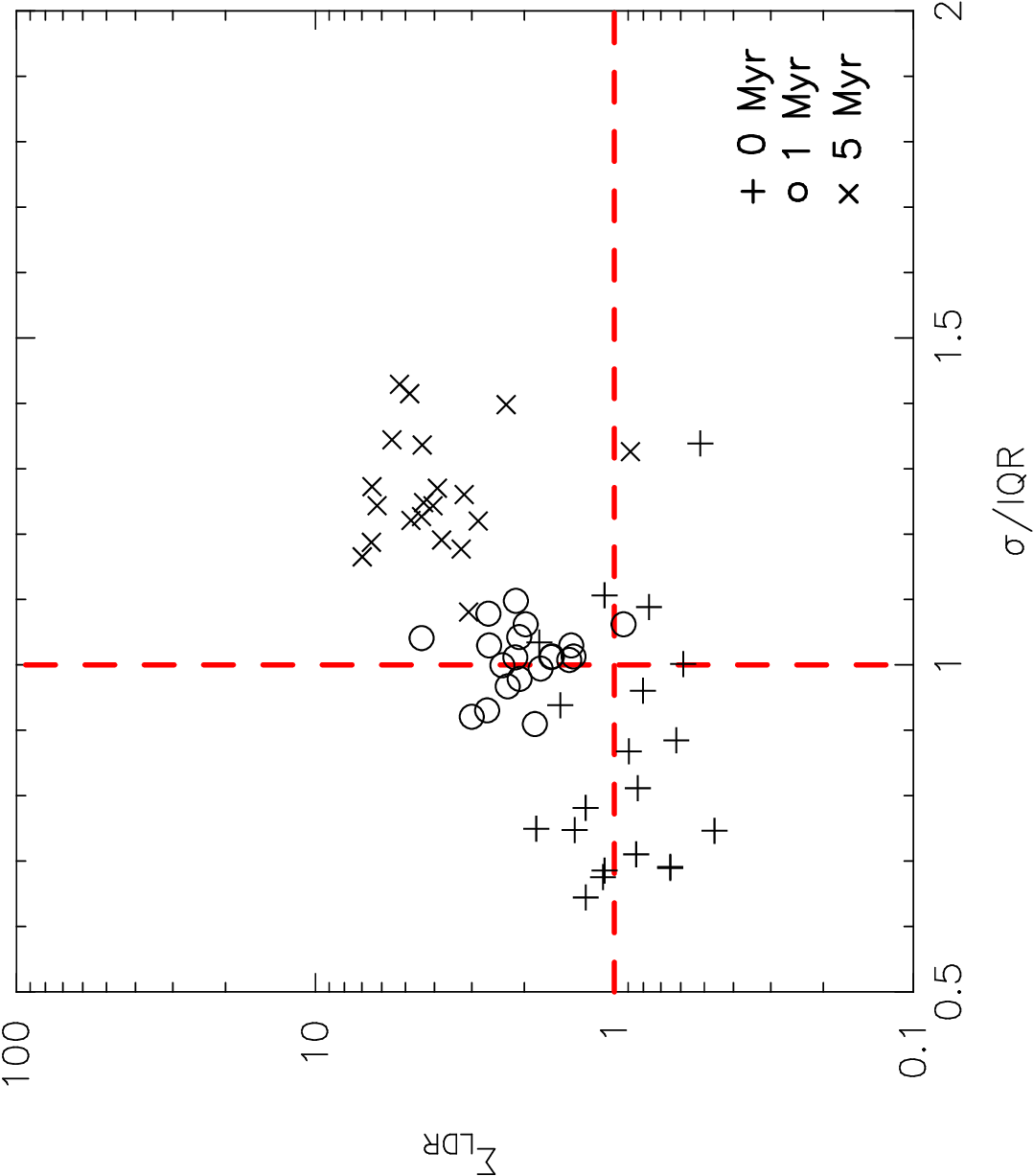}}}
\hspace*{0.3cm} 
\subfigure[$\alpha_{\rm vir} = 0.3, D = 3.0$]{\label{Sig_vd_IQR-c}\rotatebox{270}{\includegraphics[scale=0.27]{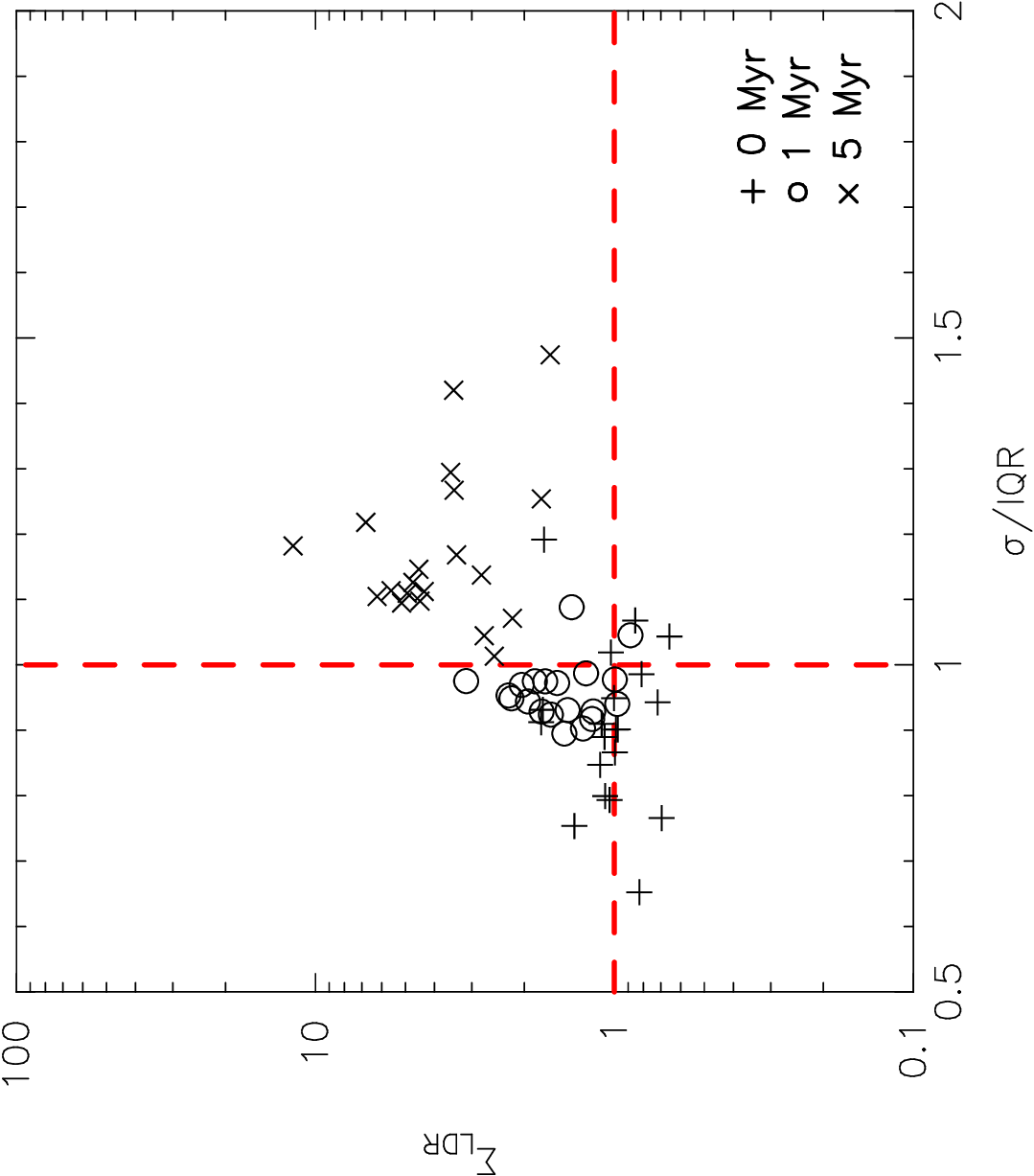}}}
%\vspace*{-0.3cm}
\hspace*{-1.cm}
\subfigure[$\alpha_{\rm vir} = 0.5, D = 1.6$]{\label{Sig_vd_IQR-d}\rotatebox{270}{\includegraphics[scale=0.27]{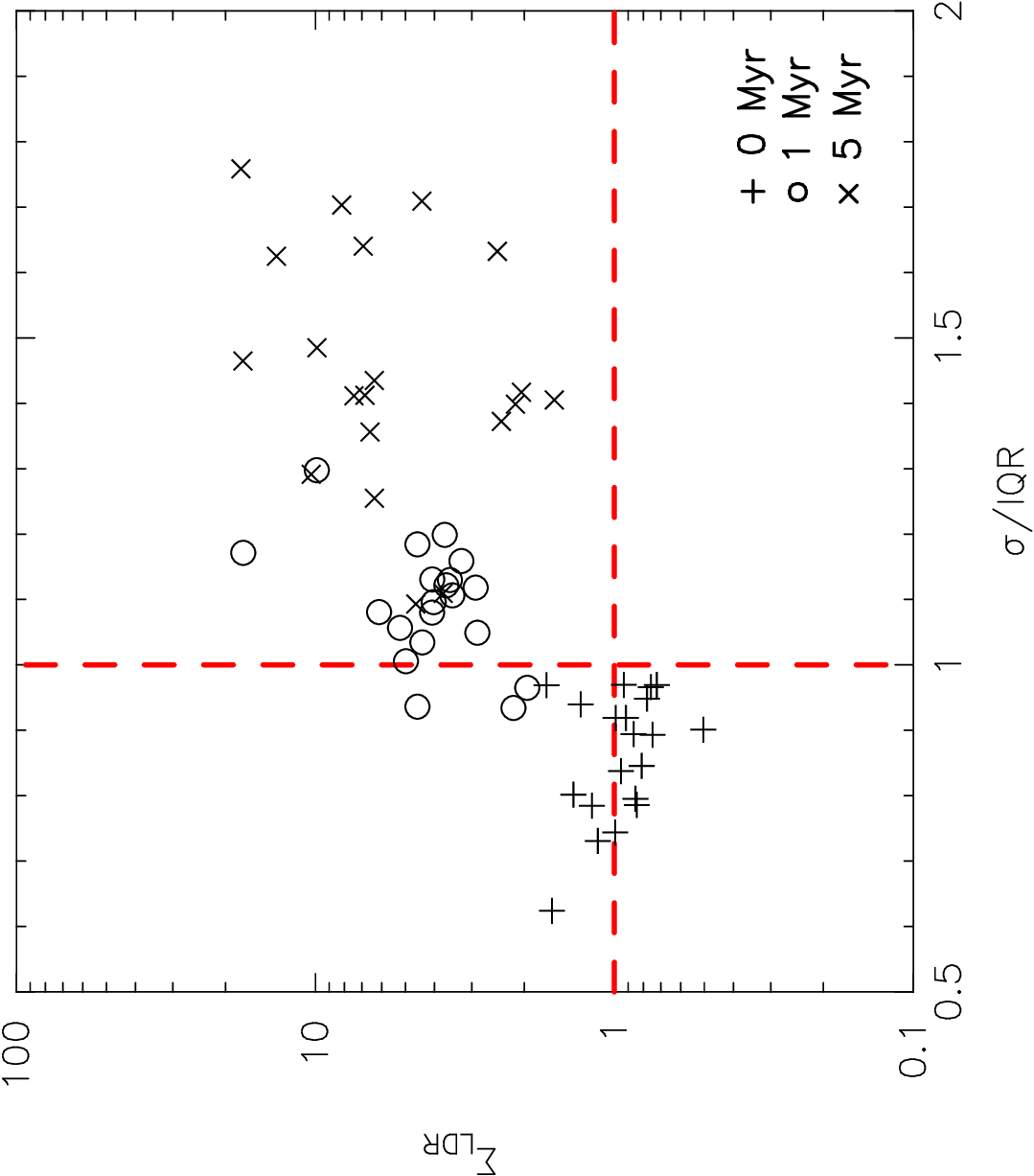}}}
\hspace*{0.3cm} 
\subfigure[$\alpha_{\rm vir} = 0.5, D = 2.0$]{\label{Sig_vd_IQR-e}\rotatebox{270}{\includegraphics[scale=0.27]{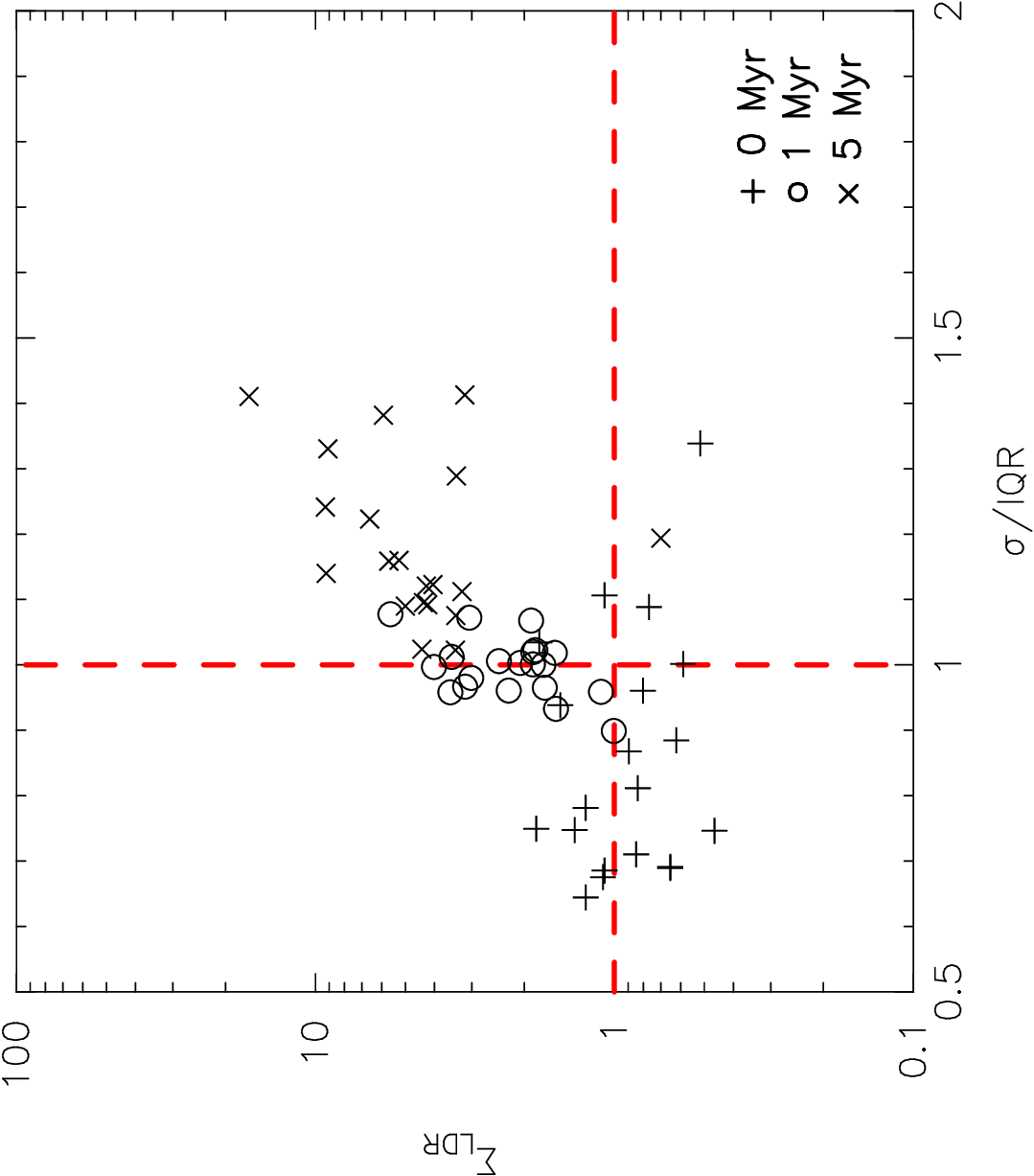}}}
\hspace*{0.3cm} 
\subfigure[$\alpha_{\rm vir} = 0.5, D = 3.0$]{\label{Sig_vd_IQR-f}\rotatebox{270}{\includegraphics[scale=0.27]{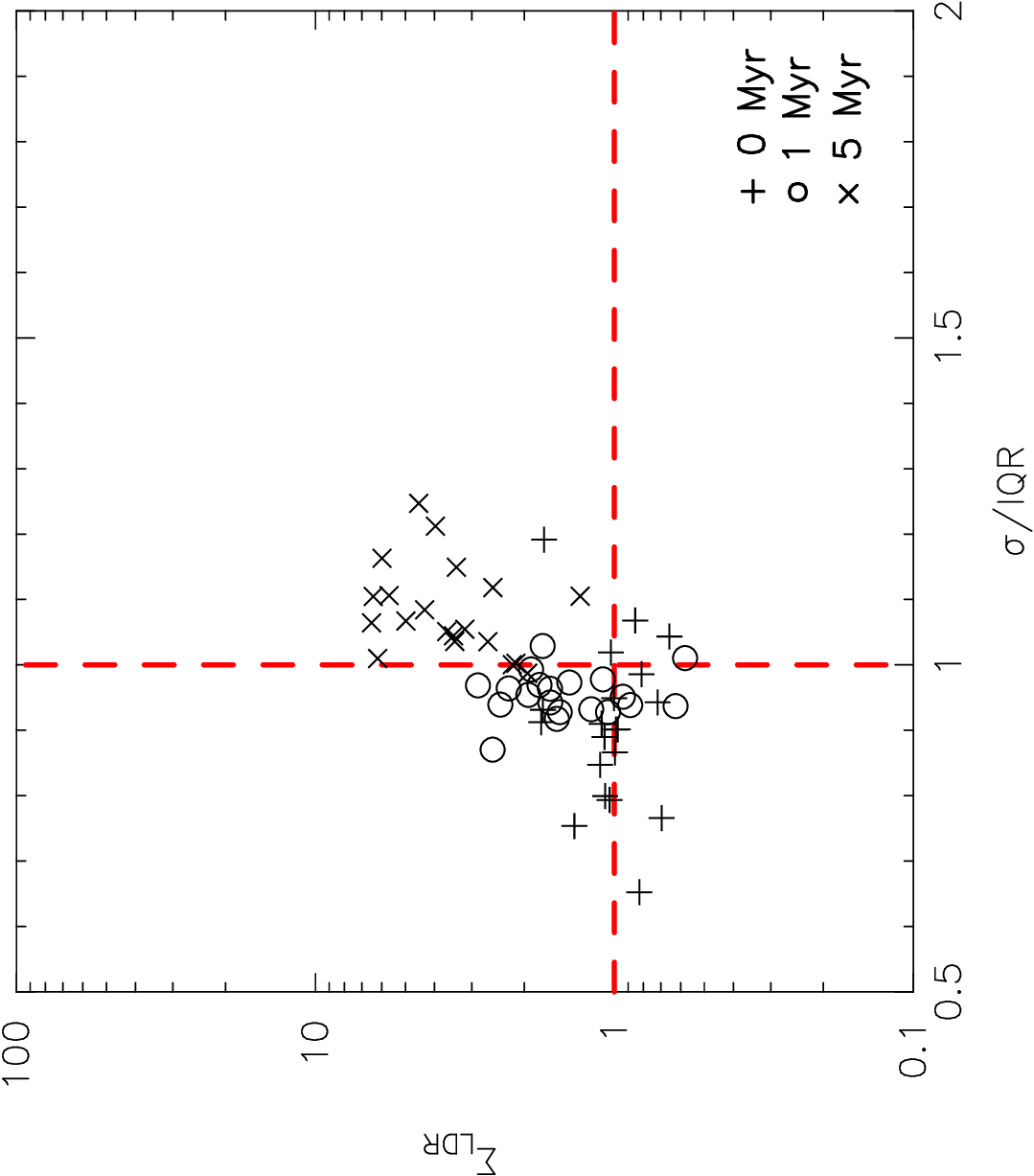}}}
\hspace*{-1.cm}
\subfigure[$\alpha_{\rm vir} = 1.5, D = 1.6$]{\label{Sig_vd_IQR-g}\rotatebox{270}{\includegraphics[scale=0.27]{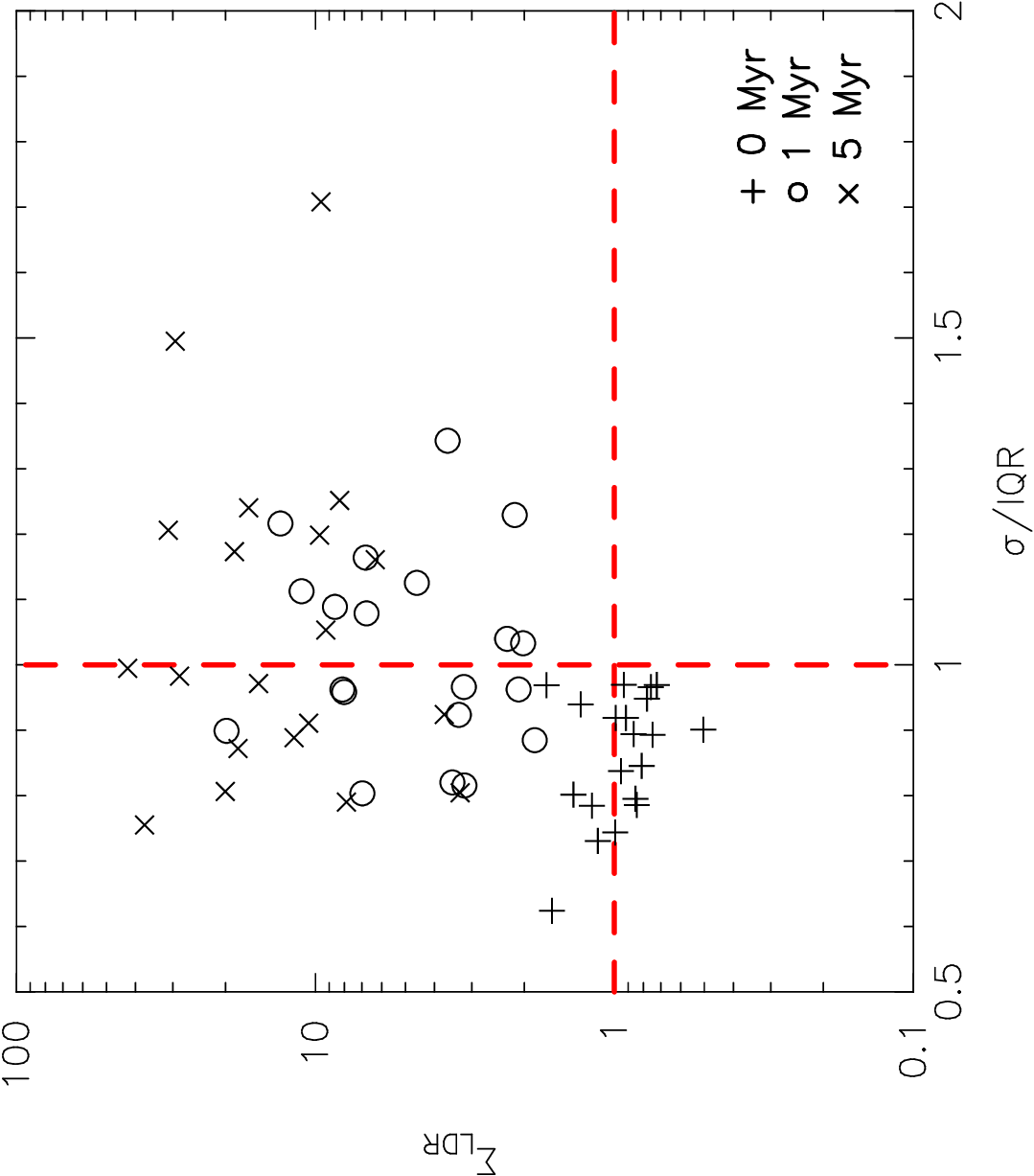}}}
\hspace*{0.3cm} 
\subfigure[$\alpha_{\rm vir} = 1.5, D = 2.0$]{\label{Sig_vd_IQR-h}\rotatebox{270}{\includegraphics[scale=0.27]{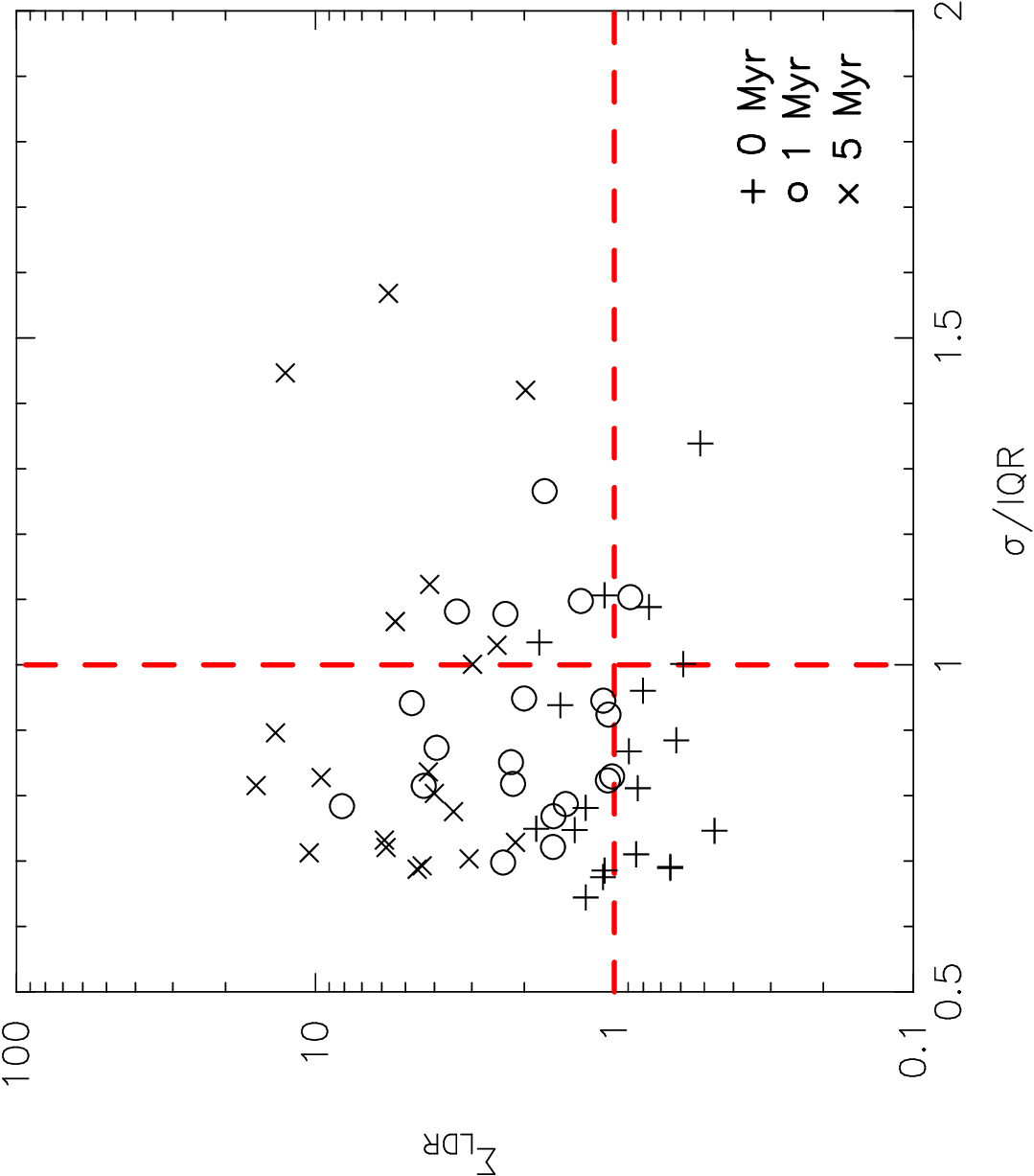}}}
\hspace*{0.3cm} 
\subfigure[$\alpha_{\rm vir} = 1.5, D = 3.0$]{\label{Sig_vd_IQR-i}\rotatebox{270}{\includegraphics[scale=0.27]{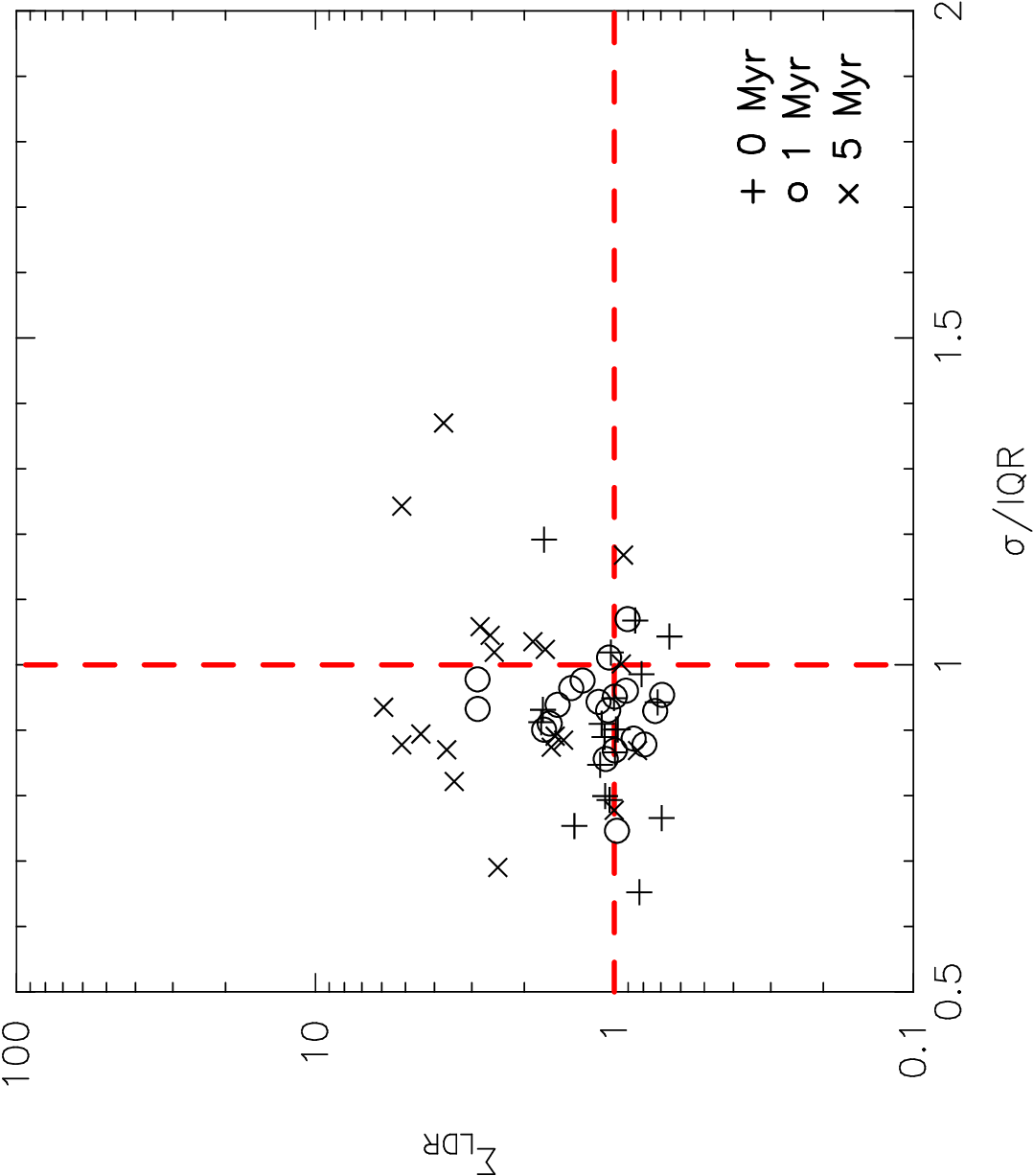}}}
\caption[bf]{Evolution of the  $\Sigma_{\rm LDR}$ local density ratio versus $\sigma$/IQR for all initial conditions in the simulations containing $N = 1500$ stars. For each set of initial conditions we show the values from 20 realisations of the same simulation, and plot values at 0\,Myr (before dynamical evolution) -- the plus signs; 1\,Myr -- the open circles; and 5\,Myr -- the crosses.  The horizontal dashed line indicates $\Sigma_{\rm LDR} = 1$ and the vertical dashed line indicates  $\sigma$/IQR = 1.  }
\label{Sig_vd_IQR}
  \end{center}
\end{figure*}

\begin{figure*}
  \begin{center}
\setlength{\subfigcapskip}{10pt}
\vspace*{-0.3cm}
\hspace*{-1.cm}\subfigure[$\alpha_{\rm vir} = 0.3, D = 1.6$]{\label{OH_Sig_vd_IQR-a}\rotatebox{270}{\includegraphics[scale=0.27]{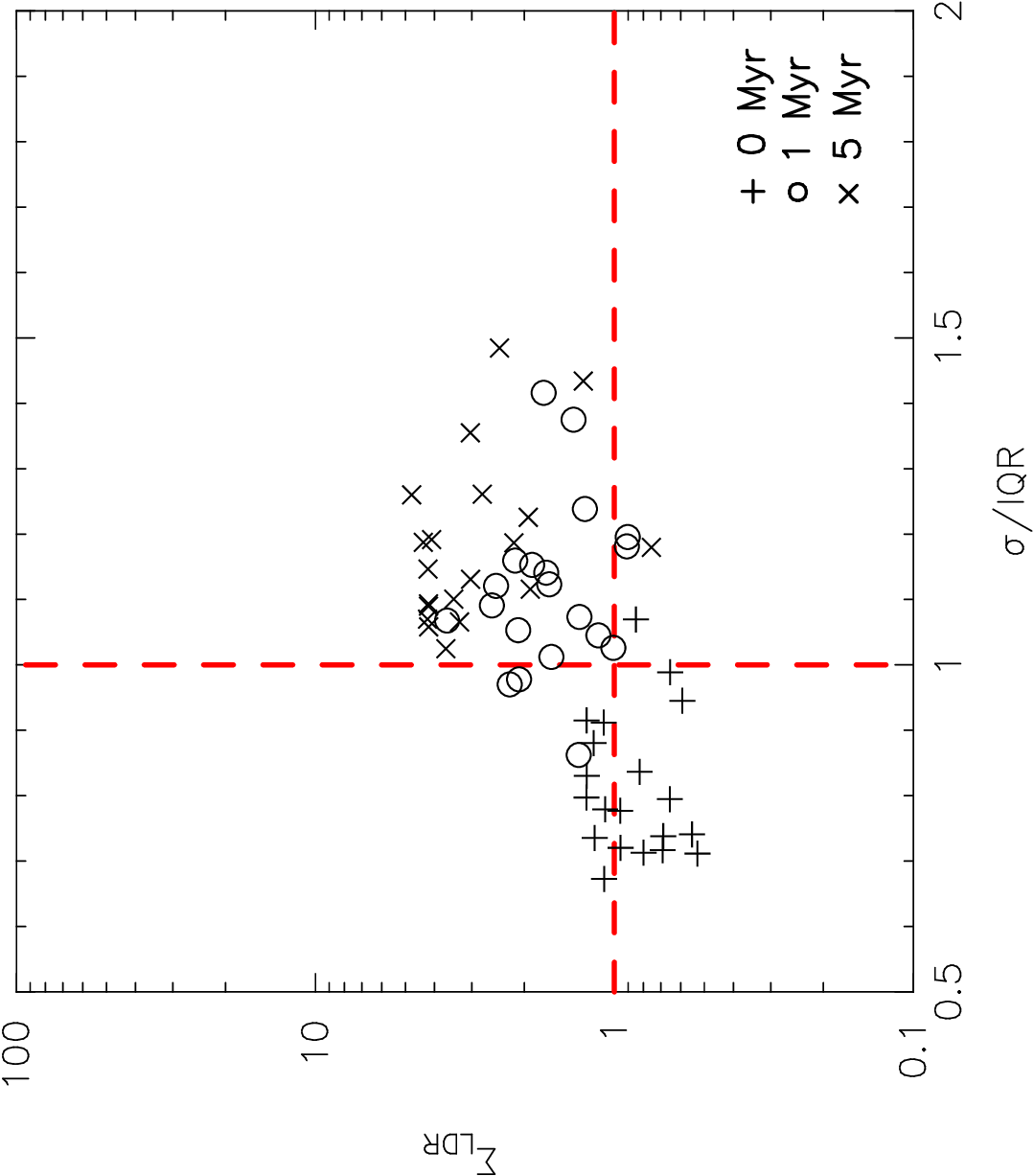}}}
\hspace*{0.3cm} 
\subfigure[$\alpha_{\rm vir} = 0.3, D = 2.0$]{\label{OH_Sig_vd_IQR-b}\rotatebox{270}{\includegraphics[scale=0.27]{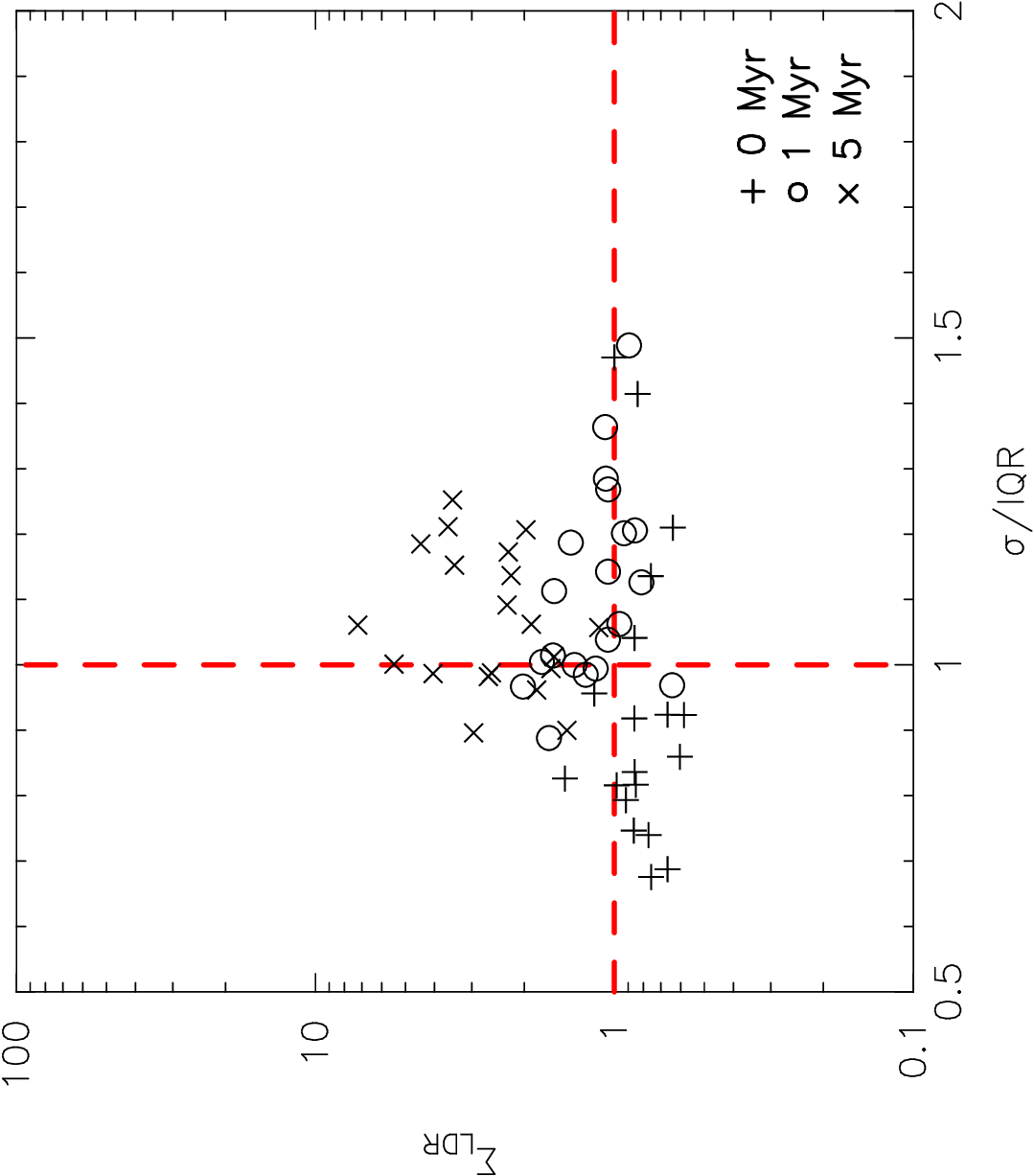}}}
\hspace*{0.3cm} 
\subfigure[$\alpha_{\rm vir} = 0.3, D = 3.0$]{\label{OH_Sig_vd_IQR-c}\rotatebox{270}{\includegraphics[scale=0.27]{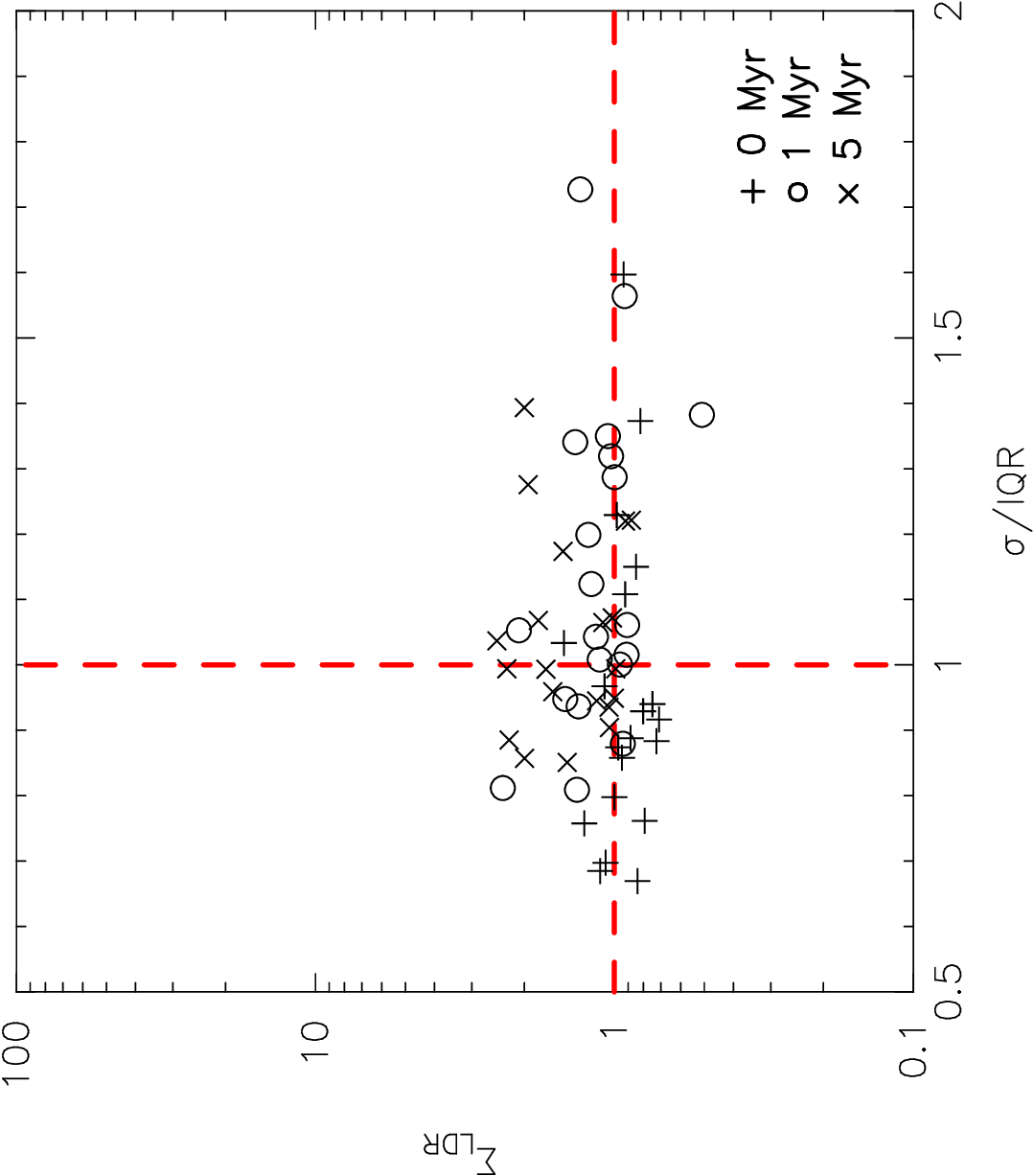}}}
%\vspace*{-0.3cm}
\hspace*{-1.cm}
\subfigure[$\alpha_{\rm vir} = 0.5, D = 1.6$]{\label{OH_Sig_vd_IQR-d}\rotatebox{270}{\includegraphics[scale=0.27]{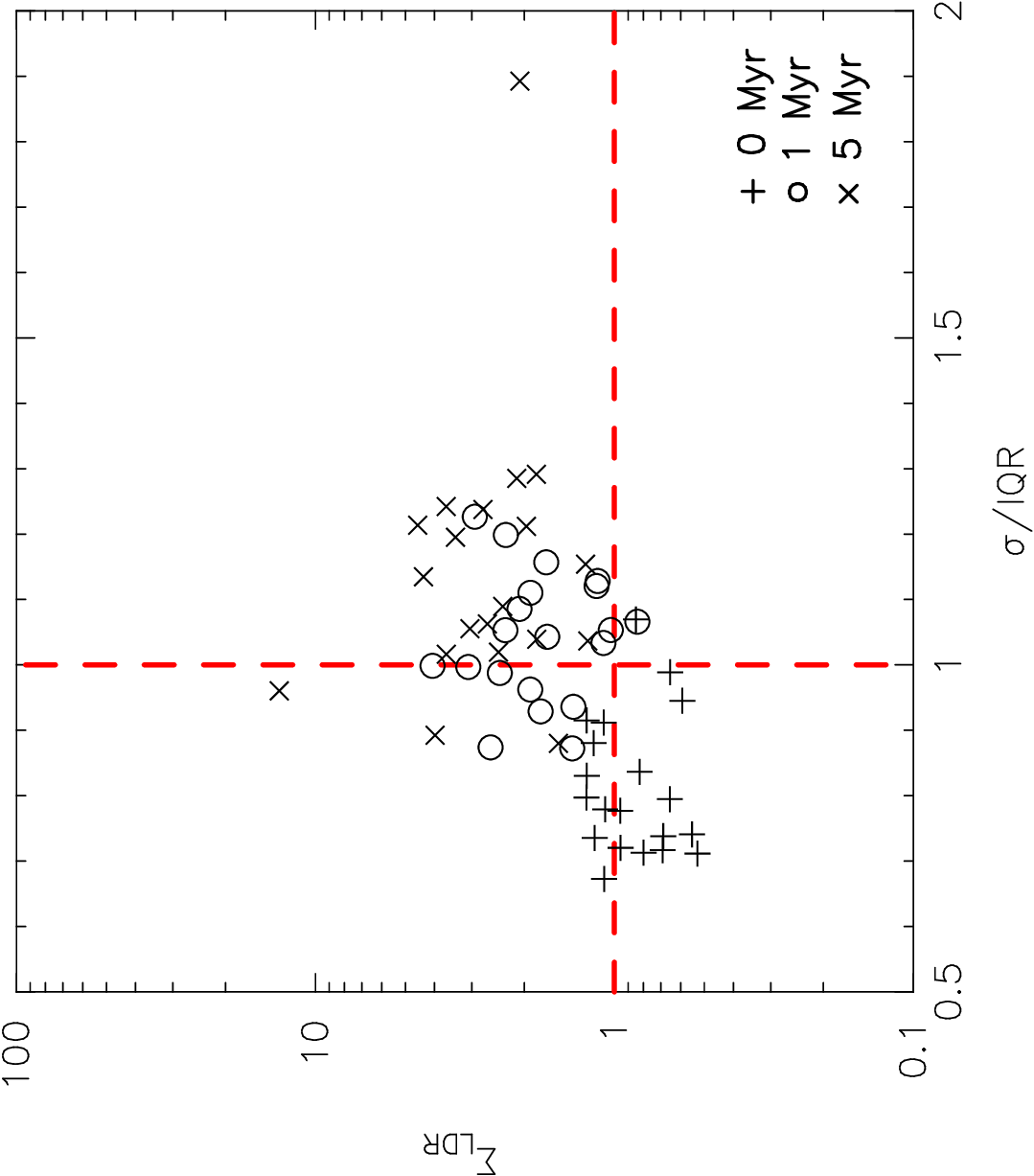}}}
\hspace*{0.3cm} 
\subfigure[$\alpha_{\rm vir} = 0.5, D = 2.0$]{\label{OH_Sig_vd_IQR-e}\rotatebox{270}{\includegraphics[scale=0.27]{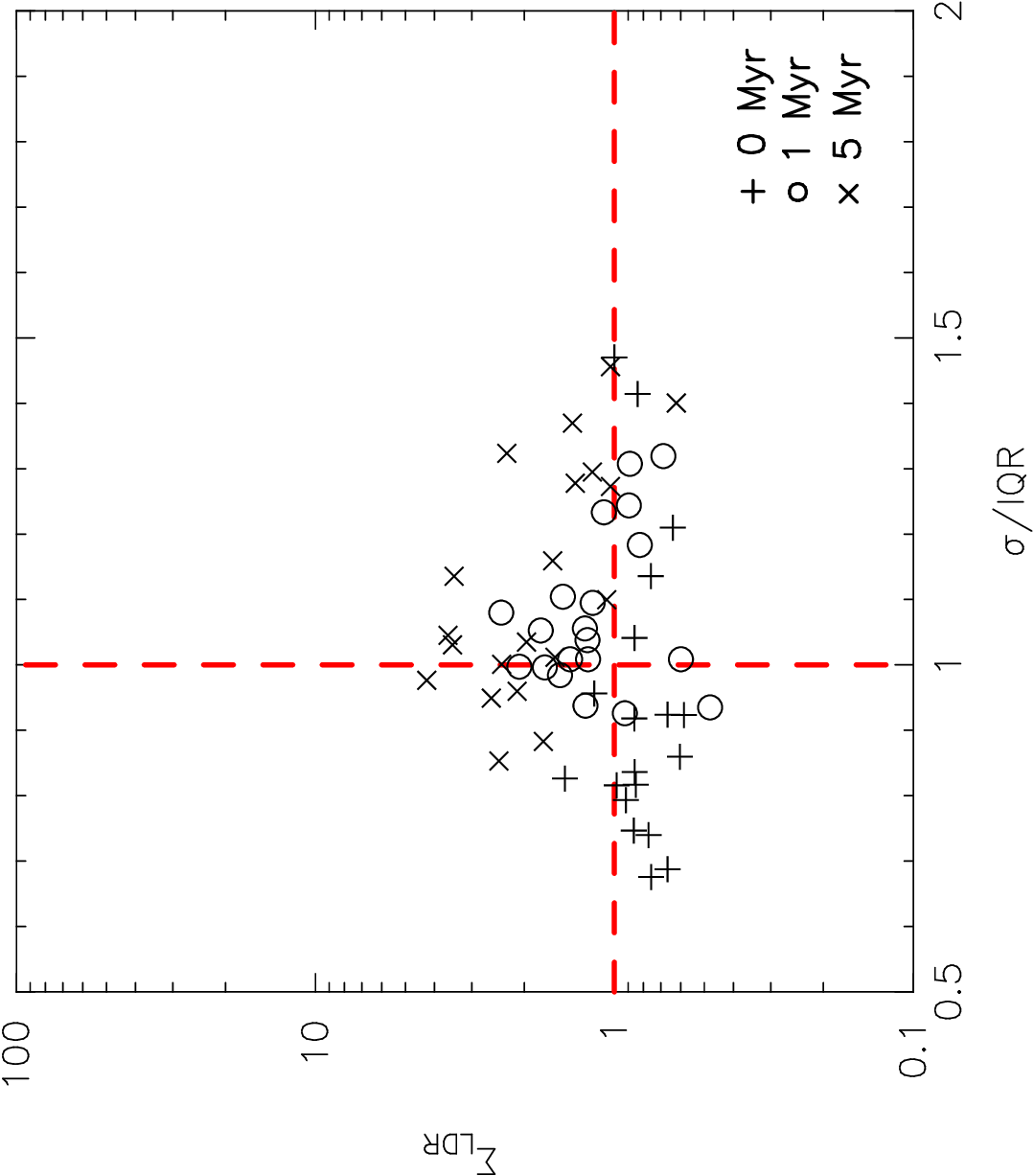}}}
\hspace*{0.3cm} 
\subfigure[$\alpha_{\rm vir} = 0.5, D = 3.0$]{\label{OH_Sig_vd_IQR-f}\rotatebox{270}{\includegraphics[scale=0.27]{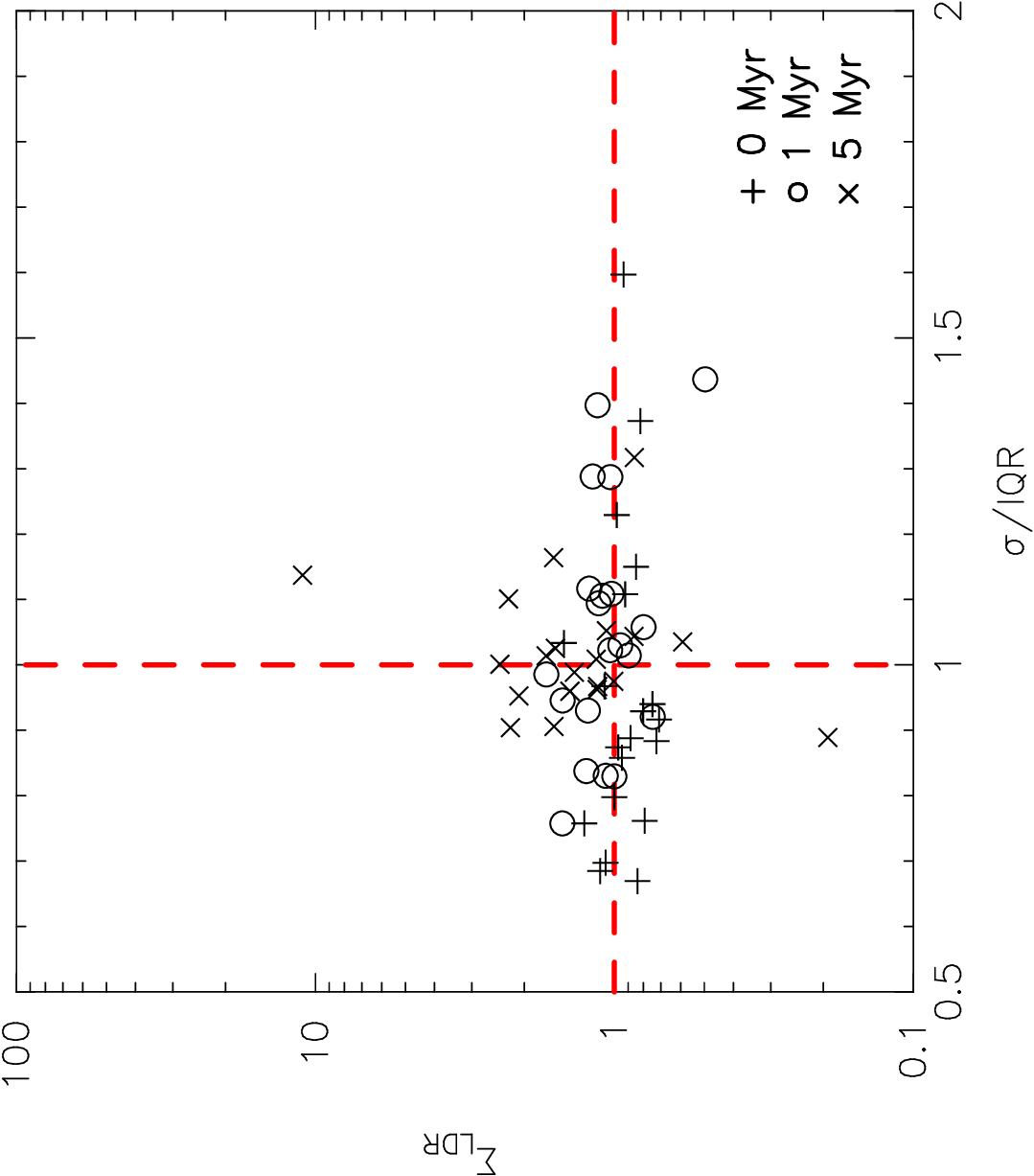}}}
\hspace*{-1.cm}
\subfigure[$\alpha_{\rm vir} = 1.5, D = 1.6$]{\label{OH_Sig_vd_IQR-g}\rotatebox{270}{\includegraphics[scale=0.27]{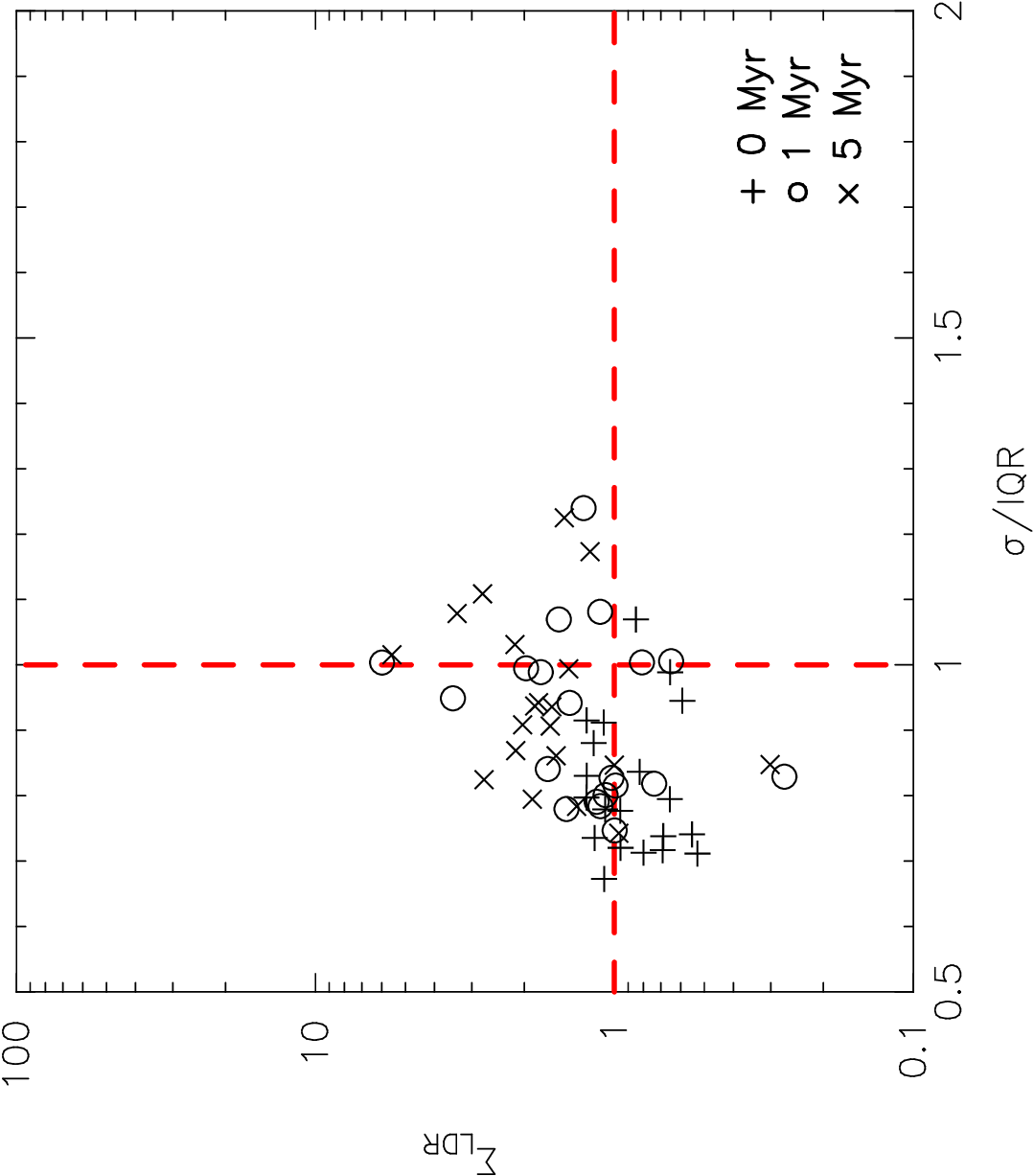}}}
\hspace*{0.3cm} 
\subfigure[$\alpha_{\rm vir} = 1.5, D = 2.0$]{\label{OH_Sig_vd_IQR-h}\rotatebox{270}{\includegraphics[scale=0.27]{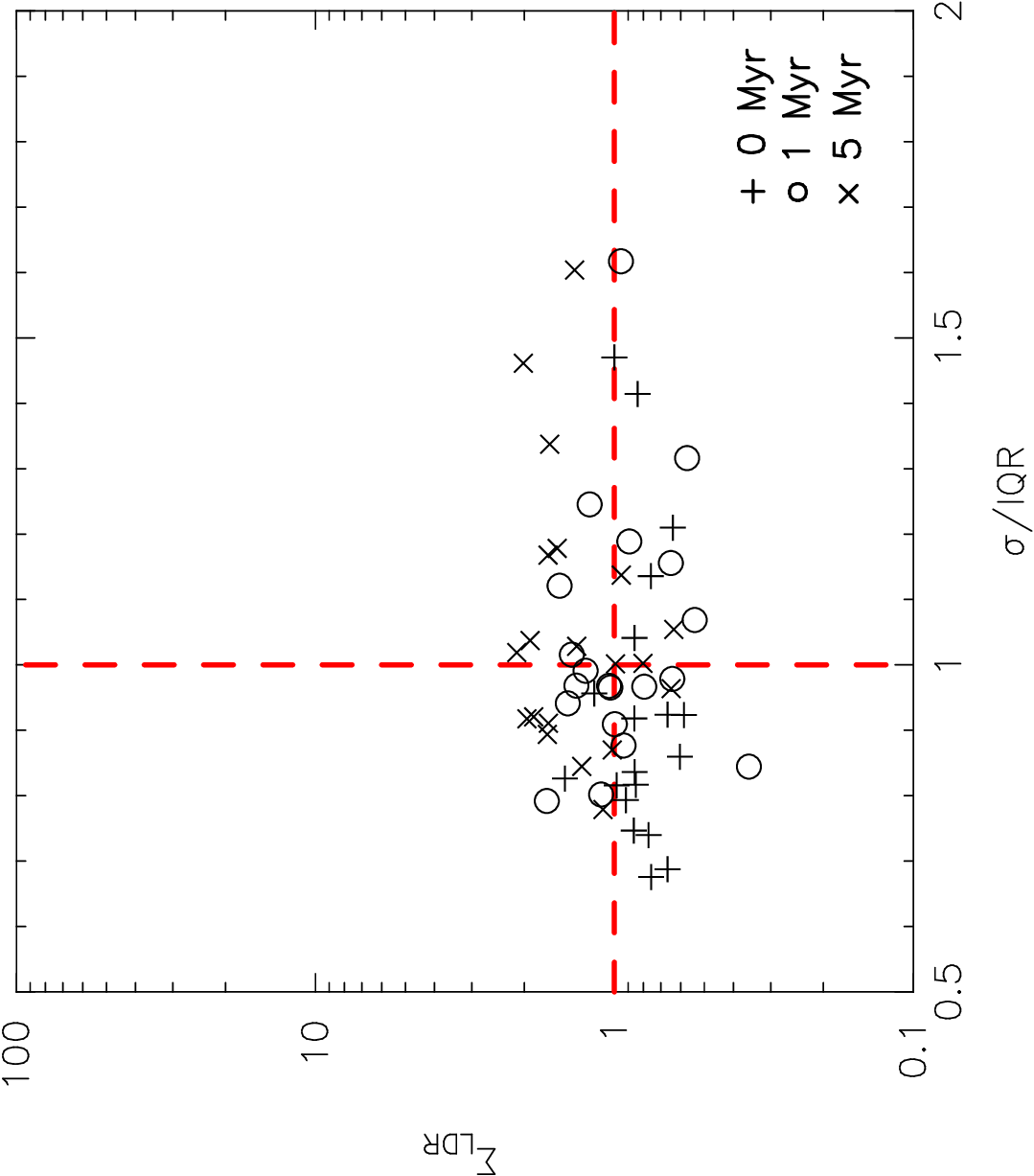}}}
\hspace*{0.3cm} 
\subfigure[$\alpha_{\rm vir} = 1.5, D = 3.0$]{\label{OH_Sig_vd_IQR-i}\rotatebox{270}{\includegraphics[scale=0.27]{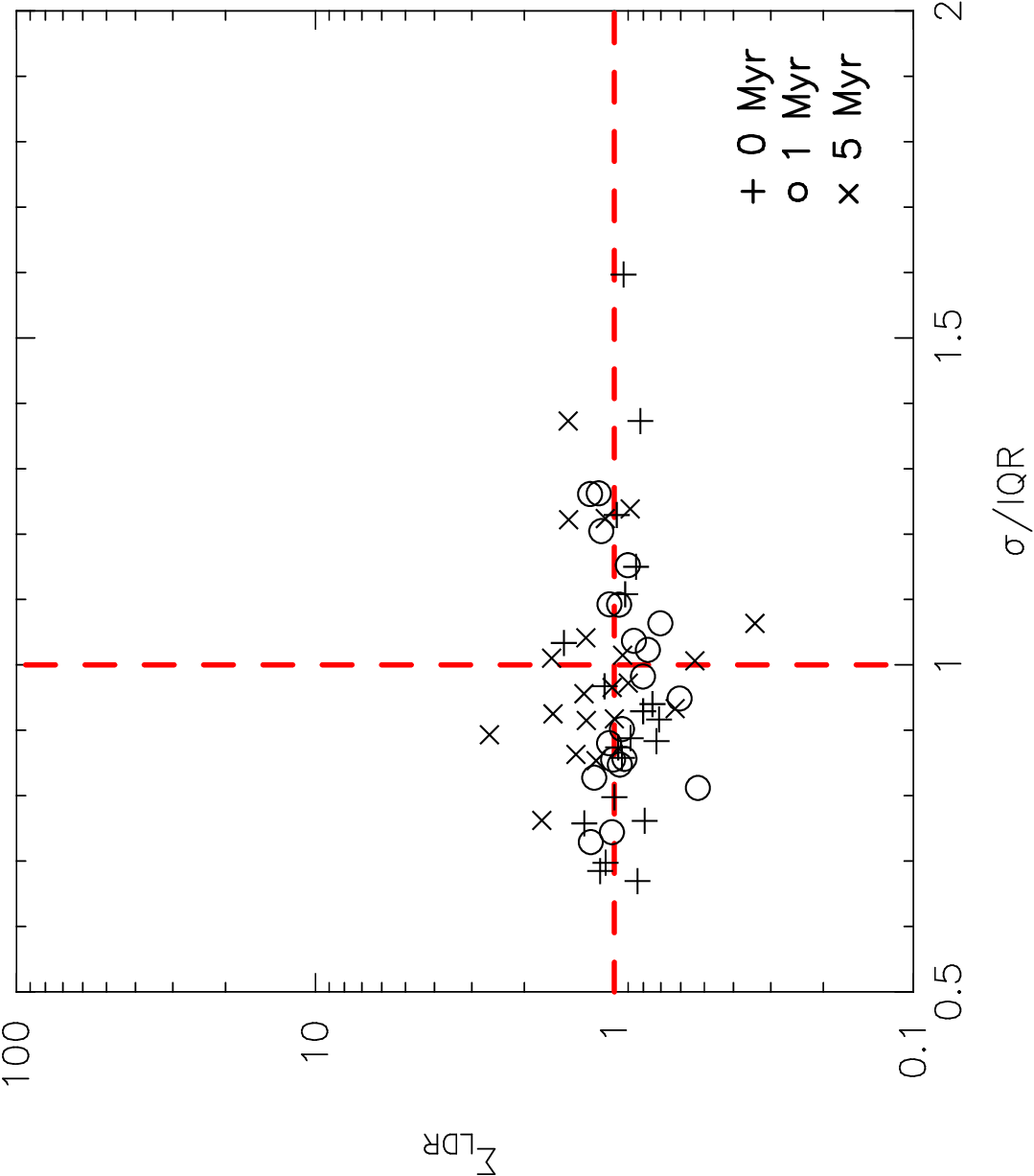}}}
\caption[bf]{Evolution of the  $\Sigma_{\rm LDR}$ local density ratio versus $\sigma$/IQR for all initial conditions in the simulations containing $N = 150$ stars. For each set of initial conditions we show the values from 20 realisations of the same simulation, and plot values at 0\,Myr (before dynamical evolution) -- the plus signs; 1\,Myr -- the open circles; and 5\,Myr -- the crosses.  The horizontal dashed line indicates $\Sigma_{\rm LDR} = 1$ and the vertical dashed line indicates  $\sigma$/IQR = 1.  }
\label{OH_Sig_vd_IQR}
  \end{center}
\end{figure*}

In \citet{Parker14b}, we demonstrated that a combination of diagnostics for the spatial distribution of star-forming regions can help distinguish between different initial conditions. Specifically, we found that the combination of the spatial structure of the region, as defined by the $\mathcal{Q}$-parameter \citep{Cartwright04,Cartwright09b} in tandem with the mass segregation ratio $\Lambda_{\rm MSR}$ \citep{Allison09a}, and the local surface density ratio $\Sigma_{\rm LDR}$ \citep{Maschberger11,Kupper11,Parker14b} evolve very differently if a region is subvirial as opposed to supervirial. These measures also enable us to place constraints on the initial density of a star-forming region \citep[see also][]{Wright14,Parker14e}.

The $\mathcal{Q}$-parameter defines the amount of spatial substructure present in a region, by dividing the mean length $\bar{m}$ of a minimum spanning tree (MST) that connects all points in a distribution via the shortest possible path length, by the mean separation between the points, $\bar{s}$. $\mathcal{Q}<0.7$ indicates a hierarchical substructured distribution, and $\mathcal{Q}>0.9$ indicates a smooth, centrally concentrated distribution. $\Lambda_{\rm MSR}$ compares the MST of a chosen subset of stars to the average MSTs of randomly chosen stars, and $\Lambda_{\rm MSR} >> 1$ indicates significant mass segregation. $\Sigma_{\rm LDR}$ takes the median surface density of a chosen subset of stars and compares this to the median surface density for the entire distribution and $\Sigma_{\rm LDR} >> 1$ indicates that the chosen subset of stars reside in areas of higher than average surface density. Despite this, there still exists degeneracy in this technique under certain initial conditions (e.g. initial densities lower than 100\,M$_\odot$\,pc$^{-3}$), and the addition of a further diagnostic(s) would be helpful. 

\subsubsection{$\mathcal{Q}$-parameter versus $\sigma$}

In Fig.~\ref{Q_vd} we plot the spatial structure of the region, as measured by the $\mathcal{Q}$-parameter, against the measured velocity dispersion about the mean radial velocity, $\sigma$ for the regions with $N = 1500$ stars, fractal dimension $D = 1.6$ and either subvirial (panel a) or supervirial (panel b) velocities.

If we consider the subvirial ($\alpha_{\rm vir} = 0.3, D = 1.6$) regions, in panel (a) we see that the dynamical evolution of these regions follows a distinct path. Interactions erase the initial substructure, so $\mathcal{Q}$ rapidly increases to $\mathcal{Q} > 0.8$ and the regions are no longer substructured. At the same time, the velocity dispersion increases slightly in the first 1\,Myr, before decreasing again after 5\,Myr. At 5\,Myr the $\mathcal{Q}$-parameter has increased further, populating distinct areas of the $\mathcal{Q} - \sigma$ plot at these times. 

The initially supervirial regions ($\alpha_{\rm vir} = 1.5$) retain their substructure, and the velocity dispersions do not change significantly (despite appearing to decrease slightly, as shown in Figs.~\ref{vd_lines},~\ref{vdA_med}~and~\ref{vdC_med}). Furthermore, because primordial substructure is preserved in the supervirial expansion, there is little distinction between different times in the simulations.

However, because the velocity dispersion depends on the number of stars in a region, two regions with very different $\sigma$ may have similar virial ratios due to their different total mass. For this reason, it is more useful to use a diagnostic that does not vary as a function of region mass.

\subsubsection{$\mathcal{Q}$-parameter versus $\sigma /{\rm IQR}$}

In Section~\ref{results:all} we noted that the statistical velocity dispersion about the mean, $\sigma$, increased with respect to the dispersion defined by the IQR. We take the ratio of these quantities, $\sigma$/IQR and plot this as a function of the  $\mathcal{Q}$-parameter in Fig.~\ref{Q_vd_IQR}. In all panels, the black points show the values determined from using every star in each simulation, whereas the grey symbols are for bound stars only. In this plot we see that the (sub)virial simulations, especially those with low fractal dimensions (more substructure), follow a distinct evolutionary path in $\mathcal{Q} - \sigma /{\rm IQR}$ space, in that initial conditions that lead to the collapse of a region and the formation of a cluster tend towards high $\mathcal{Q}$ and $\sigma /{\rm IQR}$ ratios (panels a -- f). The supervirial simulations never attain high $\sigma /{\rm IQR}$ ratios and these plots are degenerate with time (panels g -- i).

The evolution in $\mathcal{Q} - \sigma /{\rm IQR}$ space is more pronounced when we consider all the stars in the simulation, rather than just the bound stars. This is likely due to the fact that unbound stars, which are ejected from the clusters, are included in the calculation of $\sigma$, but not in the IQR, and so the ratio of these quantities increases. Observationally, it has been difficult to determine the full population of a star-forming region (including ejected stars), but the advent of \emph{Gaia} and associated surveys may allow observers to trace back ejected stars to their birth regions using a combination of proper motion velocities \citep{Allison12} and chemical tagging \citep{Bland-Hawthorn10}. Combining this information with the $\mathcal{Q} - \sigma /{\rm IQR}$ plot \citep[as well as the $\mathcal{Q} - \Sigma_{\rm LDR}$ plot,][]{Parker14b,Parker14e}, will place constraints on the initial conditions of a star-forming region, such as the initial virial ratio and amount of spatial structure.

However, when the star-forming regions start with low-density initial conditions ($<$100\,M$_\odot$\,pc$^{-3}$, Fig.~\ref{Q_OHvd_IQR}), the evolution of $\mathcal{Q} - \sigma /{\rm IQR}$ in the first 10\,Myr is minimal compared to the more dense initial conditions in  Fig.~\ref{Q_vd_IQR}. In that sense, it is only possible to rule out dense, supervirial initial conditions using the $\mathcal{Q} - \sigma /{\rm IQR}$ plot when confronted with observations of a region with age $<$10\,Myr. However, low-density regions will eventually evolve towards the same values of $\mathcal{Q} - \sigma /{\rm IQR}$. In general, we find that regions with initial median densities of a factor 10 lower than our fiducial high-density simulations evolve over timescales a factor 10 longer. In a future paper we will explore the long-term evolution of high-N regions and determine which fraction of them remain as bound open clusters.

\subsubsection{$\Sigma_{\rm LDR}$ versus $\sigma /{\rm IQR}$}

Finally, we plot the evolution of the local surface density ratio, $\Sigma_{\rm LDR}$ against $\sigma /{\rm IQR}$, in Fig.~\ref{Sig_vd_IQR}. This plot can potentially distinguish between substructured subvirial initial condtions, and substructured supervirial simulations (compare panels (a) and (g)). This is principally because the  $\sigma /{\rm IQR}$ ratio does not increase for the supervirial initial conditions whilst $\Sigma_{\rm LDR}$ does. Again, simulations with lower densities do not evolve as quickly, and in Fig.~\ref{OH_Sig_vd_IQR} we show that this plot is also degenerate for the low-density regions on timescales less than 10\,Myr. However, we reiterate that combining these diagnostics with further measures of spatial structure are likely to be more fruitful than using velocity dispersions alone.

\section{Discussion}
\label{discuss}

The results of our $N$-body simulations can be summarised as follows. Firstly, the radial velocity dispersion appears to be unsuitable as the sole indicator of the initial, or even present-day virial state of a stellar system. This is because for a bound cluster in virial equilibrium, after several Myr of evolution  the velocity dispersion routinely exceeds the value estimated assuming a region of that mass and radius is in virial equilibrium (the so-called virial mass velocity dispersion estimate). 

Secondly, the regions that have supervirial initial conditions remain supervirial throughout, which is shown by the velocity dispersion measurement greatly exceeding the estimate from the virial mass. However, if (sub)virial regions also have velocity dispersions that appear supervirial, distinguishing between genuinely supervirial regions and those that are evolved (sub)virial regions becomes non-trivial. 

Finally, in regions undergoing violent relaxation, the most massive stars attain larger velocity dispersions than the average stars in the region. This is not due to their being in close binary systems, which would inflate the velocity dispersion \citep{Gieles10,Cottaar12b,Cottaar14a}; we have checked and very few of the massive stars  (which are all initially single) end up in close binary systems. Previous work \citep{Allison10,Parker14b,Caputo14} has shown that violent relaxation leads to strong dynamical mass segregation. If this mass segregation were a signature of energy equipartition we would expect the most massive stars to be moving more slowly than the average star, which is the opposite to what we see in the simulations. We will explore this problem further in a future paper.

Our work is not the first to examine the evolution of the radial velocity dispersion in $N$-body simulations. Earlier studies often focussed on the reaction of the velocity dispersion of the stars to the removal of a background gas potential \citep[e.g.][]{Goodwin97,Kroupa03b,Adams06,Goodwin06,Baumgardt07,Proszkow09,Moeckel10}. However, recent observational \citep{Rochau10,Cottaar12a} and theoretical work \citep{Smith11,Kruijssen12a,Moeckel12,Parker13a} has shown that the reaction of dense star-forming regions with a more realistic treatment of the gas potential to gas expulsion is minimal. It is more likely that the dynamical interactions within the most dense regions drive the evolution, rather than the removal of gas.

In our simulations, we have ignored the (likely minimal) effect of gas removal although we do include supervirial simulations in our analysis which would be the dynamic outcome if gas expulsion is important, and simply focused on the evolution of the velocity dispersion for different initial conditions. Typically, we find that no region with $N = 1500$ stars has a velocity dispersion larger than $\sigma =  1.5$\,km\,s$^{-1}$, irrespective of the initial virial ratio, and regions with fewer stars ($N = 150$) have much lower maximum dispersions, typically $\sim 0.5$\,km\,s$^{-1}$.  (Note that these values are specific to our simulations. In the following section we discuss how these values can be compared to observed star-forming regions.) However, there is little variation in the velocity dispersion values for regions with similar numbers of stars, and as we have discussed, the virial mass estimate can often be misleading. 

Given this, can we use the radial velocity dispersion to place any constraints on the initial conditions, or future evolution of a star-forming region? We have seen in Fig.~\ref{Q_vd} that combining the velocity dispersion with an accurate determination of the structure of the region (using the $\mathcal{Q}$-parameter) can help when the regions have high local densities ($> 10^3$M$_\odot$\,pc$^{-3}$), as initial substructure is erased and the velocity dispersion stays relatively constant. However, when the initial density is lower ($\sim 10^2$M$_\odot$\,pc$^{-3}$), structure is retained for longer, and the $\mathcal{Q} - \sigma$ plot becomes degenerate.

Dense, violent initial conditions lead to the ejection of stars and subsequently inflates the statistical dispersion, $\sigma$. Because of this, the ratio of $\sigma$ to the IQR dispersion increases for these initial conditions. When this ratio is shown as a function of the $\mathcal{Q}$-parameter (Fig.~\ref{Q_vd_IQR}), then a clear signature is apparent in the evolution.

\subsection{Comparison with observations}

Based on our results, what can we say about observed velocity dispersions in nearby star-forming regions? Firstly, it is worth noting that in our simulations we have access to the radial velocity of every star, and knowledge of whether that star is in a binary or not \citep[our regions were set up with no binaries initially, but wide systems can form in these substructured regions with correlated stellar velocities,][]{Kouwenhoven10,Moeckel10,Parker14d}. Spectroscopic binaries inflate the velocity dispersion due to their fast orbital motion, and have to be corrected for in observed dispersions using either multi-epoch data \citep{Geller08,Gieles10} or removing this motion using a maximum likelihood technique \citep{Odenkirchen02,Cottaar12b,Cottaar14a}. 

\citet{Furesz08} measured a radial velocity dispersion of $\sigma = 3.1$\,km\,s$^{-1}$ for the Orion Nebula Cluster (ONC), suggesting that the region is extremely supervirial, as derived by \citet{Olczak08}, who estimate $\sigma_{\rm vir} = 1.6$\,km\,s$^{-1}$. Since \citet{Olczak08} use the half-mass radius to estimate $\sigma_{\rm vir}$ their conclusion will not be seriously affected by the cluster boundary issues we have detailed here and given the large difference between the measured and virial velocity dispersions they find it is likely the case that the ONC is not in virial equilibrium. However, we note that in our simulations that undergo cool-collapse, the measured velocity dispersion always exceeds $\sigma_{\rm vir}$ even when adopting the half-mass radius. 

More recently, \citet{Foster15} observed NCG\,1333 and find a velocity dispersion for pre-stellar cores of $\sigma = 0.5$\,km\,s$^{-1}$ (consistent with being subvirial as they determine $\sigma_{\rm vir} = 1.1$\,km\,s$^{-1}$ for the gas), whereas the young stars have a larger velocity dispersion of $\sigma = 0.92 \pm 0.12$\,km\,s$^{-1}$, consistent with virial equilibrium ($\sigma_{\rm vir} = 0.79 \pm 0.20$\,km\,s$^{-1}$). If the pre-stellar cores form with lower velocities, and then interact as they form stars, then our cool-collapse dynamical evolution model readily explains this difference. If this were to be the case, we predict that the stars in NGC\,1333 would have a large $\mathcal{Q}$-parameter and potentially a high $\Sigma_{\rm LDR}$ ratio -- two further signatures of dynamical evolution. Data on the masses of stars in this region are not publicly available, however, an analysis of the structure using data from \citet{Gutermuth08} suggests $\mathcal{Q} = 0.91$, consistent with strong dynamical evolution.

\citet{Jeffries14} present observations of the $\gamma^2$~Velorum region, which appears to consist of two spatially coincident (at least along the line of sight) sub-clusters with very different kinematic properties. One sub-cluster has a velocity dispersion of $\sigma = 0.33$\,km\,s$^{-1}$ and the second has $\sigma = 1.7$\,km\,s$^{-1}$. \citet{Jeffries14} found the second cluster to be supervirial while the first was estimated to have a virial velocity dispersion of $\sigma_{\rm vir} = 0.27$\,km\,s$^{-1}$, which is consistent with being in virial equilibrium, especially when taking into account our finding that systems in virial equilibrium may exhibit radial velocity dispersions slightly above the virial velocity dispersion. The two populations are offset by about 2\,km\,s$^{-1}$ in radial velocity. None of our simulated regions display this degree of kinematic substructure, despite some of the supervirial regions evolving into spatially distinct `binary' or `double' clusters \citep[see e.g.\,\,fig.~4b from][for an example of such a system]{Parker14a}. We suggest that the two populations in $\gamma^2$~Vel formed as kinematically distinct star-formation events, and are either chance superposition, or the result of a collision, between two sub-clusters with different initial conditions \citep[as recently advocated by][]{Mapelli15}.

\section{Conclusions}
\label{conclude} 

We present $N$-body simulations of the dynamical evolution of star-forming regions with a wide range of initial conditions in order to investigate the behaviour of the radial velocity dispersion. Our conclusions can be summarised as follows.

(i) Using the velocity dispersion to estimate the virial state of a region becomes degenerate as the region dynamically evolves to a relaxed configuration. Comparison of the radial velocity dispersion to the virial velocity dispersion can suggest that a region is supervirial when its full phase-space information would imply that it is actually in virial equilibrium (Fig.~\ref{vir_rat}). This is especially true at later ages ($>$5\,Myr, or $\sim$50 crossing times). We postulate that this is because a dynamically evolved region is never exactly in virial equilibrium, but rather fluctuates about equilibrium. Nevertheless, this is a potential pitfall when analysing observational data.

(ii) Supervirial regions display a velocity dispersion that is well in excess of that estimated using the virial mass, $\sigma_{\rm vir}$  at very early times ($<$2\,Myr in the dense simulations, corresponding to $\sim$20 crossing times), and so regions that are strongly supervirial initially should not suffer from this degeneracy.

(iii) In star-forming regions undergoing violent relaxation (virial ratio $\alpha_{\rm vir} = 0.3 - 0.5$, fractal dimension $D = 1.6 - 2.0$), the most massive stars attain higher velocity dispersions than the velocity dispersion of the whole region. This implies that the most massive stars are moving faster than the average star, even if they have dynamically mass segregated. This does not appear to be due to these stars being in close binary systems, but rather that the massive stars have `decoupled' from the rest of the cluster, as argued by \citet{Allison11}.

(iv) Combining a measure of a star-forming region's structure -- i.e.\,\,the $\mathcal{Q}$-parameter -- with the velocity dispersion can be used to infer  the initial conditions of a star-forming region, but only if that region is dynamically evolved. We therefore recommend using the structural information detailed in \citet{Parker14b} in tandem with, or instead of, the velocity dispersion whenever it is available to determine the initial conditions.     

%%%%%%%%%%%%%%%%%%%%%%%%%%%%%%%%%%%%%%%%%%%%%%%%%%%%%%%%%%%%%%%%%%%%%%
\section*{Acknowledgements}

We thank the anonymous referee for their comments and suggestions, which have significantly improved the paper. RJP acknowledges support from the Royal Astronomical Society in the form of a research fellowship, and from the European Science Foundation (ESF) within the framework of the ESF `Gaia Research for European Astronomy Training' exchange visit programme (exchange grant 4994). NJW also acknowledges support from the Royal Astronomical Society in the form of a research fellowship.

\bibliographystyle{mn2e}
\bibliography{general_ref}

\label{lastpage}

\end{document}